\documentclass[journal,10pt,a4paper,twocolumn,twoside]{IEEEtran}

\usepackage{tikz-qtree}
\usepackage{subcaption}
\usepackage{caption}
\usepackage[numbers]{natbib}
\usepackage[english]{babel}
\usepackage[utf8x]{inputenc}
\usepackage{graphicx}
\usepackage{array}
\usepackage{hyperref}
\usepackage{amsmath}
\usepackage{amssymb}

\usepackage{braket}
\usepackage{balance}
\usepackage{amsthm}

\usepackage{multirow}
\usepackage{booktabs}

\usepackage{enumitem}

\let\Contentsline\contentsline
\renewcommand

\newcommand{\eqdef}{\stackrel{\triangle}{=}}

\hypersetup{colorlinks=true,citecolor=black,filecolor=black,linkcolor=black,urlcolor=black}

\usepackage{tikz}
\usetikzlibrary{arrows, calc, decorations.markings, positioning}
\makeatletter
\newenvironment{timeline}[6]{%
    \newcommand{\startyear}{#1}
    \newcommand{\tlendyear}{#2}
    \newcommand{\yearcolumnwidth}{#3}
    \newcommand{\rulecolumnwidth}{#4}
    \newcommand{\entrycolumnwidth}{#5}
    \newcommand{\timelineheight}{#6}
    \newcommand{\templength}{}
    \newcommand{\entrycounter}{0}
    \long\def\ifnodedefined##1##2##3{%
        \@ifundefined{pgf@sh@ns@##1}{##3}{##2}%
    }
    \newcommand{\ifnodeundefined}[2]{%
        \ifnodedefined{##1}{}{##2}
    }
    \newcommand{\drawtimeline}{%
        \draw[timelinerule] (\yearcolumnwidth+5pt, 0pt) -- (\yearcolumnwidth+5pt, -\timelineheight);
        \draw (\yearcolumnwidth+0pt, -10pt) -- (\yearcolumnwidth+10pt, -10pt);
        \draw (\yearcolumnwidth+0pt, -\timelineheight+15pt) -- (\yearcolumnwidth+10pt, -\timelineheight+15pt);
        \pgfmathsetlengthmacro{\templength}{neg(add(multiply(subtract(\startyear, \startyear), divide(subtract(\timelineheight, 25), subtract(\tlendyear, \startyear))), 10))}
        \node[year] (year-\startyear) at (\yearcolumnwidth, \templength) {\startyear};
        \pgfmathsetlengthmacro{\templength}{neg(add(multiply(subtract(\tlendyear, \startyear), divide(subtract(\timelineheight, 25), subtract(\tlendyear, \startyear))), 10))}
        \node[year] (year-\tlendyear) at (\yearcolumnwidth, \templength) {\tlendyear};
    }
    \newcommand{\entry}[2]{%
        \pgfmathtruncatemacro{\lastentrycount}{\entrycounter}
        \pgfmathtruncatemacro{\entrycounter}{\entrycounter + 1}
        \ifdim \lastentrycount pt > 0 pt%
            \node[entry] (entry-\entrycounter) [below of=entry-\lastentrycount] {##2};
        \else%
            \pgfmathsetlengthmacro{\templength}{neg(add(multiply(subtract(\startyear, \startyear), divide(subtract(\timelineheight, 25), subtract(\tlendyear, \startyear))), 10))}
            \node[entry] (entry-\entrycounter) at (\yearcolumnwidth+\rulecolumnwidth+10pt, \templength) {##2};
        \fi
        \ifnodeundefined{year-##1}{%
            \pgfmathsetlengthmacro{\templength}{neg(add(multiply(subtract(##1, \startyear), divide(subtract(\timelineheight, 25), subtract(\tlendyear, \startyear))), 10))}
            \draw (\yearcolumnwidth+0.5pt, \templength) -- (\yearcolumnwidth+7.5pt, \templength);
            \node[year] (year-##1) at (\yearcolumnwidth, \templength) {##1};
        }
        \draw ($(year-##1.east)+(2.5pt, 0pt)$) -- ($(year-##1.east)+(7.5pt, 0pt)$) -- ($(entry-\entrycounter.west)-(5pt,0)$) -- (entry-\entrycounter.west);
    }
    \newcommand{\plainentry}[2]{
        \pgfmathtruncatemacro{\lastentrycount}{\entrycounter}
        \pgfmathtruncatemacro{\entrycounter}{\entrycounter + 1}
        \ifdim \lastentrycount pt > 0 pt%
            \node[entry] (entry-\entrycounter) [below of=entry-\lastentrycount] {##2};
        \else%
            \pgfmathsetlengthmacro{\templength}{neg(add(multiply(subtract(\startyear, \startyear), divide(subtract(\timelineheight, 25), subtract(\tlendyear, \startyear))), 10))}
            \node[entry] (entry-\entrycounter) at (\yearcolumnwidth+\rulecolumnwidth+10pt, \templength) {##2};
        \fi
        \ifnodeundefined{invisible-year-##1}{%
            \pgfmathsetlengthmacro{\templength}{neg(add(multiply(subtract(##1, \startyear), divide(subtract(\timelineheight, 25), subtract(\tlendyear, \startyear))), 10))}
            \draw (\yearcolumnwidth+2.5pt, \templength) -- (\yearcolumnwidth+7.5pt, \templength);
            \node[year] (invisible-year-##1) at (\yearcolumnwidth, \templength) {};
        }
        \draw ($(invisible-year-##1.east)+(2.5pt, 0pt)$) -- ($(invisible-year-##1.east)+(7.5pt, 0pt)$) -- ($(entry-\entrycounter.west)-(5pt,0)$) -- (entry-\entrycounter.west);
    }
    \begin{tikzpicture}
        \tikzstyle{entry} = [%
            align=left,%
            text width=\entrycolumnwidth,%
            node distance=18pt,%
            anchor=west]
        \tikzstyle{year} = [anchor=east]
        \tikzstyle{timelinerule} = [%
            draw,%
            decoration={markings, mark=at position 1 with {\arrow[scale=1.5]{latex'}}},%
            postaction={decorate},%
            shorten >=0.2pt]
        \drawtimeline
}
{
    \end{tikzpicture}
    \let\startyear\@undefined
    \let\tlendyear\@undefined
    \let\yearcolumnwidth\@undefined
    \let\rulecolumnwidth\@undefined
    \let\entrycolumnwidth\@undefined
    \let\timelineheight\@undefined
    \let\entrycounter\@undefined
    \let\ifnodedefined\@undefined
    \let\ifnodeundefined\@undefined
    \let\drawtimeline\@undefined
    \let\entry\@undefined
}
\makeatother

\begin{document}

\newtheorem{theo}{Theorem}
\newtheorem{theor}{Theorem}
\newtheorem{cor}{Corollary}
\newtheorem{lem}{Lemma}
\newtheorem{prop}{Proposition}
\newtheorem{ins}{Insight}
\newtheorem{remark}{Remark}
\newtheorem{exmp}{Example}

\theoremstyle{definition}
\newtheorem{defin}{Definition}
\newtheorem{ass}{Assumption}
\newtheorem{rem}{Remark}

\title{{How Deep the Theory of Quantum Communications Goes: 
Superadditivity, Superactivation and Causal Activation}}
\author{Seid Koudia, Angela~Sara~Cacciapuoti$^*$,~\IEEEmembership{Senior Member,~IEEE,} Kyrylo Simonov, Marcello~Caleffi,~\IEEEmembership{Senior Member,~IEEE}
    \thanks{S. Koudia is with \textit{FLY: Future Communications Laboratory}, Department of Physics \textit{Ettore Pancini}, University of Naples Federico II, Naples, 80126 Italy. A.S. Cacciapuoti and M. Caleffi are with \textit{FLY: Future Communications Laboratory}, Department of Electrical Engineering and Information Technology (DIETI), University of Naples Federico II, Naples, 80125 Italy. K. Simonov is independent researcher. E-mail: \href{mailto:seid.koudia@unina.it}{seid.koudia@unina.it} \href{mailto:angelasara.cacciapuoti@unina.it}{angelasara.cacciapuoti@unina.it}, \href{mailto:kyrylo.simonov@univie.ac.at}{kyrylo.simonov@univie.ac.at}, \href{mailto:marcello.caleffi@unina.it}{marcello.caleffi@unina.it}. Web: \href{http://www.quantuminternet.it}{www.quantuminternet.it}.}
	\thanks{A.S. Cacciapuoti and M. Caleffi are also with the Laboratorio Nazionale di Comunicazioni Multimediali, National Inter-University Consortium for Telecommunications (CNIT), Naples, 80126, Italy.}
	\thanks{This work was partially supported by project xxx.}
	\thanks{$^*$Corresponding author.}
}

\maketitle

\begin{abstract}
In the theory of quantum communications, a deeper structure has been recently unveiled, showing that the capacity does not completely characterize the channel ability to transmit information due to phenomena -- namely, superadditivity, superactivation and causal activation -- with no counterpart in the classical world. Although how deep goes this structure is yet to be fully uncovered, it is crucial for the communication engineering community to own the implications of these phenomena for understanding and deriving the fundamental limits of communications. Hence, the aim of this treatise is to shed light on these phenomena by providing the reader with an easy access and guide towards the relevant literature and the prominent results from a communication engineering perspective.
\end{abstract}


\section{Introduction}
\label{Sec:1}

\IEEEPARstart{T}{ransmitting} data reliably over noisy communication channels is one of the key applications of information theory, and it is well understood for channels modelled by classical physics. Initiated by Shannon's seminal work \cite{27}, the study of communication channels involving the exchange of classical data led to over time the establishment of the field of classical Shannon theory. The greatest achievement of the latter is the realization that any noisy communication channel can be modeled as a stochastic map connecting input signals selected by a sender -- say Alice -- who operates at one end of the channel, to the corresponding output accessible to the receiver -- say Bob. Shannon stressed that the performance of this communication channel is gauged by a single quantity, the so-called \textit{capacity} of the channel.

Nevertheless, information is not just an abstract mathematical notion. Instead, it exhibits an intrinsic relationship with the physical channel nature, which poses fundamental limits on the possibility of processing or transferring it. This is where quantum theory comes into play in the study of communication channels \cite{29}. As a matter of fact, any two parties wishing to exchange information should encode it in the state of some system acting as information carrier. Whenever the system exhibits a quantum nature -- such as a photonic pulse propagating through an optical fiber -- the propagation of the information carrier as well as the overall processing must follow the principles and the laws of quantum mechanics. Accordingly, as a generalization of channels in Shannon theory, \textit{quantum channels} are introduced, linking the initial states of quantum information carriers controlled by Alice with their output states manipulated by Bob.

One surprising quantum effect, which can be resourceful for this paradigmatic shift from classical to quantum communications, is quantum entanglement. This new type of correlations, with no classical counterpart, can boost the communication capabilities drastically. In fact, despite that an entangled state shared between Alice and Bob -- alone -- does not provide any communication possibilities \cite{33}, when used to \textit{assist} a quantum channel, it can enhance the performance by doubling the classical capacity as in \textit{quantum superdense coding} \cite{31}. Or, even more surprising, it can enable the transfer of quantum information with the transmission of two classical bits as in \textit{quantum teleportation} \cite{33,31,32}.

\begin{table*}[t]
	\centering
    \begin{tabular}{| p{0.15\textwidth} | p{0.35\textwidth} | p{0.45\textwidth}|}
		\toprule
		\textbf{} & \textbf{Classical Communications} & \textbf{Quantum Communications}\\
		\midrule
		\textbf{non-zero-capacity channels} & $n$ uses of a communication channel \textit{do not} transmit more than $n$ times the amount of information that can be transmitted with a single channel use (\textbf{additivity}) & - $n$ uses of a communication channel can transmit \textit{more} than $n$ times the amount of information that can be transmitted with a single channel use (\textbf{superadditivity}) \\
		& & - channels combined in a quantum trajectory can transmit more information with respect to a classical placement of the same channels (\textbf{causal activation})\\
        \midrule
        \textbf{zero-capacity channels} & \textit{can not} transmit information, regardless of the number of uses and/or the placement of these channels & \textit{can} transmit information either with a classical placement of different channels (\textbf{superactivation}) or by combining the channels in a quantum trajectory (\textbf{causal activation}) \\
		\bottomrule
	\end{tabular}
	\caption{Classical vs quantum communications. Superadditivity, superactivation, and causal activation can enable an unparalleled boost of the capacity of a quantum channel, which is not achievable in classical communications.}
	\label{Tab:Intro}
	\hrulefill
\end{table*}

\begin{figure*}[t]
    \centering
    \includegraphics[width=1\textwidth]{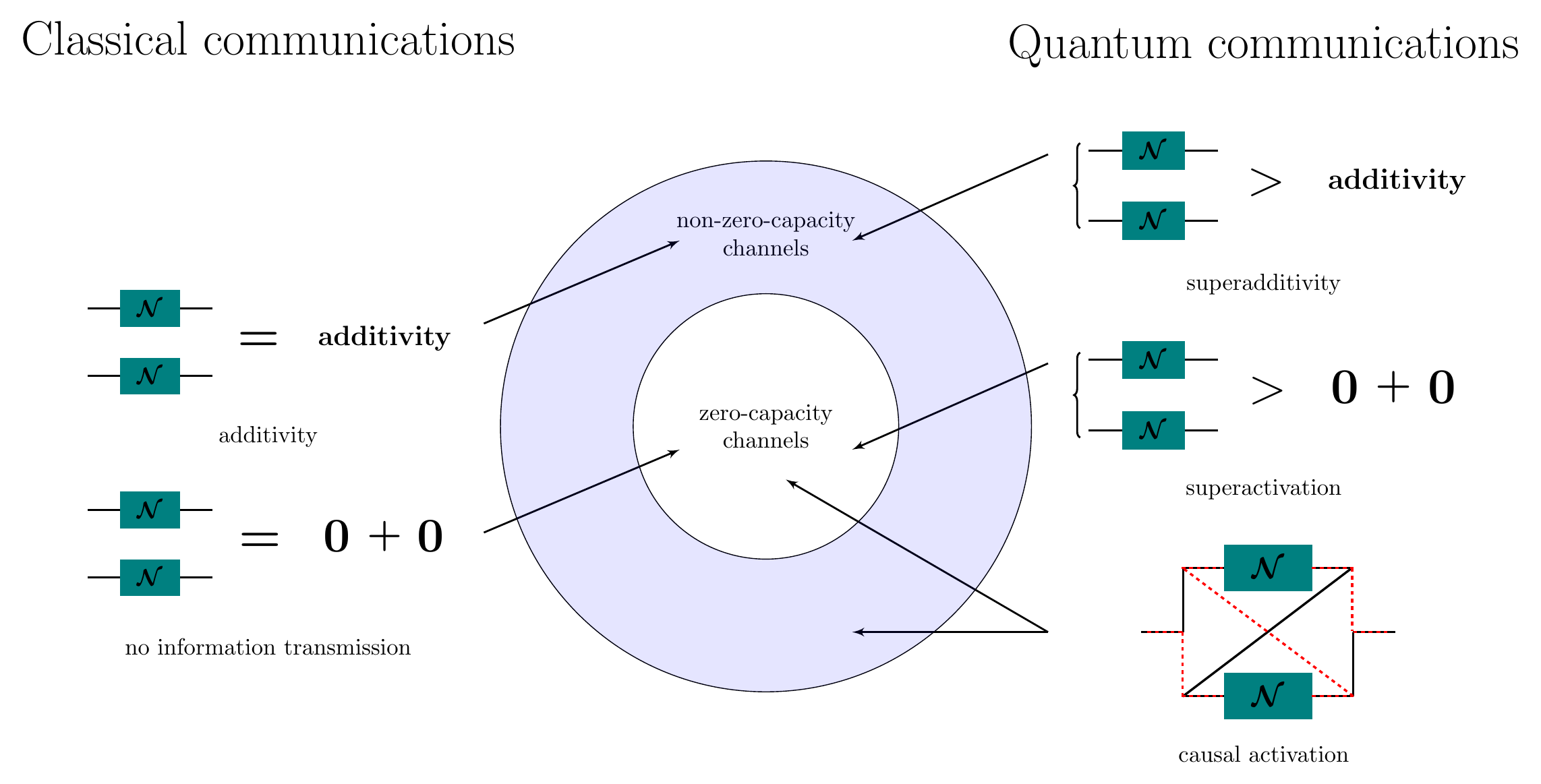}
    \caption{Pictorial representation of non-zero vs zero-capacity channels highlighting the different phenomena -- namely, superadditivity, superactivation, and causal activation -- affecting the fundamental notion of channel capacity in ways with no counterpart in the classical Shannon theory.}
    \label{NewFig:2}
    \hrulefill
\end{figure*}

However, quantum Shannon theory has more to offer, as summarized in Table~\ref{Tab:Intro} and pictorially represented in Figure~\ref{NewFig:2}. Indeed, a proper channel encoder allowed to encode the information -- either classical or quantum -- into entangled states enhances the performance achievable thorough a quantum channel. This potential gain is referred to as \textit{superadditivity} of the quantum channel capacity, and such a topic constituted a long and hot debate in the quantum communications community \cite{6,15,24,26}.

Even more astonishing, there exists pairs of channels that, although they do not have individually the ability to transmit any amount of quantum information, are able to transmit information when used together on entangled inputs. This is known as the \textit{superactivation} phenomenon \cite{4,14,16}, which shows that the quantum capacity is a strongly non-additive quantity.

Both the superadditivity and the superactivation
phenomena, which have no counterpart in the classical Shannon theory, induce an yet to be solved question on how different noisy channels interact and enhance each other's capabilities, as we will highlight and discuss in the following.

But the marvels of the quantum realm are not by any means limited to the unconventional phenomena of superadditivity and superactivation. Indeed, quantum Shannon theory deals with information encoded in quantum carriers, but still considers the propagation of information through classical trajectories, so that the path taken by messages in space is always well-defined, i.e., where channels are in definite causal order. 

Counter-intuitively, quantum mechanics allows quantum particles to propagate simultaneously among multiple space-time trajectories. This ability enables a quantum information carrier to propagate through a \textit{quantum trajectory} \cite{36,18,3,13,9}. An important setup is given by a quantum trajectory where the constituting communications channels are combined in a quantum superposition of different orders, so that the causal order of the channels become indefinite. This unconventional placement of the channels is theoretically and experimentally implemented through the \textit{quantum switch}, which is a supermap resulted from an extension of quantum mechanics under the name of process matrix formalism \cite{39,40} or before this, quantum combs \cite{37,38}.

The superposition of trajectories and the quantum switch supermap have proved to be able to describe powerful setups for the transmission of classical/quantum information \cite{19}. As instance, whenever Alice and Bob are restricted to use quantum channels with zero-classical-capacity, no classical information can be sent throughout any classical configuration of the channels, neither parallel or sequential. Conversely, a \textit{causal activation}\footnote{The term causal activation was coined in \cite{18} to distinguish the phenomenon of activating vanishing capacities of quantum channels with indefinite causal order of channels from the known phenomenon of superactivation \cite{13}.} of the classical capacity\footnote{Indeed, causal activation occurs also for quantum capacities, as discussed in Section~\ref{Sec:6}.} occurs when the channels are placed in a quantum configuration through the quantum switch, and non-vanishing information can be transmitted from Alice to Bob.

The unconventional phenomenon of \textit{causal activation} led researchers to work toward the extension of quantum Shannon theory for modelling coherent superposition of quantum channels \cite{3} as well as superposition of their causal orders \cite{13} as a communication resource. This extension should not come as a surprise. Indeed, also within the ``\textit{classical}'' quantum Shannon theory, phenomena such as \textit{superadditivity} and \textit{superactivation} prove that the communication potential of a channel strictly depends on the context in which it is used. Hence, this shows that genuine quantum phenomena play a paramount role for future communications, and they should be fully understood and harnessed to achieve unprecedented information transfer capacities.

\subsection{Outline and Contribution}
\label{Sec:1.1}

As mentioned above, \textit{superadditivity}, \textit{superactivation}, and \textit{causal activation} are all phenomena affecting the fundamental notion of channel capacity -- as introduced by Claude Shannon with his seminal work \cite{27} -- in ways with no counterpart in the classical Shannon theory. Unfortunately, the existing literature is \textit{prepared by} and \textit{prepared for} the physics community. This still leads to a fundamental gap between the literature and the communications engineering community.

The aim of this paper is precisely to bridge this gap, by introducing the most novel, astonishing and intriguing properties of quantum communications, which can:
\begin{itemize}
    \item provide a capacity gain for both classical and quantum information through the \textit{superadditivity} phenomenon,
    \item provide a non-null capacity for quantum information through the \textit{superactivation} phenomenon,
    \item provide both a capacity gain (when the individual channels exhibit non-null capacity) or a non-null capacity (when the individual channels are zero-capacity channels) for both classical and quantum information through the \textit{causal activation} phenomenon, by exploiting the genuine quantum placement of quantum channels provided by quantum trajectories.
\end{itemize}

Stemming from the discussion above, in the following we shed the light on the notions of superadditivity and superactivation of quantum channel capacities, as well as the more recently discovered phenomenon of causal activation of different capacities, that accompanies the propagation of information along quantum trajectories, with the objective of allowing the reader:

\begin{enumerate}[label=\roman*)]
    \item to own the implications of these phenomena for understanding and deriving the fundamental limits of communications; 
    \item to grasp the challenges as well as to appreciate the marvels arising with the paradigmatic shift from designing classical communications to design quantum communications.
\end{enumerate}

Through the manuscript, the nature of these phenomena and, in particular, the differences among the resources responsible for these advantages are elaborated. In fact, the understanding of these phenomena is a key to grasp how different resources can be distributed through quantum networks \cite{marcello} more efficiently, and how they can be used optimally in the engineering of a near-term Quantum Internet \cite{Kimble_2008,Pirandola2016UniteTB,CalCacBia2018,CacCalTafCatGherBia2020,CalChaCuoHasCac2020,CacCalRodLaj2020,Wehner2018QuantumIA,lajos,Razavi2012}. Indeed, due to the fast grow of both fields, such an understanding will serve the quantum engineering and the communications engineering communities alike to have an easy access and guide towards the relevant literature and to the prominent results, which will be of paramount importance for designing efficient communication protocols.

To the best of authors' knowledge, a tutorial of this type is the first of its own.

\begin{figure}[t]
    \centering
    \begin{minipage}[c]{1\columnwidth}
        \usetikzlibrary{trees}
        \tikzstyle{every node}=[thick,anchor=west]
       \begin{tikzpicture}[level distance=2cm, grow via three points={one child at (0.3,-0.55) and
        two children at (0.3,-0.5) and (0.3,-1.15)},
        edge from parent path={([xshift=0.0mm] \tikzparentnode.south west) |- (\tikzchildnode.west)},
        growth parent anchor=south west,
  edge from parent/.style = {draw, -latex}]
  \node {}
    child { node {\small I. Introduction}
      child[xshift=0.1cm] { node {\small I-A. Outline and Contribution}}
    }		
    child [missing] {}
    child { node {\small II. Preliminaries}
      child[xshift=0.1cm] { node {\small II-A. Quantum background}}
      child[xshift=0.1cm] { node {\small III-B. From Classical Capacity to Quantum Capacities}}
      child [xshift=0.1cm]{ node {\small II-C. Operational Definition of Quantum Channel Capacities}}
      child[xshift=0.1cm] { node  {\small II-D. Classical Capacity of Quantum Channels}}
      child [xshift=0.1cm]{ node {\small II-E. Quantum Capacity of Quantum Channels}}
      child [xshift=0.1cm]{ node {\small II-F. Bibliographic Notes}}
    }
    child [missing] {}
    child [missing] {}				
    child [missing] {}				
    child [missing] {}
    child [missing] {}				
    child [missing] {}
    child { node {\small  III. Quantum Marvels}
      child [xshift=0.1cm]{ node {\small III-A. Superadditivity}}
      child [xshift=0.1cm]{ node  {\small III-B. Superactivation}}
      child [xshift=0.1cm]{ node {\small III-C. Causal Activation}}
    }
    child [missing] {}				
    child [missing] {}				
    child [missing] {}
    child { node {\small IV. Superadditivity of Quantum Channel Capacities}
      child [xshift=0.1cm]{ node {\small IV-A. Superadditivity of Holevo Information}}
      child [xshift=0.1cm]{ node  {\small IV-B. Superadditivity of Coherent Information}} 
      child[xshift=0.1cm] { node {\small IV-C. Superadditivity of Classical and Quantum Capacities}}
    }
    child [missing] {}				
    child [missing] {}				
    child [missing] {}
    child { node {\small V. Superactivation of Quantum Channel Capacities}
      child [xshift=0.1cm] { node {\small V-A. Classes of Zero Capacity Channels}}
      child [xshift=0.1cm] { node  {\small V-B. Superactivation of Quantum Capacity}}
      child[xshift=0.1cm] { node {\small V-C. Non-convexity of Quantum Capacity }}
      child[xshift=0.1cm] { node {\small V-D. Classical Capacity }}
    }
    child [missing] {}				
    child [missing] {}				
    child [missing] {}
    child [missing] {}
    child { node {\small VI. Causal activation of Quantum Channel Capacities }
      child [xshift=0.1cm]{ node {\small VI-A. Quantum Switch}}
      child [xshift=0.1cm]{ node  {\small VI-B. Causal Activation of Holevo Information}}
      child [xshift=0.1cm]{ node {\small VI-C. Causal Activation of Quantum Capacity }}
    }
    child [missing] {}				
    child [missing] {}				
    child [missing] {}
    child { node {\small VII. Conclusions and Future Perspectives }
    child[xshift=0.1cm] { node {\small VII-A. Summary}}
    child [xshift=0.1cm]{ node  {\small VII-B. Open Problems}}
    }
    child [missing] {}
    child [missing] {}
    child { node {\small  VIII. Appendices  }
    child [missing] {}};
\end{tikzpicture}
    \end{minipage}
    \caption{Paper Structure}
    \label{fig:content}
    \hrulefill
\end{figure}
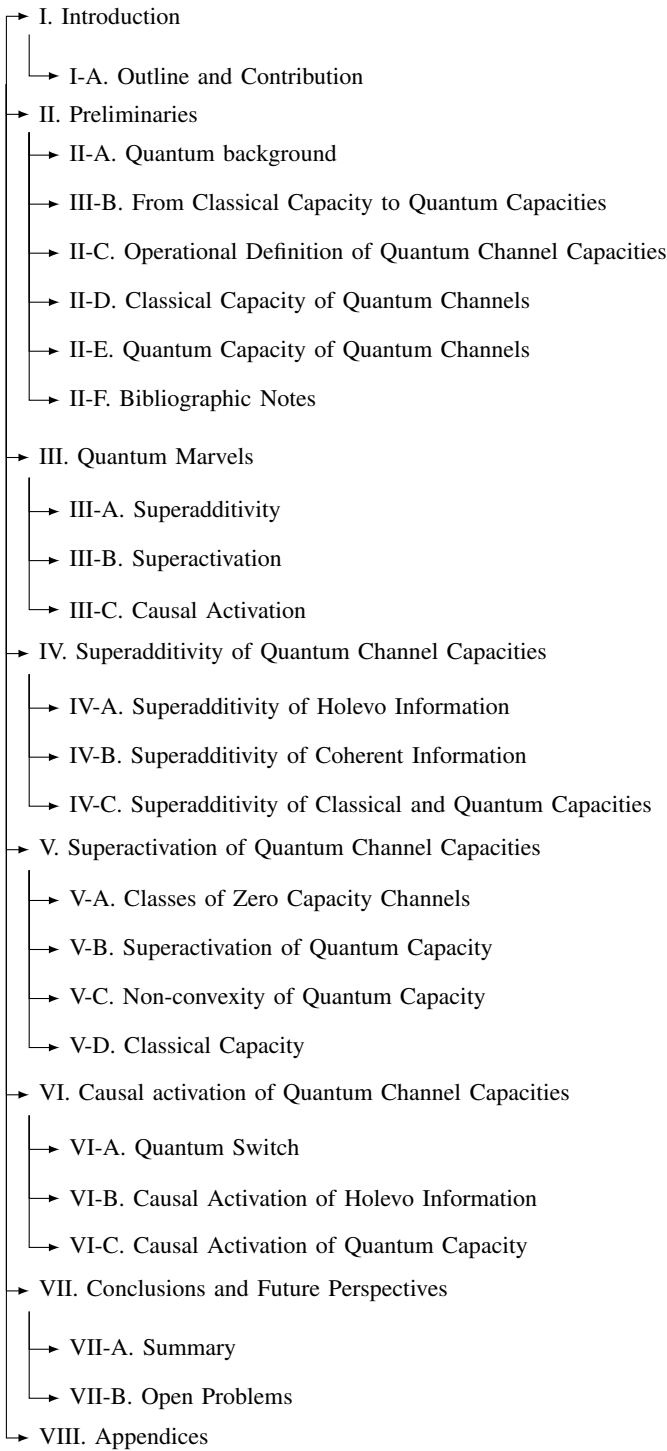

The paper is structured as depicted in Figure~\ref{fig:content}. Specifically, 
in Section~\ref{Sec:2}, we provide the reader -- by assuming a basic background of classical Shannon theory -- with a concise description of the preliminaries needed to understand and to formally characterize these phenomena. Then
in Section~\ref{Sec:3}, we conduct an informal description of the three unconventional phenomena -- superactivation, superadditivity and causal activation -- from a communication engineering perspective. In Section~\ref{Sec:4}, we first discuss the superadditivity phenomenon for one-shot capacities -- i.e., Holevo information and coherent information -- and then we generalize our discussions to regularized capacities. Continuing further our discussions, in Section~\ref{Sec:5} we detail the superactivation phenomenon for quantum capacities, and we point out the rationale behind being it restricted to quantum information. In Section~\ref{Sec:6}, we discuss the causal activation phenomenon for different capacities, ranging from Holevo information through coherent information to classical and quantum regularized capacities. Finally, we conclude our tutorial in Section~\ref{Sec:7}. Specifically, we first summarizing the differences and similarities between the communication advantages of these three phenomena, in terms of resources enabling these advantages. Then, we discuss the challenges and open problems arising with the engineering of these phenomena from a communication engineering perspective. Supplementary material is included in Appendices~\ref{App:0.1}-\ref{App:4} with the aim of providing the reader outside the specialty of the article with an easy-to-consult summary of some definitions and results.

\begin{table*}
	\centering

    \begin{tabular}{| p{0.6\textwidth} |p{0.1\textwidth}|}
		\toprule
		    \textbf{Notion} & \textbf{Appearance} \\
		\midrule
		    Quantum bit & Appendix~\ref{AppA:1}  \\
        \midrule
            Superposition & Appendix~\ref{AppA:1}   \\ 
        \midrule
            Unitary transformation &  Appendix~\ref{AppA:2}  \\
        \midrule
            Projective measurement &  Appendix~\ref{AppA:3}\\
        \midrule
            Mixed state &  Appendix~\ref{AppA:4}   \\
        \midrule
            Pure state &  Appendix~\ref{AppA:4}   \\
        \midrule
            Density matrix &  Appendix~\ref{AppA:4} \\
        \midrule
            Positive operator-valued measure (POVM) &  Appendix~\ref{AppA:5}\\
        \midrule
            Entangled state &  Appendix~\ref{AppA:6}  \\
        \midrule
            Quantum channel &  Appendix~\ref{App:1}  \\
        \midrule
            Completely positive trace-preserving (CPTP) map &  Appendix~\ref{App:1}  \\
        \midrule
            Kraus representation &  Appendix~\ref{AppB:A}  \\
        \midrule
            Isometric extension (Stinespring dilation) &  Appendix~\ref{AppB:B}  \\
        \midrule
            Choi state &  Appendix~\ref{AppB:C}  \\
        \midrule
            (Anti-)Degradability of a quantum channel  &  Appendix~\ref{App:2}  \\
        \midrule
            Von Neumann entropy &  Appendix~\ref{App:3}  \\
        \midrule
            Entropy of exchange &  Appendix~\ref{App:3}  \\
        \midrule
            Holevo information &  Appendix~\ref{App:3}  \\
        \midrule
            Quantum mutual information &  Appendix~\ref{App:3}  \\
        \midrule
            Conditional von Neumann entropy &  Appendix~\ref{App:3}  \\
        \midrule
            Entropy of exchange &  Appendix~\ref{App:3}  \\
        \midrule
            Codeword &  Appendix~\ref{App:4}  \\
        \midrule
            Rate of the code &  Appendix~\ref{App:4}  \\
        \bottomrule
	\end{tabular}
	\caption{Basic notions and sections of the manuscript where their formal mathematical definitions are defined or introduced.}
	\label{Tab:App}
	\hrulefill
\end{table*}

\section{Preliminaries}
\label{Sec:2}

Ever since its almost 100-year history, quantum mechanics has not only strikingly challenged our view of Nature. Its novel counter-intuitive concepts without classical counterparts \cite{Jammer1966TheCD} have found their applications in a plethora of branches of science and engineering, and they revolutionized them. This has turned quantum mechanics from a formalism built to describe certain unexplained physical phenomena (e.g., black-body radiation and photoelectric effect) and fit experimental data to a machinery that can be used in developing technologies that rely upon quantum effects.

Here, we provide a concise introduction to concepts and formalism needed to present and to discuss the phenomena of superadditivity, superactivation and causal activation. The basic notions and the notation adopted throughout the paper are summarized in Table~\ref{Tab:App} and Table~\ref{Tab:00}, respectively, along with the indication of the section of the manuscript in which the corresponding concept is formally defined or introduced.

\begin{table*}
	\centering

    \begin{tabular}{| p{0.15\textwidth} |   p{0.60\textwidth}|p{0.1\textwidth}|}
		\toprule
		    \textbf{Symbol} & \textbf{Definition} &\textbf{Appearance} \\
        \midrule
            $C(\mathcal{N})$ & The classical capacity of the quantum channel $\mathcal{N}$ & Section~\ref{Sec:2.4}\\
        \midrule
            $Q(\mathcal{N})$ & The quantum capacity of the quantum channel $\mathcal{N}$ & Section~\ref{Sec:2.5}\\
		\midrule
		    $\mathcal{H}$ &  Hilbert space&  Appendix~\ref{App:0.1}\\
        \midrule
            $\ket{\psi}$ & Pure state&  Appendix~\ref{App:0.1}\\ 
        \midrule
            $A^\dagger$  & Conjugate transpose of an operator A &  Appendix~\ref{App:0.1}\\
        \midrule
            $\{M_i\}$   & Elements of a measurement setup POVM &  Appendix~\ref{App:0.1}\\
        \midrule
            $\rho$ & Density operator &  Appendix~\ref{App:0.1}\\
        \midrule
            $\otimes$  & Tensor product &  Appendix~\ref{App:1}\\
        \midrule
            $\mathcal{L}(\mathcal{H})$  & The set of density operators on the Hilbert space $\mathcal{H}$ &  Appendix~\ref{App:1}\\
        \midrule
            $\mathcal{N}(\cdot)$ & A quantum channel &  Appendix~\ref{App:1}\\
        \midrule
            $\mathcal{N}(\cdot)=\sum_iK_i\cdot K_i^{\dagger}$ & Kraus decomposition of the channel $\mathcal{N}$ &  Appendix~\ref{App:1}\\
        \midrule
            $\mathcal{U}_{\mathcal{N}}$  & The isometric extension of the channel $\mathcal{N}$ &  Appendix~\ref{App:1}\\
        \midrule
            $\mathcal{N}^c$  & The complementary channel of the channel $\mathcal{N}$ &  Appendix~\ref{App:1}\\
        \midrule
            $\tilde{\Phi}^{BA'}_{\mathcal{N}}$  & The Choi state of the channel $\mathcal{N}$ &  Appendix~\ref{App:1}\\
        \midrule
            $S(\rho)$  & The von Neumann entropy of the density operator $\rho$ &  Appendix~\ref{App:3}\\
        \midrule
            $H(X)$   & The Shannon entropy of the random variable $X$ &  Appendix~\ref{App:3}\\
        \midrule
            $\{p_x,\rho_x\}$  & An ensemble of quantum states &  Appendix~\ref{App:3}\\
        \midrule
            $\chi(\{p_x,\rho_x\},\mathcal{N})$  & The Holevo information of the channel $\mathcal{N}$ with the input ensemble $\{p_x,\rho_x\}$ &  Appendix~\ref{App:3}\\
        \midrule
            $I(X:Y)$  & The mutual information between the random variables $X$ and $Y$ &  Appendix~\ref{App:3}\\
        \midrule
            $I(\rho,\mathcal{N})$   & The quantum mutual information between the output of the channel $\mathcal{N}$ and the input state $\rho$ &  Appendix~\ref{App:3}\\
        \midrule
            $I_c(\rho,\mathcal{N})$   & The coherent information of the channel $\mathcal{N}$ with respect to the input state $\rho$ &  Appendix~\ref{App:3}\\
        \midrule
            $S(A|B)$  & The conditional von Neumann entropy between quantum systems $A$ and $B$ &  Appendix~\ref{App:3}\\
		\bottomrule
	\end{tabular}
	\caption{Adopted notation and section of the manuscript where the notation is defined or introduced.}
	\label{Tab:00}
	\hrulefill
\end{table*}

\subsection{Quantum background}
\label{Sec:2.1}

\subsubsection*{A.1) The quantum bit}
What makes quantum mechanics attractive from a communications engineering perspective? First of all, its very principles offer a novel way to treat information when encoded in a quantum system. Classically, two mutually exclusive states -- i.e., $0$ and $1$ -- can be encoded in a bit, which is in only one of these states at any time. Conversely, suppose now that two states $\ket{0}$ and $\ket{1}$ of a quantum two-level system (for example, the polarization of a photon) are used to encode them\footnote{Above we utilized the bra-ket notation usually adopted for quantum state. For a proper introduction to this notation, we refer the reader to Appendix~\ref{App:0.1}.}. In this case, the \textit{superposition principle} -- the corner-stone of quantum mechanics -- allows to go beyond bit's classical behavior, since the system can be in both states simultaneously. Hence, we can introduce the quantum bit (qubit) whose state $\ket{\psi}$ encodes more than simply the states $\ket{0}$ and $\ket{1}$, since it can be in a superposition of them as follows:
\begin{equation}
    \label{eq:2.1}
    \ket{\psi}= \alpha\ket{0}+\beta\ket{1}
\end{equation}
with $\alpha, \beta \in \mathbb{C}$, known as \textit{amplitudes}, satisfying $|\alpha|^2+|\beta|^2=1$. Hence, a qubit can encode not only classical information (the states $\ket{0}$ and $\ket{1}$) but also quantum information manifested in the coherence (carried by the \textit{complex} amplitudes $\alpha$ and $\beta$) it can possess. This type of information has no classical counterpart. An important consequence of the superposition principle is a new way of processing and encoding information \footnote{An illustration of this feature is provided by the Elitzur-Vaidman bomb testing problem: we are supposed to have a bunch of bombs that are activated by a sensor absorbing a photon. Since some sensors have a defect and do not absorb photons, we have to select the working bombs from the bunch. Classically, there is no way to find out whether a bomb works properly without making it actually explode by shining light on the sensor. However, if a photon before reaching the sensor hits a half-silvered mirror, the superposition principle allows to distinguish -- probabilistically, with a success ratio as high as $33\%$ -- between the working and faulty bombs and selects some of the working ones without explosion \cite{Elitzur-93, Penrose-94}.}, which can be exploited to significantly increase the security of communications and even to exchange information without actual transmission of the information carrier between the parties \cite{Salih-13, Vaidman-19}. A rigorous definition of the qubit is given in Appendix~\ref{App:0.1}.

\vspace{6mm}
\subsubsection*{A.2) Quantum measurement}
In order to retrieve data from a qubit, one has to perform a measurement of the corresponding degree of freedom encoding the information (for example, polarization of the photon). For a superposed state of a qubit, the result of the measurement is probabilistic due to the Born's rule of quantum mechanics. For instance, for the qubit given in \eqref{eq:2.1}, one obtains state $\ket{0}$ with the probability $|\alpha|^2$ and state $\ket{1}$ with the probability $|\beta|^2$, hence retrieving at most one bit of information. Crucially, the measurement causes the state of the qubit to collapse to the measured state. Indeed, if the measurement of the qubit given in \eqref{eq:2.1} has revealed the state $\ket{0}$, any further measurement will reveal the same outcome regardless of the initial superposition. This means that the measurement irreversibly alters the state of the qubit, which loses thus the coherence previously existing between the two states $\ket{0}$ and $\ket{1}$. A formal definition of quantum measurements is given in Appendix~\ref{App:0.1}.

\vspace{6mm}
\subsubsection*{A.3) No-cloning}
Classical communication protocols rely on the ability to copy the information and to transmit it to many different users. This fundamental assumption is widely exploited through the whole protocol stack \cite{Jessica-2022}. Conversely, quantum information cannot be copied or cloned, as stated by the \textit{no-cloning theorem} \cite{Wootters-1982}. In simpler terms, quantum information cannot be multicasted or broadcasted, in contrary to classical information. Consequently, the no-cloning theorem poses drastic unconventional challenges for the design of quantum networks, as most of the known classical protocols fail to be extended to the quantum paradigm \cite{Jessica-2022}. Fortunately, a non-trivial caveat to some of the restrictions would rely on the notion of entanglement and its astounding advantages. 

\vspace{6mm}
\subsubsection*{A.4) Entanglement}
The superposition principle leads to a number of intriguing genuinely quantum phenomena, including the celebrated \textit{entanglement} \cite{Schrdinger2005DieGS}. Entanglement is a sort of correlations between parties of some (joint) system, which have no classical counterpart. In his seminal paper \cite{Bell5}, John Bell has established constraints on correlations between two systems that cannot be broken by classical correlations. These constraints can be formalized in a form of inequalities for the statistical properties of outcomes of measurements performed on the joint system (the most famous form of Bell inequalities is also known as the CHSH inequalities). It has been shown that quantum entanglement can violate such inequalities. This makes entanglement a invaluable resource that might beat classical resources in different communications contexts. Although it remained until the end of the last century the question of what entanglement is useful for, eventually entanglement has been harnessed to outperform classical communication protocols and to provide security for quantum key distribution \cite{Ekert-1991,Pironio-2009}\footnote{See Appendix~\ref{App:0.1} for an overview of the basic quantum-informational notions.}. Specifically, \textit{quantum superdense coding} \cite{32} came against what was previously known in information theory to be a coding bound for classical information. Classically, if a sender -- say Alice -- wants to communicate a two-bit message to a receiver -- say Bob -- she has to use twice a single-bit classical channel. The same still holds even if Alice and Bob are connected by a quantum channel transmitting classical bits encoded within qubits. Conversely, if Alice and Bob share a-priori entanglement, a two-bit classical message can be sent through a single use of a quantum channel. Furthermore, this protocol has proved not to just outperform the performance of classical communication protocols, but also to be extremely secure \cite{32,Ekert-1991,Pironio-2009}. But there is more to it. A qubit can never be transmitted using only classical channels, as these latter can not preserve the genuine quantum coherence \cite{resource}. Luckily, the \textit{quantum teleportation protocol} -- the dual of superdense coding -- allows for the transmission of an unknown qubit state using a two-bit classical channel \cite{31}, by exploiting again entanglement as a fundamental resource. The design of these two protocols challenged the classical notions of information theory and classical communications, and it opened the door towards a new era of quantum communications. 

\vspace{6mm}
\subsubsection*{A.5) Quantum channels}
In communications engineering, information is usually encoded according to the physical medium that carries it. This physical medium is usually modeled as a classical channel, which does not take into account the quantum mechanical properties of the physical system carrying the information. Conversely, in quantum communications, quantum channels model the physical medium by considering the quantum mechanical properties of the physical carrier as well as its quantum interactions with the physical environment. The rationale for this is to keep track of the coherence present in the physical carrier, and to harness its advantages in encoding classical and quantum information alike. Indeed, quantum channels -- with particular instances given by optical fibers and free-space carrying quantum light -- might be seen as transformations of a given quantum mechanical system state, inducing its evolution from an initial state (input of the channel) to the final state (output of the channel). Accordingly, classical channels might be seen as a particular class of quantum channels where quantum coherences are completely absent. This paradigm shift from classical to quantum channels affects the very same concept of capacity, the quantity characterizing communication channels performance, as introduced in the following paragraph.

\vspace{6mm}
\subsubsection*{A.6) Channel Capacity}
Capacity is an intrinsic property of communication channels -- be them classical or quantum -- which measures the maximum rate at which information can be reliably transferred between Alice and Bob. The capacity establishes the ultimate boundary between communication rates that are achievable in principle and those which are not. Indeed, when quantum effects are involved, there does not exist a \textit{single notion} of capacity to evaluate the performance of a quantum channel. Rather, there exist \textit{multiple, nonequivalent} definitions of capacities \cite{7,Wilde}, as introduced in Section~\ref{Sec:2.2} and described in details with Sections~\ref{Sec:2.4} and \ref{Sec:2.5}.

\subsection{From Classical Capacity to Quantum Capacities}
\label{Sec:2.2}

When a communication channel is used to communicate messages between two parties, Alice and Bob, it is fundamental to assess the channel capacity -- namely, the maximal amount of information Alice and Bob could reliably transfer by choosing appropriate encoding and decoding operations\footnote{See Appendix~\ref{App:4} for a concise overview and a formal description of the encoding/decoding operations.}.

\begin{figure}[t]
    \centering
    \includegraphics[width=1\columnwidth]{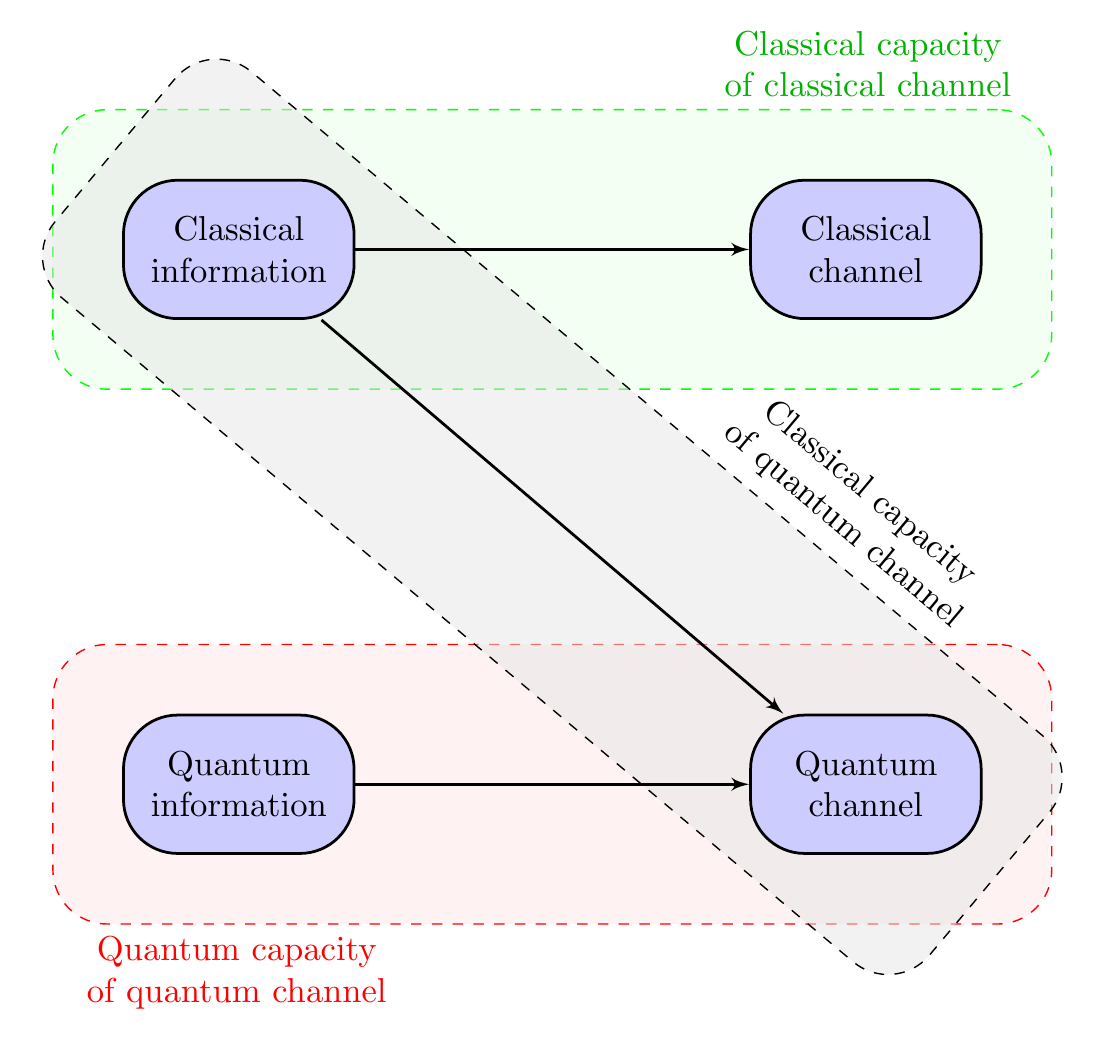}
    \caption{Classical vs Quantum Capacity. The capacity of a channel measures the maximum rate at which information can be reliably transferred between communication parties through such a channel. A classical channel can be used to send classical information only and, therefore, it is fully characterized by its classical capacity. A quantum channel can transmit either classical or quantum information, and the corresponding rates are bounded by its classical and quantum capacities, respectively.}
    \label{NewFig:1}
    \hrulefill
\end{figure}

What is meant by reliable is that there is an infinitesimally vanishing probability that the message, sent by Alice, arrives with any alteration to Bob \cite{41}. The condition of vanishing error probability is generally imposed in the asymptotic limit where infinitely long codes are allowed. In this setup, an explicit closed-form expression for classical channel capacities exists, which depends on the noise model given through the conditional probability $p(y|x)$ characterizing the channel, where $x$ and $y$ denote the input and the output messages, respectively. Accordingly, the classical capacity is expressed as \cite{41}:
\begin{equation}
    C = \max_{p(x)} I(X:Y)
    \label{eq:2.2}
\end{equation}
where the maximization is over all probability distributions on x, and where $I(X:Y)$, defined in \eqref{eq:app.3.7b} reported in Appendix~\ref{App:3}, denotes the mutual information between the input and output random variables $X$ and $Y$.

\vspace{6pt}

Surprisingly and contrary to the classical case, extending this framework to quantum channels leads to the introduction of different capacities, depending on which context -- i.e., depending on whether Alice and Bob are exchanging classical, private or quantum information -- the quantum channel is used for \cite{7,Wilde}, as shown in Figure~\ref{NewFig:1}.

In the following, we will restrict our attention on: i) the classical capacity $C(\cdot)$ over quantum channels, and ii) the quantum capacity $Q(\cdot)$ over quantum channels. A general scheme for classical/quantum capacity is shown in Figure~\ref{Fig:05}. Specifically, the former capacity $C(\cdot)$ deals with the transmission of classical information through a quantum information carrier, by assuming the presence of proper classical-to-quantum encoder $\mathcal{E}$ and decoder $\mathcal{D}$, whereas the latter capacity $Q(\cdot)$ requires the availability of quantum-to-quantum encoder/decoder for allowing the transmission of quantum information.

Furthermore, for each of the mentioned capacities, we are going to distinguish between one-shot capacities $\chi(\cdot)$ and $I_c(\cdot)$ and (regularized) capacities $C(\cdot)$ and $Q(\cdot)$. Specifically, the one-shot capacity restricts the encoder $\mathcal{E}$ to generate states that are separable\footnote{We refer the reader to Appendix~\ref{App:0.1} for a proper definition of separable states.} over multiple uses of the channel, whereas the (regularized) capacity is achieved by relaxing this constraint and hence allowing the encoder to generate entangled states. Clearly, it results $\chi(\mathcal{N}) \leq C(\mathcal{N})$ and $I_c(\mathcal{N}) \leq Q(\mathcal{N})$ for any quantum channel $\mathcal{N}$ \cite{Wilde}.

In what follows, we are going to give in Section~\ref{Sec:2.3} the operational definitions of the quantum capacities used in Section~\ref{Sec:4}, without making reference to the explicit structure of the channels. Afterwords, in Section~\ref{Sec:2.4} and \ref{Sec:2.5} we are going to review the important quantum coding theorems for memoryless channels, which express the capacities in terms of explicit entropic quantities.

\subsection{Operational Capacity Definition for a Quantum Channel}
\label{Sec:2.3}

\begin{figure}[t]
    \centering
    \includegraphics[width=1\columnwidth]{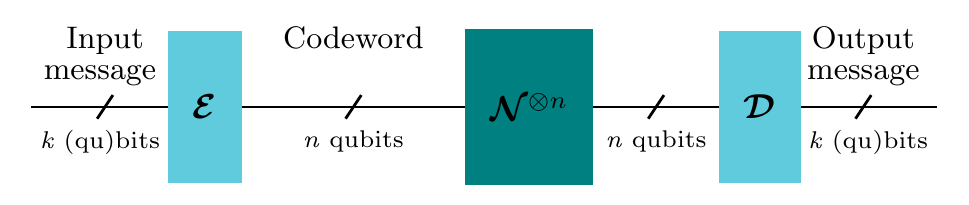}
    \caption{Operational scheme for the capacity over quantum channel $\mathcal{N}$, with the encoder $\mathcal{E}^{k\rightarrow n}$ and the decoder $\mathcal{D}^{n\rightarrow k}$ depending on the nature of the message, i.e., classical or quantum. The tensor product $\mathcal{N}^{\otimes n}$ denotes $n$ uses of channel $\mathcal{N}$, achievable either with a parallel placement (in space) of $n$-times channel $\mathcal{N}$ or, equivalently, with $n$ independent uses of such a channel in time.}
    \label{Fig:05}
    \hrulefill
\end{figure}

The classical/quantum capacity of a quantum channel\footnote{We refer the reader to Appendix~\ref{App:1} for a concise introduction to quantum channels, and to \cite{7,Cariolaro-15,0,Singh2021} for an in-depth treatise of quantum channel capacities.} $\mathcal{N}$ is the maximum achievable rate at which information encoded in quantum carriers can be transferred reliably from Alice to Bob. As in classical Shannon theory, the ratio $\frac{k}{n}$ is what measures the rate, where $k$ is the number of exchanged bits/qubits of information between the sender and the receiver, and $n$ is the number of uses of the communication channel.

Similarly to classical Shannon theory, the reliability condition requires that, at the asymptotic use of the channel (i.e., when $n\rightarrow \infty$), the fidelity\footnote{In a nutshell, the fidelity $F$ is a measure, with values between $0$ and $1$ of the distinguishability of two arbitrary quantum states $\rho$ and $\sigma$, and it is generally defined as \cite{0,Wilde} $F(\rho,\sigma)=\mathrm{Tr}(\sqrt{\sqrt{\rho}\sigma\sqrt{\rho}})^2$ with $\operatorname{Tr}(\cdot) = \sum_{i} (\cdot)_{ii}$ denoting the trace.} operator. $F$ between the channel input/output -- or, correspondingly when it comes to classical communications, the error probability\footnote{In this case, the fidelity and the probability of error are linked through the probability of successfully decoding the message, which is expressed in terms of the trace distance between the input and output states of the noisy channel $\mathcal{N}$ \cite{0}.} -- can be made arbitrarily close to one -- or, correspondingly, arbitrary close to zero.

\begin{figure*}[t]
    \begin{minipage}[c] {0.49\textwidth}
        \centering
        \includegraphics[width=1\columnwidth]{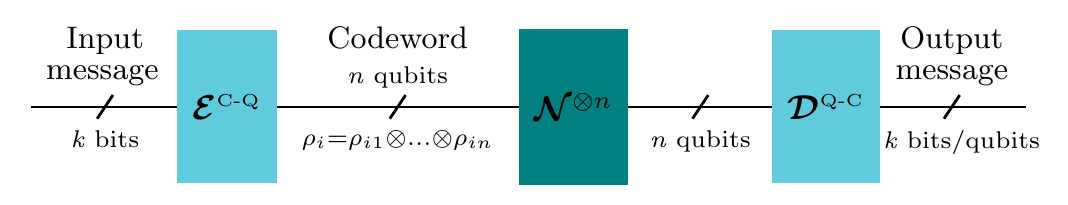}
        \subcaption{A scheme for the one-shot capacity $\chi(\mathcal{N}^{\otimes n})$ of the quantum channel $\mathcal{N}$ through $n$ uses of the channel $\mathcal{N}$. A set of classical messages in an alphabet $\mathcal{M}$ is encoded by a classic-to-quantum encoder $\mathcal{E}^{C-Q}$ constrained to separable codewords, namely, $\rho_i = \rho_{i1} \otimes \cdots \otimes \rho_{in}$ for the $i$-th codeword. After transmission, a quantum-to-classic decoder $\mathcal{D}^{Q-C}$ is applied to retrieve the classical message. The decoder is a measurement given by the optimal POVM, which is allowed to act collectively on the joint output state in order to obtain a set of classical messages.}
        \label{Fig:06-a}
    \end{minipage}
    \hspace{0.02\textwidth}
    \begin{minipage}[t]{0.49\textwidth}
        \centering
        \includegraphics[width=1\columnwidth]{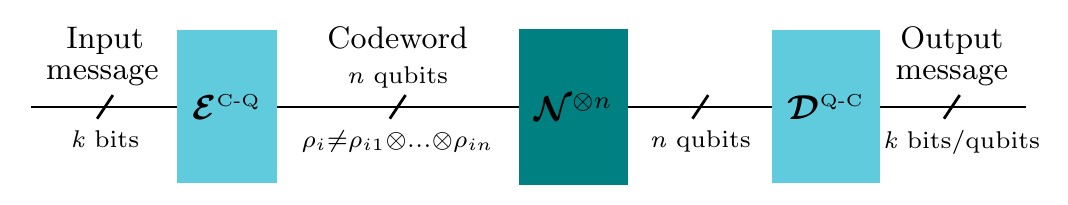}
        \subcaption{A scheme for the classical capacity $C(\mathcal{N})$ of quantum channel $\mathcal{N}$ through $n$ uses of the channel $\mathcal{N}$. The encoder $\mathcal{E}^{C-Q}$ is not restricted to separable codewords, rather, it is allowed to encode the classical information into entangled input states $\rho_i\neq\rho_{i1}\otimes\cdots\otimes\rho_{in}$. Similarly to the scheme of the one-shot capacity, the decoder is allowed to perform entangling measurements.}
        \label{Fig:06-b}
    \end{minipage}
    \caption{One-shot vs. regularized classical capacity from the encoder perspective.}
    \label{Fig:06}
    \hrulefill
\end{figure*}

Henceforth, the classical/quantum capacity of a quantum channel can be given in an operational way, depicted in Figure~\ref{Fig:05}, as:
\begin{align}
    &\lim_{\epsilon\rightarrow0}\limsup_{n\rightarrow\infty}\big\{ \frac{k}{n}: \exists \mathcal{E}^{k\rightarrow n},\quad \exists  \mathcal{D}^{n\rightarrow k},\quad\nonumber\\
    &\min_{m\in\mathcal{M}}F\big(\ket{m} ,\mathcal{D}^{n\rightarrow k}\circ\mathcal{N}^{\otimes n}\circ \mathcal{E}^{k\rightarrow n}(\ket{m})\big)>1-\epsilon   \big\}
    \label{eq:2.3}
\end{align}
with the fidelity measuring the distinguishability between the input symbol $\ket{m}$ and the output symbol $\mathcal{D}^{n\rightarrow k}\circ\mathcal{N}^{\otimes n}\circ \mathcal{E}^{k\rightarrow n}(\ket{m})$. $\mathcal{E}^{k\rightarrow n}$ and $\mathcal{D}^{n\rightarrow k}$ denote the encoder and the decoder, mapping the\footnote{Where $k=\log_2 d$, with $d$ being the dimension of the message Hilbert space, in case of quantum capacity.} $k$-qubits/bits message that Alice wants to share with Bob into a $n$-qubits code-word sent through the quantum channel as described in Appendix~\ref{App:4}.

Importantly, classical information could be encoded in the orthogonal basis of the Hilbert space, whereas quantum information must be encoded in the span of the orthogonal basis of the Hilbert space due to the genuine quantum coherence. Intuitively, when decoding information encoded in the Hilbert space, we can retrieve more classical information than quantum. By oversimplifying, the rationale for this can be understood in terms of no-cloning theorem, which allows classical information to be copied whereas quantum information cannot. Accordingly, for any given channel $\mathcal{N}$, the quantum capacity $Q(\mathcal{N})$ is upper bounded by the classical capacity $C(\mathcal{N})$ \cite{Caruso_2014}.

Expressions for capacities of quantum channels, in terms of entropic functions, have been provided by sophisticated coding theorems.  While we refer the reader to \cite{7,Wilde,0,56,57,61,65,63,64} for detailed review of different notions of quantum channel capacities, for both channels with and without memory, and entanglement-assisted capacities, we focus here on memoryless channels and their unassisted capacities. 

\subsection{Classical Capacity of Quantum Channels}
\label{Sec:2.4}

The expression of the classical capacity of an arbitrary quantum channel $\mathcal{N}$ has been formalized by the Holevo-Schumacher-Westermoreland (HSW) coding theorem with reference to the one-shot capacity\footnote{Also known as \textit{Holevo information} of channel $\mathcal{N}$ in the literature \cite{Wilde,Caruso_2014,0}. Accordingly, the two terms will be interchangeably used in the following.} $\chi(\mathcal{N})$ \cite{68,66}. The expression resembles Shannon's formula given in \eqref{eq:2.2} for the classical capacity of classical channels, as it can be expressed in terms of a maximization of the Holevo information\footnote{See Appendix~\ref{App:3} for a concise definition of Holevo information.} $\chi\big(\mathcal{N},\{p_x,\rho_x\},\mathcal{N}\big)$ over the set of input ensembles $\{p_x,\rho_x\}$ encoding the classical messages. Formally, it is given by:
\begin{equation}
    \chi(\mathcal{N}) = \max_{\{p_x,\rho_x\}}\chi\big(\{p_x,\rho_x\},\mathcal{N}\big)
    \label{eq:2.4} 
\end{equation}
and the maximization can be taken always over pure input states, restricting so the search space.

The operational meaning of the HSW theorem is that, given an ensemble $\{p_x,\rho_x\}$ and an integer satisfying $N \leq 2^{n\chi\big(\{p_x,\rho_x\},\mathcal{N}\big)}$, one can choose $N$ $n$-qubits codewords $\rho_1, \rho_2, \dots, \rho_N$ in separable product form $\rho_i=\rho_{i1}\otimes \dots \otimes \rho_{in}$ and an associated decoding measurement setup, allowing Bob to discriminate between the $N$ output states $\mathcal{N}^{\otimes n}(\rho_{i}) = \mathcal{N}(\rho_{i1})\otimes\dots\otimes \mathcal{N}(\rho_{in})$ arbitrarily well in the asymptotic limit of $n$. The positive operator-valued measure (POVM)\footnote{See Appendix~\ref{App:0.1} for the definition and an example of POVM.} assigned for the measurement setup is allowed to be an entangling measurement that operates collectively on the $n$-qubits output of each codeword.

As mentioned in Section~\ref{Sec:2.2}, if we unrestrict the encoder $\mathcal{E}$ from mapping messages only to product states as in Figure~\ref{Fig:06-a}, and we rather allow it to produce entangled codewords as in Figure~\ref{Fig:06-b}, we obtain the classical capacity $C(\mathcal{N})$ of the quantum channel $\mathcal{N}$.

In the HSW coding theorem, this is achieved by adopting a block coding strategy, which, for any $n > 1$, allows Alice to use $n$ copies of the channel as a single \textit{extended} channel $\mathcal{N}^{\otimes n}$ with associated Holevo information $\chi(\mathcal{N}^{\otimes n})$, where the maximum is taken over all input ensembles, including entangled states\footnote{Since product states are allowed as well, it is clear that $\chi(\mathcal{N}^{\otimes n}) \geq n \chi(\mathcal{N})$ and, hence, $C(\mathcal{N}) \geq \chi(\mathcal{N})$ as discussed in Section~\ref{Sec:4.1}.}, for the $n$ elementary channels. As a result, the capacity $C(\mathcal{N})$ of $\mathcal{N}$ can be obtained by taking the limit $n\rightarrow \infty$ over the associated rate $\frac{\chi(\mathcal{N}^{\otimes n})}{n}$. This is known as the \textit{regularization} procedure of the capacity, and it allows the capacity $C(\mathcal{N})$ to be written as:
\begin{equation}
    C(\mathcal{N})= \lim_{n\rightarrow\infty} \frac{1}{n}\chi(\mathcal{N}^{\otimes n})
    \label{eq:2.5}
\end{equation}
As it appears, the capacity $C(\mathcal{N})$ is not easily computed in general \cite{68}, as it requires a maximization over an unbounded number of uses of the channel. Indeed, a single-letter formula of the capacity is known only for few types of quantum channels, e.g., the depolarizing channel \cite{5}.

\subsection{Quantum Capacity of Quantum Channels}
\label{Sec:2.5}

Similarly to the HSW theorem, the quantum capacity theorem -- widely known as the LSD theorem \cite{0,Caruso_2014} -- expresses the quantum capacity $Q(\mathcal{N})$ in terms of a regularization of the one shot capacity\footnote{Also known as \textit{coherent information} of channel $\mathcal{N}$ in the literature \cite{Wilde,Caruso_2014,0}. Accordingly, the two terms will be interchangeably used in the following.} $I_c(\mathcal{N})$. The latter quantity expresses the maximal achievable rate through the quantum channel when the quantum-to-quantum encoder is constrained to generate separable codewords only.

Formally, the one-shot quantum capacity $I_c(\mathcal{N})$ is expressed in terms of the coherent information\footnote{See Appendix~\ref{App:3} for a concise definition of the coherent information.} of channel $\mathcal{N}$ with respect to the arbitrary state $\rho$ as:
\begin{equation}
    I_c(\mathcal{N})=\max_\rho I_c(\rho,\mathcal{N})
    \label{eq:2.6}
\end{equation}
where the maximization is taken over all possible input quantum states.

As already mentioned, the one-shot capacity $I_c(\mathcal{N})$ does not fully characterize the quantum capacity $Q(\mathcal{N})$, which is the maximum achievable rate, for which the fidelity of the transmitted state is arbitrarily large, i) over asymptotically many uses of the channel $\mathcal{N}$, and ii) with the encoder allowed to generate entangled codewords. Likewise to the classical capacity, when a block coding strategy is used the quantum capacity can be expressed as \cite{74,69,75}:
\begin{equation}
    Q(\mathcal{N})=\lim_{n \to \infty}\frac{1}{n}I_c(\mathcal{N}^{\otimes n})
    \label{eq:2.7}
\end{equation}

Of course, the quantum capacity $Q(\mathcal{N})$ is a non-tractable quantity in general. The rationale for this is because \eqref{eq:2.7} involves maximizing the coherent information over an unbounded number of channel uses. In fact, entanglement across channel uses can even increase the coherent information from zero to non-zero. One might think that only a finite number of channel uses might be sufficient to calculate the capacity, as imposing a cut-off in the number of uses of the channel. It turns out this is completely wrong, as it has been shown that whatever value of $n$ we fix, we can always find a channel with vanishing coherent information $I_c(\mathcal{N}^{\otimes n})$, nonetheless, the quantum capacity $Q(\mathcal{N})$ is non-vanishing \cite{12}.

\label{Sec:2.6}
\subsection{Bibliographic Notes}
One of the earliest uses of quantum information is classical communications over quantum channels.  This research was initiated by the early work of Holevo \cite{58}, in which the Holevo bound on classical capacity was established. Later on, a lower bound on the Holevo information of a channel was provided independently by Schumacher and Westmoreland \cite{68}, and Holevo \cite{66}. Classical communications in one shot setting has been studied by a number of authors, including Hayashi \cite{Hayashi,Hayashi_2007}, Renes and Renner \cite{Renes_2011}, Wang and Renner \cite{Wang_2012}, Datta et al. \cite{Datta_2013}, Mathews and Wehner \cite{Matthews_2014}, Wilde \cite{Wilde_2013}.

The quest for determining a quantum capacity in the Shannon's sense was raised by Shor \cite{shor95}. Different notions of quantum communications were established since then. The one adapted in this tutorial is based on entanglement transmission which was defined by Schumacher \cite{Schumacher1996SendingQE}. The notion of subspace transmission was proposed by Barnum et al. \cite{Barnum_2000}. Devetak \cite{69} gave the definition of entanglement generation. Kretschman and Werner \cite{Kretschmann_2004} showed that the capacities derived from these variations are all equal. The coherent information of asymptotic uses of a quantum channel was derived by Schumacher \cite{Schumacher1996SendingQE} as an upper bound on quantum capacity, Barnum et al. \cite{49}, Schumacher and Nielsen \cite{51}. The coherent information as a lower bound on the quantum capacity was established by Loyd \cite{74}, Shor \cite{75}, Devetak \cite{69}. Another proof for the achievability of the coherent capacity was provided by Hayden et al., using the decoupling lemma \cite{76}, which was initiated by Schumacher and Westmoreland \cite{Schumacher2002decoupling}. 
The one-shot setting of quantum capacity was treated in many papers, including Buscemi and Datta \cite{Buscemi_2010}, Datta and Hsieh \cite{Datta_20133}, Wang et al. \cite{Wang_2019}, Kiavansh et al. \cite{kianvash2020bounding}.

\section{Quantum Marvels}
\label{Sec:3}

\begin{figure*}[t]
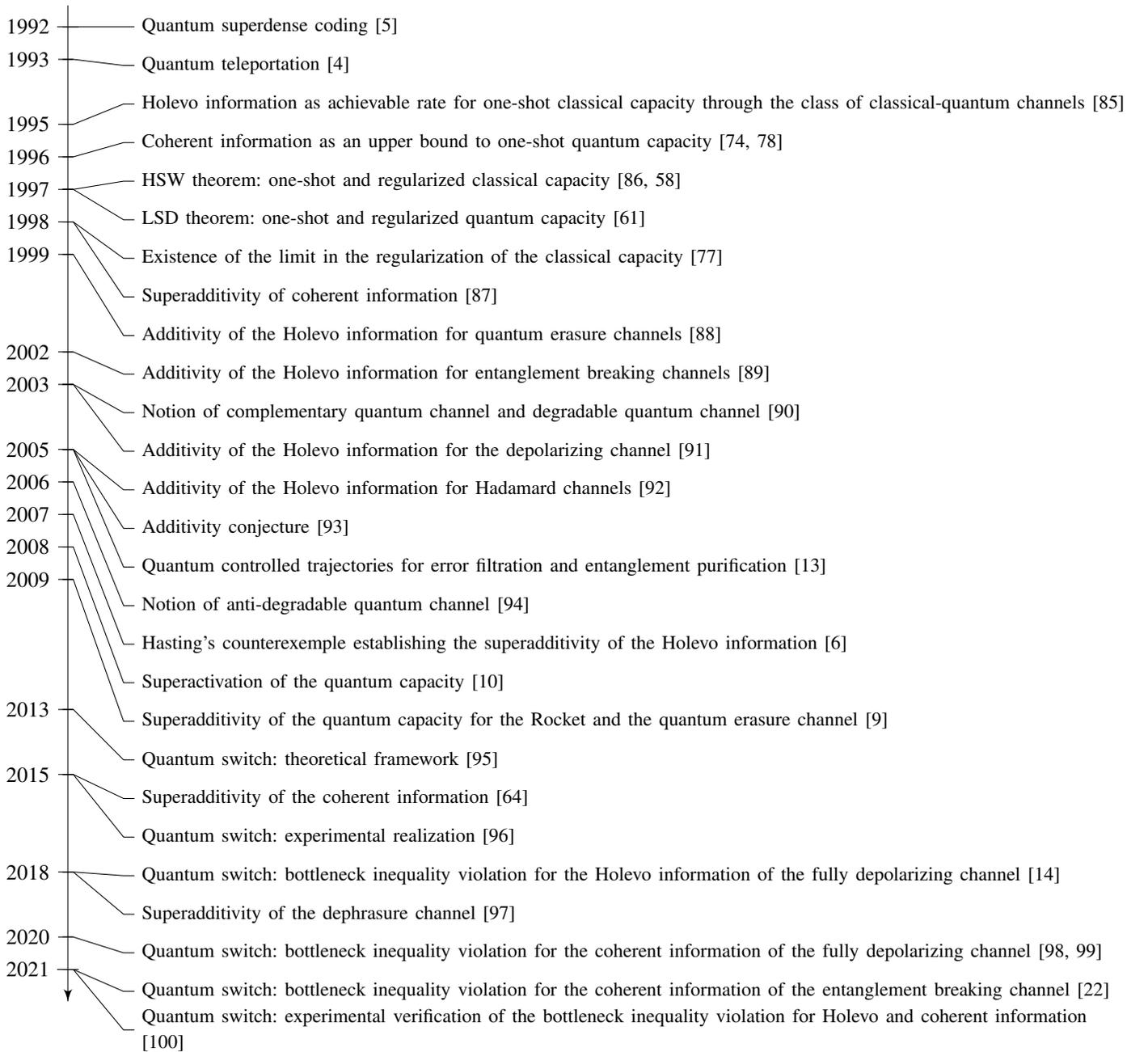

    \centering
    \begin{timeline}{1992}{2021}{0.05\textwidth}{0.05\textwidth}{0.89\textwidth}{0.55\paperheight}
        \entry{1992}{\small Quantum superdense coding \cite{32}}
        \entry{1993}{\small Quantum teleportation \cite{31}}
        \entry{1995}{\small Holevo information as achievable rate for one-shot classical capacity through the class of classical-quantum channels \cite{hausladen}}
        \entry{1996}{\small Coherent information as an upper bound to one-shot quantum capacity \cite{Schumacher1996SendingQE,51} }
        \entry{1997}{\small HSW theorem: one-shot and regularized classical capacity \cite{Holevo-capacity,68}}
        \plainentry{1997}{\small LSD theorem: one-shot and regularized quantum capacity \cite{74}}
        \entry{1998}{\small Existence of the limit in the regularization of the classical capacity \cite{49}}
        \plainentry{1998}{\small Superadditivity of coherent information \cite{53}}
        \entry{1999}{\small Additivity of the Holevo information for quantum erasure channels \cite{80}}
        \entry{2002}{\small Additivity of the Holevo information for entanglement breaking channels \cite{Shor-EBC}}
        \entry{2003}{\small Notion of complementary quantum channel and degradable quantum channel \cite{Devetak2003TheCO}}
        \entry{2003}{\small Additivity of the Holevo information for the depolarizing channel \cite{King2003TheCO}}
        \entry{2005}{\small Additivity of the Holevo information for Hadamard channels \cite{King2005PropertiesOC}}
        \entry{2005}{\small Additivity conjecture \cite{77}}
        \plainentry{2005}{\small Quantum controlled trajectories for error filtration and entanglement purification \cite{36} }
        \entry{2006}{\small Notion of anti-degradable quantum channel \cite{Caruso_2006}}
        \entry{2007}{\small Hasting's counterexemple establishing the superadditivity of the Holevo information  \cite{6}}
        \entry{2008}{\small Superactivation of the quantum capacity \cite{4}}
        \entry{2009}{\small Superadditivity of the quantum capacity for the Rocket and the quantum erasure channel \cite{26}}
        \entry{2013}{\small Quantum switch: theoretical framework  \cite{83}}
        \entry{2015}{\small Superadditivity of the coherent information \cite{12}}
        \entry{2015}{\small Quantum switch: experimental realization \cite{Procopio_2015}}
        \entry{2018}{\small Quantum switch: bottleneck inequality violation for the Holevo information of the fully depolarizing channel \cite{18}}
        \entry{2018}{\small Superadditivity of the dephrasure channel \cite{LedLeuSmi-18}}
        \entry{2020}{\small Quantum switch: bottleneck inequality violation for the coherent information of the fully depolarizing channel \cite{10,11}}
        \entry{2021}{\small Quantum switch: bottleneck inequality violation for the coherent information of the entanglement breaking channel \cite{19}}
        \entry{2021}{\small Quantum switch: experimental verification of the bottleneck inequality violation for Holevo and coherent information \cite{8}}
    \end{timeline}
    \caption{Timeline for milestones on superadditivity, superactivation and causal activation of quantum channels.}
    \label{Fig:00}
    \hrulefill
\end{figure*}

In this section we present the three dazzling phenomena of superadditivity, superactivation and causal activation. An easy-access guide towards the literature related to these phenomena and the prominent results as \textit{timeline of the milestones} is provided with Figure~\ref{Fig:00}.

\subsection{Superadditivity}
\label{Sec:3.1}

\begin{figure}[t]
   \centering
    \includegraphics[width=0.9\columnwidth]{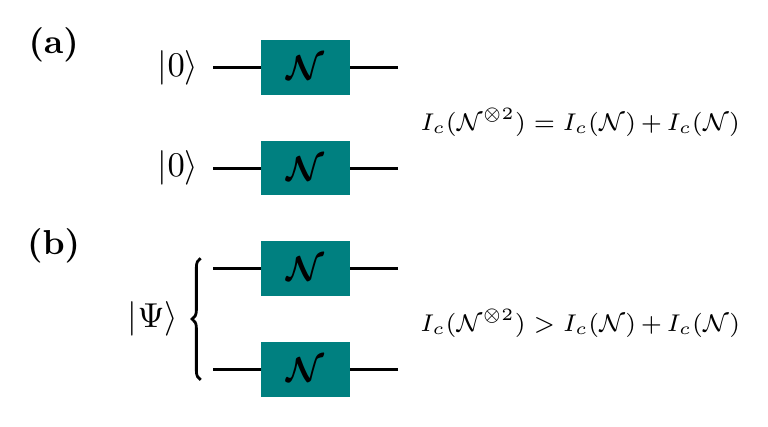}
    \caption{A scheme showing superadditivity of the one-shot quantum capacity of channel $\mathcal{N}$. (a) When two instances of the channel (this is formally given by the tensor product $\mathcal{N}^{\otimes 2} \eqdef \mathcal{N}\otimes\mathcal{N}$) are used on separable inputs such as $\ket{0}\otimes\ket{0}$, the coherent information of the two channels together $I_c({\mathcal{N}^{\otimes2}})$ is the sum of the two individual coherent information $I_c(\mathcal{N})+I_c(\mathcal{N})$. (b) Conversely, when the two instances of the channel are used on an entangled state $\ket{\Psi}$, superadditivity of the coherent information occurs and the joint coherent information $I_c({\mathcal{N}^{\otimes2}})$ exceeds the sum of individual coherent information $I_c(\mathcal{N})+I_c(\mathcal{N})$.}
    \label{Fig:02}
    \hrulefill
\end{figure}

As mentioned above, entanglement has no longer been considered only as a foundational concept that breaks the operational causal explanations of correlations formulated in terms of Bell inequalities \cite{Bell5}. It started rather to be considered as a tool with wider applications in different areas of communication engineering. And researchers are continuing to dig for other surprises of quantum phenomena within the field. 

Astoundingly, it was found that -- contrary to classical communications\footnote{In the following and in agreement with the literature \cite{nielsen_chuang_2010}, we refer to communications through classical channels as \textit{classical communications}, whereas we denote communications through quantum channels as \textit{quantum communications}. In the latter case, whether the quantum channels will be used to transmit quantum or classical information will be specified in the context.}, when a quantum channel is used independently multiple times, its performance in terms of \textit{coherent information}\footnote{See Section~\ref{Sec:2} for a proper introduction to the different definitions of capacities, including the coherent information -- through a quantum channel.} \cite{7,Wilde,nielsen_chuang_2010} can be non additive on the number of uses \cite{6,53}.

In other words, in classical communication scenarios, if a channel is able to transmit a bit of information, when it is used $n$ times, the amount of information that can be transmitted is no more than $n$ bits. Formally, the mutual information between the output $\mathbf{Y}^n$ and the input $\mathbf{X}^n$ random variables on $n$ uses of a classical channel $\{p_i(y|x)\}_{i=1}^n$ is always bounded by n times the single letter capacity of the channel:
\begin{equation}
    \label{eq:3.1}
    I(\mathbf{X}^n,\mathbf{Y}^n)\leq \sum_{i=1}^n I(X_i,Y_i)
\end{equation}
In other words, the use of correlated codewords, jointly sampled, in transmitting information does not provide any communication benefit with respect to the use of uncorrelated codewords, sampled from a product distribution. 

In contrast, in quantum communication scenarios, a quantum channel that can transmit a certain amount of information (classical or quantum), when used $n$-times it can send more than $n$ times that amount of information. This is extremely against classical additive logic of $2=1+1$. Indeed, in the quantum domain, superadditivity can happen and it results\footnote{We adopted -- in analogy with the superactivation literature \cite{4,boche-0} -- such an expression to better summarize the superadditivity phenomenon.}:
\begin{equation}
    \label{eq:3.2}
    2 > 1+1 \nonumber
\end{equation}

This unconventional phenomenon requires the use of entanglement to encode messages, which in turn can be either classical or quantum. This is known in the literature as the \textit{superadditivity} of quantum channel capacities, and it is depicted in Figure~\ref{Fig:02}.  The figure illustrates that when Alice and Bob use multiple instances of a quantum channel $\mathcal{N}$\footnote{In this paper, we focus on channels that do not exhibit memory effects over many uses. In other words, we are interested in noisy channels where multiple uses of the same channel over time or the use of multiple copies of the same channel in parallel are equivalent. This is due to the assumption of noise between different uses being independent.} to communicate messages encoded in separable input states, the coherent information of the two channels together $I_c(\mathcal{N}^{\otimes 2})$ is equal to the sum of the two individual coherent information $I_c(\mathcal{N})+I_c(\mathcal{N})$. This is trivial in classical communications\footnote{Multiple uses of classical channels allow error correction/diversity strategies, but not superadditivity of the channel capacity.}. On the contrary, when Alice and Bob use the channel the same way as before -- but encoding messages in entangled states -- the overall coherent information $I_c(\mathcal{N}^{\otimes 2})$ exceeds the sum of the coherent information of individual channels, in the form
\begin{equation}
    \label{eq:3.3}
    I_c(\mathcal{N}^{\otimes 2}) > I_c(\mathcal{N})+I_c(\mathcal{N})
\end{equation}
Indeed, a similar behaviour has been observed for the Holevo information $\chi(\mathcal{N}^{\otimes 2})>\chi(\mathcal{N})+\chi(\mathcal{N})$ \cite{6}.

This phenomenon shows how entanglement can be considered as a key factor for unravelling the unconventional potential of quantum theory when it comes to quantum communication. Equally, it highlights that this potential is not limited to quantum messages, given that quantum communications can boost the classical information transmission rates as well, as shown in Section~\ref{Sec:4.1}. 

\subsection{Superactivation}
\label{Sec:3.2}
\begin{figure}[t]
    \centering
    \includegraphics[width=0.9\columnwidth]{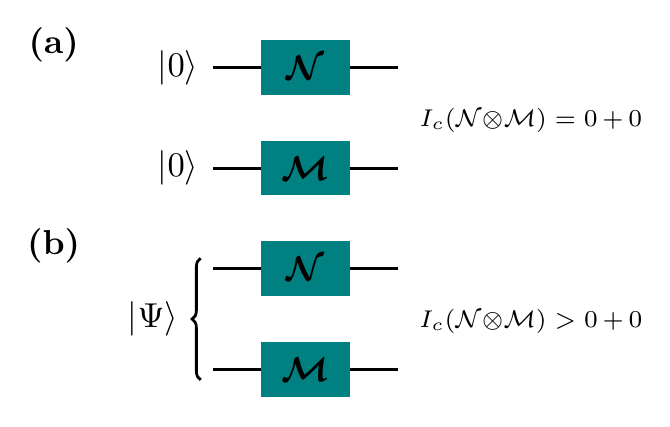}
    \caption{A scheme showing superactivation of the one-shot quantum capacity of two \textit{zero capacity} quantum channels $\mathcal{N}$ and $\mathcal{M}$. (a) When the two channels are used on separable inputs such as $\ket{0}\otimes\ket{0}$ encoding the quantum message, the coherent information of the two channels together $I_c(\mathcal{N}\otimes\mathcal{M})$ is the sum of the two individual capacities $I_c(\mathcal{N})$ and $I_c(\mathcal{M})$, and hence it is identically zero. (b) When the two channels are used on an entangled state $\ket{\Psi}$ properly encoding the quantum message, superactivation of the capacity occurs and the joint coherent information $I_c(\mathcal{N}\otimes\mathcal{M})$ can be greater than zero, allowing the two channels to transmit a non-vanishing amount of quantum information.}
    \label{Fig:03}
    \hrulefill
\end{figure}

More surprisingly, our rather simplistic understanding of nature is broken by quantum logic, when it comes to the phenomenon of \textit{superactivation} \cite{4}. This is when two different quantum channels that cannot transmit any amount of information separately -- i.e., \textit{zero capacity} channels \cite{85} -- when properly used together, they can transmit information. In classical information logic the relation $2 \cdot 0= 0 + 0 $ holds, whereas this is not the case when it comes to quantum information, where the relation\footnote{Also denoted as $0+0 > 0$ in the relevant literature \cite{4,boche-0}, meaning that a proper use of two different zero-capacity channels gives a non-null capacity.}: 
\begin{equation}
    \label{eq:3.4}
    2 \cdot 0 > 0+0 \nonumber
\end{equation}
is possible.

The superactivation phenomenon, as we discuss in more details in Section~\ref{Sec:5}, relies on entanglement \cite{4}\footnote{We must note that there exist protocols achieving superactivation by exploiting shared randomness between sender/receiver rather than entangled codewords \cite{Cubitt_2011,boche2018secret}.}. This is depicted in Figure~\ref{Fig:03}. This scheme shows that when the two zero capacity channels $\mathcal{N}$ and $\mathcal{M}$ -- with no ability of transferring quantum information -- are used on separable inputs encoding a quantum message, the coherent information of the two channels together is the sum of the two individual coherent information. Hence, the overall channel $\mathcal{N}\otimes\mathcal{M}$ does not allow transmission of any quantum information. On the other hand, if the quantum message is wisely encoded in an entangled state given as joint input to the channels, the overall channel $\mathcal{N} \otimes \mathcal{M}$ gains potential for the transmission of quantum information. Accordingly, the overall coherent information $I_c(\mathcal{N}\otimes\mathcal{M})$ satisfies the following inequality:
\begin{equation}
    \label{eq:3.5}
    I_c(\mathcal{N}\otimes\mathcal{M}) > I_c(\mathcal{N})+I_c(\mathcal{M})=0
\end{equation}

We note that -- as for superadditivity -- entanglement plays a fundamental role in enabling unparalleled phenomena in quantum communications. We further note that -- conversely to superadditivity -- no superactivation phenomenon is known to exist for quantum channels conveying classical information \cite{Gyongyosi_2012}, as discussed in Section~\ref{Sec:5}. This shows that quantum communications represent an heterogeneous communication paradigm, where the communication potential of a channel depends on the information nature of the message.

\subsection{Causal activation}
\label{Sec:3.3}

\begin{figure*}
    \begin{minipage}[c] {0.49\textwidth}
        \centering
        \includegraphics[width=1\columnwidth]{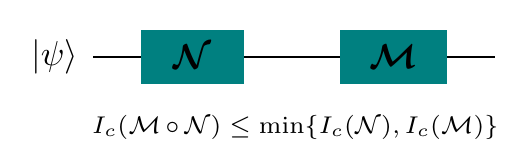}
        \subcaption{A classical sequential trajectory where the information carrier prepared in a certain state $\ket{\psi}$ undergoes the transformation $ \mathcal{M}\circ \mathcal{N}$, in which channel $\mathcal{N}$ is acting on the carrier before channel $\mathcal{M}$. Both the quantum and classical capacity of this scheme are upper bounded by the \textit{bottleneck inequality} given in \eqref{eq:app.3.9}, i.e., by the minimum of the capacities of each of the two concatenated channels.}
        \label{Fig:04-a}
    \end{minipage}
    \hspace{0.02\textwidth}
    \begin{minipage}[c] {0.49\textwidth}
        \centering
        \includegraphics[width=1\columnwidth]{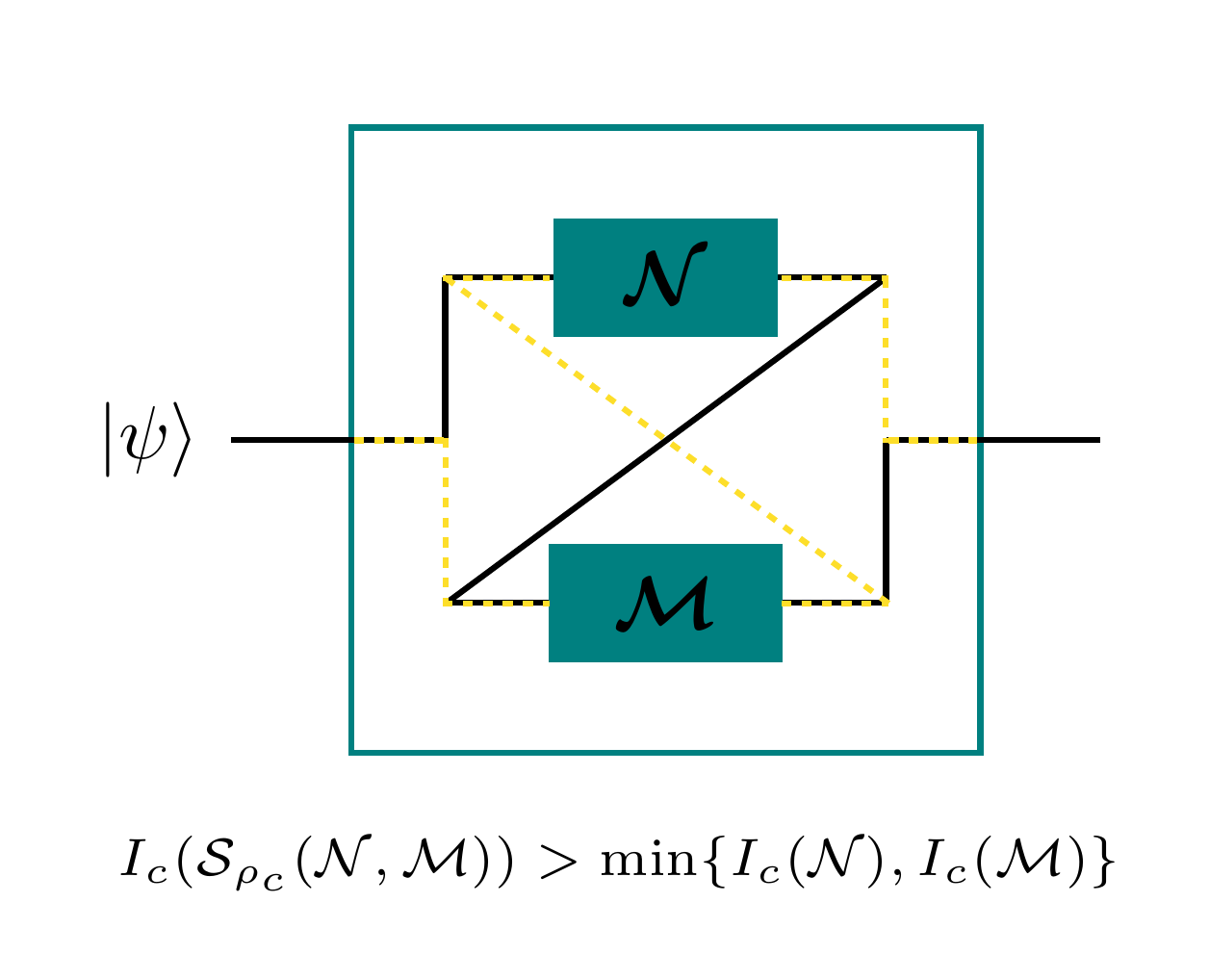}
        \subcaption{A quantum trajectory, which is a coherent superposition of the two classical sequential trajectories $\mathcal{N} \circ \mathcal{M}$ and $\mathcal{M} \circ \mathcal{N}$. This placement of channels is neither equivalent to a sequential trajectory in which the channels are timelike separated, nor equivalent to a parallel placement where the channels are spacelike separated. The overall coherent information $I_c\big(\mathcal{S}_{\rho_c}(\mathcal{N},\mathcal{M})\big)$ of the equivalent channel $\mathcal{S}_{\rho_c}(\mathcal{N},\mathcal{M})$ -- with $\rho_c$ denoting the quantum system controlling the causal order between the two channels -- can violate the bottleneck inequality.}
        \label{Fig:04-b}
    \end{minipage}
    \caption{A scheme showing causal activation of the coherent information for two quantum channels $\mathcal{N}$ and $\mathcal{M}$.}
    \label{Fig:04}
    \hrulefill
\end{figure*}

Although the inception of the quantum formalism has been initiated more than a century ago \cite{nielsen_chuang_2010}, its surprises are still coming out to this day, and there is much more out there to be discovered.

Recently, quantum information theorists, investigating causality in the quantum realm, have discovered that quantum mechanics allows for causal order to be indefinite \cite{39,83}. In simpler terms, causality between events -- channels from a communications engineering perspective -- might be not fixed, as shown in Figure~\ref{Fig:04}. If so, two communication channels, say $\mathcal{N}$ and $\mathcal{M}$, instead of affecting the information carrier in a definite causal order -- i.e., either $\mathcal{M} \circ \mathcal{N}$ or $\mathcal{N} \circ \mathcal{M}$ -- they act on the carrier in a genuinely quantum superposition of causal ordering. Hence, the information carrier evolves through a \textit{quantum trajectory} \cite{Costa_2016}. One example of quantum trajectories is the \textit{quantum switch} \cite{Procopio_2015,Rubino_2017}, which is a supermap acting on a set of channels and places them in a coherent superposition of different orders, which is a genuinely quantum placement setup.

It has been both theoretically \cite{18,19,20} and experimentally \cite{8} verified that the quantum switch can be used for communications, in an outperforming way, even when it is compared to known quantum protocols.\footnote{The notion of indefinite causal ordering is still debated in the community of causal modelling. In particular it can explain some observed phenomena differently from our usual causal models, but it cannot be explained in the framework of process tensors alone even by the most general temporal process tensor \cite{Milz_2018}. This makes it rather genuinely different from the usual temporal processes that we can account for, including quantum channels with memory, be it Markovian or non-Markovian \cite{Costa_2016}. Most importantly, the quantum switch itself does not violate any causal inequality formulated in a theory-independent manner \cite{39}.} Indeed, it has been shown that there are zero capacity channels that cannot transfer any information in the usual setups, i.e., parallel or sequential setups where the order of channels is well definite. But, when used in a quantum superposition of causal orders, these channels transmit non vanishing information (either classical or quantum, depending on the setup). This phenomenon, also termed as \textit{causal activation} in literature \cite{2}, as astounding as it is, harnesses its advantage from a genuinely quantum coherence between causal orders.

Indeed, causal activation should be regarded as a new way of placing communication channels \cite{3}, with no similarity with classical placement, such as parallel or sequential ones. In fact, as we discuss in more details in Section~\ref{Sec:6}, whereas superadditivity and superactivation exploits quantum channels combined in a classical way, causal activation exploits a new degree of freedom, namely, the quantum placement of quantum channels.

\section{Superadditivity of Quantum Channel Capacities}
\label{Sec:4}

Here we detail one of the quantum marvel phenomena introduced in Section~\ref{Sec:3}, namely, the superadditivity.

The additivity notion is very important as many questions in quantum information theory reduces to the additivity properties of some key functions \cite{77}. In this section, we are going to discuss the additivity properties of the Holevo information and the coherent information, which are the essential elements for characterizing the capacities of quantum channels.

\subsection{Superadditivity of Holevo information}
\label{Sec:4.1}
Originally, the Holevo information was believed to be additive for all quantum channels \cite{77}, that is
\begin{equation}
    \label{eq:4.1}
    \chi(\mathcal{N}^{\otimes n})=n\chi(\mathcal{N})
\end{equation}
This would imply that the Holevo information would be a good characterization of the classical capacity in the general case, i.e., $\chi(\mathcal{N}) = C(\mathcal{N})$. This conjecture, known as the \textit{additivity conjecture}, was proved to hold for some classes of quantum channels, e.g., entanglement breaking channels \cite{Shor-EBC} or depolarizing channel \cite{King2003TheCO}. 

Surprisingly, Hastings found the existence of a counterexample to the additivity conjecture \cite{6}, demonstrating that it does not hold in the general case. He showed that, when entangled input states are used, the Holevo information is not only weakly superadditive, instead, it exhibits a strong superadditivity property. The counterexample relies on the use of two random channels $\mathcal{N}$ and $\Bar{\mathcal{N}}$:
\begin{align}
    &\mathcal{N}(\rho)=\sum_i p_i U_i^{\dagger}\rho U_i\nonumber\\
    &\Bar{\mathcal{N}}(\rho)=\sum_i p_i\Bar{U}_i^{\dagger} \rho \Bar{U}_i
    \label{eq:4.2}
\end{align}
which are complex conjugate to each other. Specifically, the channels have unitary Kraus operators $\{U_i\}_{i\in \{1,\dots,D\}}$ and their complex conjugates $\{\Bar{U}_i\}_{i\in \{1,\dots,D\}}$. Moreover, each unitary $U_i$ is randomly sampled from a certain given random distribution. Finally, the coefficients $p_i$ in \eqref{eq:4.2} are chosen randomly from another particular distribution, in such a way the minimum output entropy of the tensor product $\mathcal{N}\otimes\Bar{\mathcal{N}}$ of the two channels is strictly smaller than twice the minimum entropy of one of the channels alone. Formally, this is given by the following inequality:
\begin{equation}
    H_{min}(\mathcal{N}\otimes\Bar{\mathcal{N}})<2H_{min}(\mathcal{N})
    \label{eq:4.3}
\end{equation}
under the use of entangled input states to the channel $\mathcal{N}\otimes\Bar{\mathcal{N}}$.

This inequality proved\footnote{Indeed, the minimum entropy of a quantum channel $\mathcal{N}$ is defined as  $H_{min}(\mathcal{N})=\min_{\rho}S(\mathcal{N}(\rho))$ \cite{0}, where $S(\rho)$ denotes the von Neumann entropy of the state $\rho$ as detailed in Appendix~\ref{App:3}. The minimum output entropy is related to the Holevo information for irreducibly covariant quantum channels by $\chi(\mathcal{N})=S(\mathcal{N}(\frac{\mathcal{I}}{d}))-H_{min}(\mathcal{N})$ where $\frac{\mathcal{I}}{d}$ is the maximally mixed input state, with $d$ being the dimension of the input of the channel and $\mathcal{I}$ being the identity operator. Hence, for irreducibly covariant quantum channels, the subadditivity of the minimum entropy implies the superadditivity of the Holevo information.} the superadditivity phenomenon of the Holevo information, demonstrating that one of the most basic questions in quantum Shannon theory still remains wide open, i.e., there exists no general closed formula for classical capacity. This in turn shows our lack of deep understanding about classical information transmission over quantum channel.

Furthermore, it also implies that if Alice encodes the classical message she wants to communicate to Bob in an entangled state, this can help in increasing the classical capacity over the quantum channel linking Alice and Bob. This phenomenon has no counterpart in classical communications, where the capacity -- quantified by the mutual information between input and output of the channel -- cannot be increased even if classical correlations between subsequent input bits are exploited.

\subsection{Superadditivity of Coherent Information}
\label{Sec:4.2}

It was shown that the quantum capacity of a quantum channel is well-behaved and completely understood for the class of degradable channels, over which the coherent information is additive \cite{Devetak2003TheCO}, that is:
\begin{equation}
    I_c(\mathcal{N}_{degradable}^{\otimes n})=nI_c(\mathcal{N}_{degradable})
    \label{eq:4.4}
\end{equation}
Hence the regularization could be removed and the quantum capacity could be computed by a single optimization, similarly to classical channels.

However this is not true in general, as it was proven that for some channels -- e.g., the depolarizing channel \cite{53,54} -- the coherent information for multiple uses of the channel for some given value of $n$ is greater than $n$ times the coherent information provided by a single use of the channel. Hence, coherent information is superadditive \cite{15,53,12}. To illustrate this concept, let us consider the depolarizing channel $\mathcal{N}_D$, which transmits faithfully its input with probability $1-p$ and replaces it with probability $p$ by a maximally mixed state $\pi=\frac{I}{2}$, where $I$ is the $2\times 2$ identity matrix. Formally, this channel is given by:
\begin{align}
    \mathcal{N}_D(\rho) &=(1-p) \rho+ p \pi \nonumber \\
    &= (1-q)\rho+\frac{q}{3}(X\rho X+Y\rho Y+Z\rho Z)
    \label{eq:4.5}
\end{align}
with $q \eqdef \frac{3p}{4}$ as in \cite{Wilde}. To check whether the coherent information is superadditive for this channel, it suffices to calculate the coherent information for a single use of the channel, and then to find a code for multiple uses of the channel whose coherent information out-passes the single use case. 

\begin{figure}[t]
    \centering
    \includegraphics[width=0.9\columnwidth]{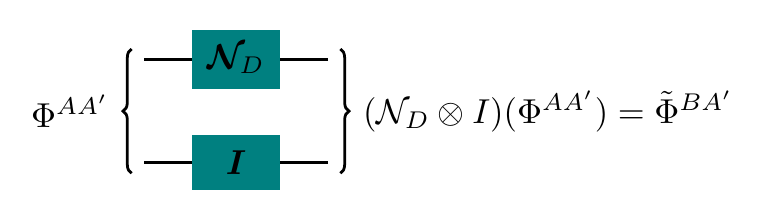}
    \caption{The figure shows the scenario used to derive the coherent information $I_c(\mathcal{N}_D)$ for the depolarizing channel $\mathcal{N}_D$. The quantum information is encoded into the maximally-entangled input state $\Phi^{AA'}$, whose part $A$ is sent through the noisy channel $\mathcal{N}_D$ and the other part $A'$ is kept as a reference, by sending it through the ideal channel $\mathcal{I}$.}
    \label{Fig:07}
    \hrulefill
\end{figure}

To this end, we note that the state maximizing the coherent information $I_c(\rho,\mathcal{N})$ in \eqref{eq:2.6} for the depolarizing channel is the maximally entangled state $\ket{\Phi}^{AA'} = \frac{1}{\sqrt{2}}(\ket{00}+\ket{11})$ \cite{Wilde,53,54}. Equivalently, this means that the coherent information of the depolarizing channel $\mathcal{N}_D$ can be obtained -- by following the scheme depicted in Figure~\ref{Fig:07} -- over the output state $\tilde{\Phi}^{BA'}$, where:
\begin{align}
    \tilde{\Phi}^{BA'} &= (\mathcal{N}_D\otimes \mathcal{I} )(\Phi^{AA'}) \nonumber \\
    &= (1-p) \Phi^{BA'} + p (\pi^{B} \otimes \pi^{A'})
    \label{eq:4.6}
\end{align}
where $\Phi^{AA'} \eqdef \ket{\Phi}^{AA'} \bra{\Phi}^{AA'}$ and $\Phi^{BA'} \eqdef \ket{\Phi}^{BA'} \bra{\Phi}^{BA'}$ denote the density matrices of the maximally entangled states. Accordingly, from \eqref{eq:app.3.8} and \eqref{eq:app.3.6} reported in Appendix~\ref{App:3}, the coherent capacity $I_c(\mathcal{N}_D)$ of the depolarizing channel for a single use is given by:
\begin{align}
    I_c(\mathcal{N}_D) &= [S(B)-S(BA')]_{\Tilde{\Phi}^{BA'}} = \nonumber \\
    &= \begin{cases}
            1-H(\Vec{q}) & \text{whenever } H(\Vec{q})\leq 1 \\
            0 & \text{otherwise}
        \end{cases}
    \label{eq:4.6bis}
\end{align}
with $\Vec{q} =(1-q,\frac{q}{3},\frac{q}{3},\frac{q}{3})$ denoting the vector of probabilities and $H(\Vec{q}) = -(1-q)\log_2 (1-q)-q\log_2 q + q \log_2 3$ denoting the Shannon entropy -- defined in \eqref{eq:app.3.2} -- of the distribution $\Vec{q}$.

\begin{figure}
    \includegraphics[width=0.9\columnwidth]{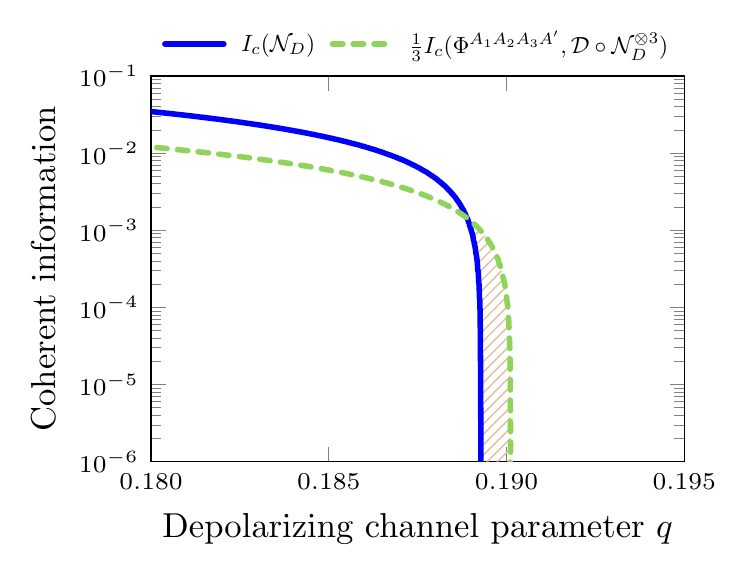}
    \caption{Coherent information for a depolarizing channel $\mathcal{N}_D$ vs. channel parameter $q$, where: i) the \textit{straight-blue} line denotes the coherent information $I_c(\mathcal{N}_D)$ achievable with a single use of depolarizing channel, and ii) the \textit{dotted-green} line denotes the coherent information achievable with three uses of the channel for the encoder output given in \eqref{eq:4.8} with proper choice of the encoder and the decoder. The plot is an illustration of the results derived in \cite{Wilde,53}}
    \label{Fig:08}
    \hrulefill
\end{figure}

A plot for the single-shot coherent capacity $I_c(\mathcal{N}_D)$ of the depolarizing channel is given in Figure~\ref{Fig:08}, where we see that it vanishes from a critical value of $q \approx 0.1893$. It is known that for antidegradability reasons, the quantum capacity $C(\mathcal{N}_D)$ of the depolarizing channel vanishes when the channel parameter $q$ satisfies $q \geq \frac{1}{4} = 0.25$ \cite{Cerf_2000} and, hence, the coherent information fully characterizes the quantum capacity of the channel. Conversely, whenever $q < \frac{1}{4}$, the coherent information does not fully characterize the quantum capacity of the channel. Consequently, the coherent capacity of multiple uses of the channel must be computed and, in the following, we will focus on a specific scenario where three uses -- instead of five as in \cite{Wilde,53,54,Shor1996QuantumEC} -- of the channel are sufficient to prove the superadditivity of the coherent information. 

Specifically, we focus on a $(3,1)$ repetition code where each qubit is transmitted with three uses of the channel $\mathcal{N}_D$, and we will show that there exist a state $\rho$ and some parametric region of the depolarizing channel $\mathcal{N}_D$ so that:
\begin{equation}
    \frac{1}{3} I_c(\rho, \mathcal{D} \circ \mathcal{N}_D^{\otimes 3} \circ \mathcal{E}) > I_c(\mathcal{N}_D)
    \label{eq:4.7}
\end{equation}
where $\mathcal{E}$ and $\mathcal{D}$ are the encoder and decoder, respectively.

Let us consider as output state of the encoder $\mathcal{E}(\rho)$ and, hence, as input to the equivalent channel, the following state:
\begin{equation}
    \ket{\Phi}^{A_1 A_2 A_3 A'} = \frac{1}{\sqrt{2}}(\ket{0000}+\ket{1111})
    \label{eq:4.8}
\end{equation}
where $A_i$ is the input to the $i$-th use of the channel and $A'$ is the reference system as in Figure~\ref{Fig:07}.

\begin{figure}
    \centering
    \includegraphics[width=0.9\columnwidth
]{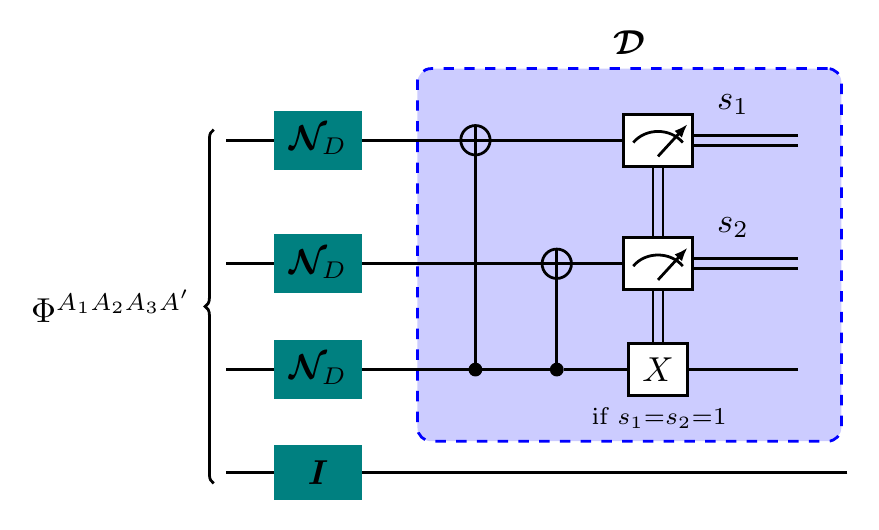}
    \caption{Scheme showing the decoder $\mathcal{D}$ used to prove the superadditivity of the coherent information for the depolarizing channel $\mathcal{N}_D$.}
    \label{Fig:09}
    \hrulefill
\end{figure}

Furthermore, let us assume we post-process the resulting state at the level of receiver with the decoder $\mathcal{D}$ shown in Figure~\ref{Fig:09}. Clearly, we have that:
\begin{equation}
    \frac{1}{3} I_c(\mathcal{N}_D^{\otimes 3}) \geq \frac{1}{3} I_c(\Phi^{A_1 A_2 A_3 A'}, \mathcal{D} \circ \mathcal{N}_D^{\otimes 3})
    \label{eq:4.9}
\end{equation}
as a result of the quantum data processing inequality \cite{Wilde}, where $\Phi^{A_1 A_2 A_3 A'} \eqdef \ket{\Phi} \bra{\Phi}^{A_1 A_2 A_3 A'}$.

Due to the convexity property of the coherent information on the receiver over classical variables \cite{Wilde,0}, the coherent information resulting from the post-processing in Figure~\ref{Fig:09} is given by the weighted average over the output of the measurements $s_1$ and $s_2$ over $B_1$ and $B_2$:
\begin{align}
    & I_c(\Phi^{A_1 A_2 A_3 A'}, \mathcal{D} \circ \mathcal{N}_D^{\otimes 3}) = \nonumber \\
    & \quad \sum_{s_1 s_2} p(s_1 s_2) I_c(\Phi^{A_1 A_2 A_3 A'}, \mathcal{D}_{s_1 s_2} \circ \mathcal{N}_D^{\otimes 3})
    \label{eq:4.10}
\end{align}
where $\mathcal{D}_{s_1 s_2}$ embeds the dependence of the post-processing on $s_1$ and $s_2$, i.e., whether there will be applied a $X$ gate on the third qubit.

For each syndrome $s_1 s_2$, there are 16 Kraus operators that can give rise to it. As an example, with probability $\frac{q^3}{27}$ each of the three channels will act as a $X$ channel, and the decoder, by measuring the first and second qubits as $00$, will keep the third qubit as unchanged. By grouping all the possibilities that give rise to a specific syndrome -- say $00$ -- we can model the overall evolution of the third qubit as going through a Pauli channel such as:
\begin{align}
     \mathcal{N}(\rho) &= q_{s_1 s_2}^I \rho + q_{s_1 s_2}^X X\rho X + q_{s_1 s_2}^Y Y\rho Y + q_{s_1 s_2}^Z Z\rho Z
     \label{eq:4.10bis}
\end{align}
characterized by the vector of probabilities $\Vec{q}_{s_1 s_2}$ with coherent information given by:
\begin{equation}
    I_c(\Phi^{A_1 A_2 A_3 A'}, \mathcal{D}_{s_1 s_2} \circ \mathcal{N}_D^{\otimes 3}) = 1 - H(\Vec{q}_{s_1 s_2})
    \label{eq:4.11}
\end{equation}

Remarkably, it has been shown that we can pick a noise parameter $q$ from the region where the coherent information of the single use of the depolarizing channel is vanishing from Figure~\ref{Fig:08}, while the coherent information in \eqref{eq:4.10} is non-vanishing. This proves \eqref{eq:4.7}, demonstrating a superadditive effect of the coherent information for the depolarizing channel.

Furthermore, it has been also demonstrated (not constructively, i.e, using random codes) that there exist channels that have vanishing coherent information for arbitrary $n$-codes, but they have a non-vanishing capacity \cite{12}. Which is even a stronger argument for the necessity of regularization for the quantum capacity. Indeed, on one hand, this means that the coherent information must be regularized over unbounded uses of the channel, hence, it cannot be used to compute the capacity in general. On the other hand, since the coherent information is additive for separable input states, additivity violation also implies that entanglement can protect information from noise in a way that is not possible classically \cite{26,LedKauDat-18}.

\subsection{Superadditivity of Classical and Quantum Capacities}
\label{Sec:4.3}

Having discussed the superadditivity of the one-shot capacities -- i.e., of the Holevo information and the coherent information -- we discuss now the superadditivity of the regularized capacities $C(\mathcal{N})$ and $Q(\mathcal{N})$.

Someone could think that some form of superadditivity for the regularized capacities might be obtained by using multiple instances of the same channel, as schematized in Figure~\ref{Fig:06-b}. However, regularized capacities -- regardless of the classical/quantum nature of the message -- over asymptotic uses of the same channel are themselves always additive. In the case of the classical capacity, this translates formally into:
\begin{equation}
    \label{eq:4.12}
    C(\mathcal{N}^{\otimes n})=nC(\mathcal{N})
\end{equation}
regardless on whether the $n$ uses of the same channel happens simultaneously in parallel or sequentially with independent uses over time. Similarly, the quantum capacity $Q(\mathcal{N})$ is additive over multiple uses of the same channel 
\begin{equation}
    \label{eq:4.13}
    Q(\mathcal{N}^{\otimes n})=nQ(\mathcal{N})
\end{equation}
This additivity property can be easily seen from the regularization of the Holevo capacity given in \eqref{eq:2.5} and from the regularization of the coherent capacity given in \eqref{eq:2.7}. 

Since the additivity is established for the use of the same channel in parallel or independently over time, it is important to understand if this holds also when different channels are considered. The answer to this question allows one to understand how different noisy channels interact and enhance each others capabilities.

Whether it is true that:
\begin{equation}
    \label{eq:4.14}
    C(\mathcal{N}\otimes \mathcal{M})\geq C(\mathcal{N})+C(\mathcal{M})
\end{equation}
is still an open problem for classical capacity of quantum channels. For instance, it can easily be noted by simple coding arguments, that the rate $C(\mathcal{N})+C(\mathcal{M})$ is always achievable by feeding the optimal code for each channel independently. The question of the superadditivity of the classical capacity relies on whether there could be a code with entangled states of the codewords, that satisfies $C(\mathcal{N}\otimes \mathcal{M})>C(\mathcal{N})+C(\mathcal{M})$. We should note that the superadditivity of the Holevo information of two channels\footnote{Already proved to exist with Hastings counterexample to the additivity conjecture \cite{6}, as mentioned in Section~\ref{Sec:4.1}.} $\mathcal{N}$ and $\mathcal{M}$ does not guarantee the superadditivity of the overall capacity of the two channels $\mathcal{N}\otimes\mathcal{M}$ when used together.

Contrary, the situation for the quantum capacity is much more understood.

The quantum capacity can be superadditive over the use of two quantum channels $\mathcal{N}$ and $\mathcal{M}$ together \cite{13}. This could be described formally by: 
\begin{equation}
    \label{eq:4.15}
    Q(\mathcal{N}\otimes\mathcal{M})> Q(\mathcal{N})+Q(\mathcal{M})
\end{equation}

Furthermore, as discussed in Section~\ref{Sec:5} the quantum capacity can exhibit a \textit{superactivation phenomena}, which constitutes a form of superadditivity over different zero-capacity channels in the sense that the quantum capacity can satisfy the following inequality:
\begin{equation}
    \label{eq:4.16}
    Q(\mathcal{N}\otimes\mathcal{M})> Q(\mathcal{N})+Q(\mathcal{M})=0
\end{equation}
for $Q(\mathcal{N})=Q(\mathcal{M})=0$ \cite{4}. The superactivation of the quantum capacity is not possible for the classical capacity for reasons that we clarify in Section~\ref{Sec:5.4}.

\section{Superactivation of Quantum Channel Capacities}
 \label{Sec:5}

\begin{figure*}
    \begin{minipage}[c] {0.49\textwidth}
        \centering
        \includegraphics[width=1\columnwidth]{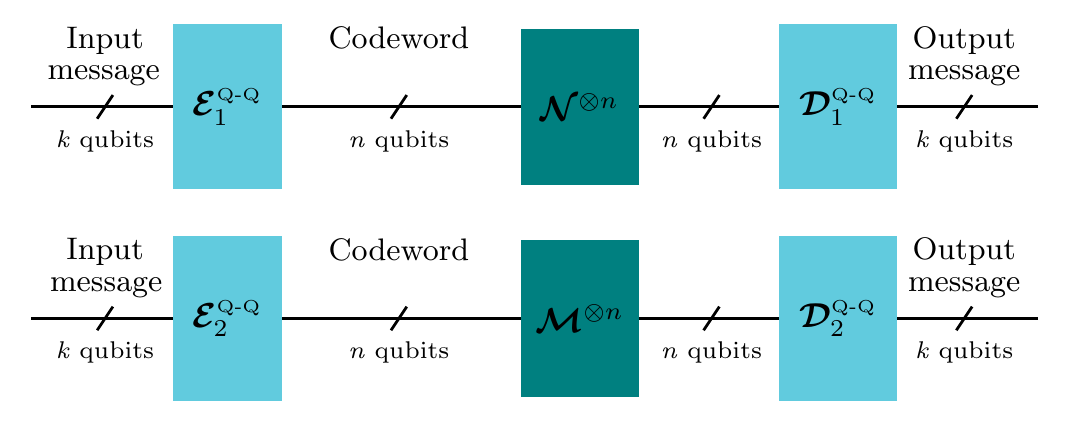}
        \subcaption{Alice and Bob attempt to separately use two zero-capacity channels $\mathcal{N}$ and $\mathcal{M}$ to transfer quantum states. Alice uses separate encoders $\mathcal{E}_1$ and $\mathcal{E}_2$ for each group of channels and Bob uses separate decoders $\mathcal{D}_1$ and $\mathcal{D}_2$. For any set of chosen encoding and decoding operations the transmission of information will fail due to the vanishing capacity of individual channels.}
        \label{Fig:10-a}
    \end{minipage}
    \hspace{0.02\textwidth}
    \begin{minipage}[c] {0.5\textwidth}
        \centering
        \includegraphics[width=1\columnwidth]{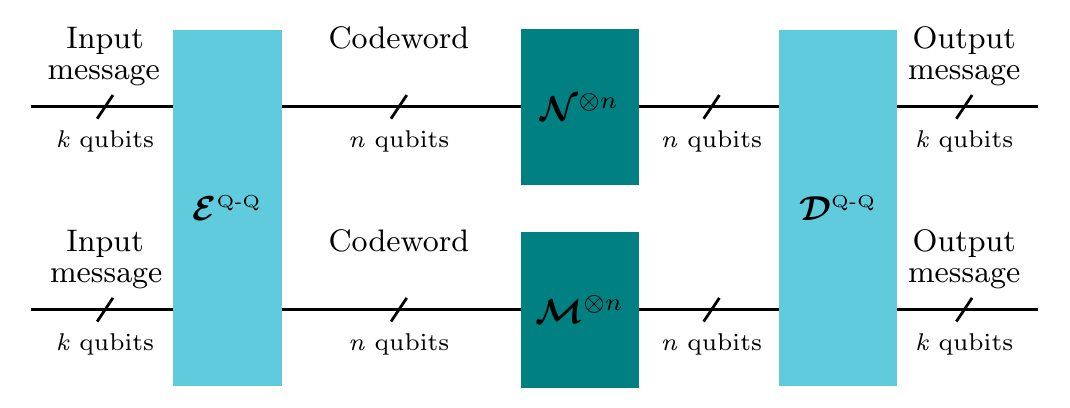}
        \subcaption{The same two channels being used for the
same task, but the} sender’s encoder $\mathcal{E}$ now has simultaneous access to the inputs of all channels being used, allowing for quantum information to get through the two channels, and the receiver’s decoding $\mathcal{D}$ is also performed jointly preserving coherence.
        \label{Fig:10-b}
        \end{minipage}
    \caption{Superactivation of the quantum capacity from the encoder perspective.}
    \label{Fig:10}
    \hrulefill
\end{figure*}

Here we detail the second marvel phenomenon introduced in Section~\ref{Sec:3}, namely, superactivation. Superactivation, as mentioned at the end of Sec~\ref{Sec:4}, is an unexpected genuinely quantum phenomenon that occurs when two zero-capacity quantum channels are used to transmit quantum information.
 
Unexpectedly, superactivation can only occur when the two cooperating quantum channels are from different families, none of which can simulate\footnote{I.e., arbitrary combinations of channels of one family cannot result in a channel from the other family \cite{4,0,Wilde}.} the other. In the next subsections, we discuss the different nonequivalent families of quantum channels known in literature. Subsequently, we provide examples of the phenomenon of superactivation for quantum channels from these families. 

\subsection{Classes of Zero-Capacity Channels}
 \label{Sec:5.1}

At least two classes of quantum channels are known to have zero capacities (whether additional classes of zero-capacity channels exist is still an open problem). The first class is known to be the family of antidegradable channels. Channels of this family, cannot transmit quantum messages due to the no-cloning theorem, which prohibits quantum information to be duplicated \cite{Caruso_2006}. As is discussed in Appendix~\ref{App:2}, antidegradable channels are self-complementary, in the sense that the environment of the channel can process its outcome to get an exact copy of the receiver. Thus, if this channel has a positive quantum capacity, it would violate the no-cloning theorem. An example of channels from this family is the \textit{50\% two-qubit erasure channel}, which faithfully transmits a two-qubit input state half of the time and outputs an erasure flag in the rest of the cases. This channel is given by \cite{0,Wilde}:
 \begin{equation}
     \mathcal{N}_{E_{50\%}}(\rho)  = \frac{1}{2}\rho+\frac{1}{2}\ket{e}\bra{e}
     \label{eq:5.1}
 \end{equation}
where $\ket{e}$ stands for the erasure flag\footnote{Mathematically, this means that $\mathcal{N}_{E_{50\%}}: \mathcal{H}_1 \otimes \mathcal{H}_2 \rightarrow  \mathcal{H}_1 \otimes \mathcal{H}_2 \oplus \operatorname{span}\{\ket{e}\}$, where $\mathcal{H}_{1,2}$ are the Hilbert spaces of the first and second qubit, respectively. Hence, this channel has a four-level input and a five-level output, where the extra output corresponds to the erasure flag.}. 

Another family which is known to have vanishing quantum capacity is the class of PPT channels. These are channels with Choi state that has a positive partial transpose\footnote{I.e., the partial transpose map $(\mathcal{I}_n \otimes \mathcal{T})$, where $\mathcal{I}_n$ is a $n$-dimensional identity map (in our case, $n=2$ for qubit), keeps the eigenvalues of the Choi state positive. See Example 5 in the Appendix~\ref{App:1} for the formal definitions of the transpose map $\mathcal{T}$ and the partial transpose map $(\mathcal{I}_n \otimes \mathcal{T})$.}, hence a PPT state. It is known that PPT states are states from which no entanglement can be distilled even asymptotically. The reason why PPT channels have zero capacity, is that no entanglement can be recovered between the sender and the receiver even at an unbounded use of the channel \cite{33,71}. A particular example of this family is the 4-dimensional Horodecki channel $\mathcal{N}_H$ given by its Kraus operators as:
\begin{align}
   & \sqrt{\frac{q}{2}}\mathrm{I}\otimes\ket{0} \bra{0}, ~\sqrt{\frac{q}{2}}Z\otimes\ket{1} \bra{1},~ \sqrt{\frac{q}{4}}Z\otimes Y, \nonumber \\
   &\sqrt{\frac{q}{4}}\mathrm{I}\otimes X,~ \sqrt{1-q}X\otimes M_0,~ \sqrt{1-q}Y\otimes M_1
   \label{eq:5.2}
\end{align}
with 
\begin{align}
   M_0 &=\begin{pmatrix}
        \frac{1}{2}\sqrt{2+\sqrt{2}} & 0\\
        0 &  \frac{1}{2}\sqrt{2-\sqrt{2}}
    \end{pmatrix} \nonumber \\
    M_1& =\begin{pmatrix}
        \frac{1}{2}\sqrt{2-\sqrt{2}} & 0\\
        0 &  \frac{1}{2}\sqrt{2+\sqrt{2}}
    \end{pmatrix}
    \label{eq:5.3}
\end{align}
and $I,X,Y,Z$ denoting the $2\times2$ generating matrices of the Pauli group.

\subsection{Superactivation of Quantum Capacity}
\label{Sec:5.2}
 
Superactivation is a strong superadditivity phenomenon that occurs when two channels, having vanishing individual quantum capacities $Q(\mathcal{N})=Q(\mathcal{M})=0$ belonging to different classes of zero capacity channels, are used together. These channels might gain potentially a quantum capacity enabling them to communicate quantum information, in such a way that:
\begin{equation}
    Q(\mathcal{N}\otimes\mathcal{M})>Q(\mathcal{N})+Q(\mathcal{M})=0
    \label{eq:5.4}
\end{equation}
As a result, we say that quantum capacity has been \textit{activated} \cite{4}. The phenomenon of superactivation is schematized in Figure~\ref{Fig:10}. 

In this context, it has been shown that, when a quantum channel is used together with a classical channel to transmit quantum information, this configuration does not increase the quantum capacity \cite{Bennett_1996}. This research area has been extended to symmetric side quantum channels \cite{Smith_2008}, whose use together with an arbitrary channel $\mathcal{N}$ exhibits the following single-letter expression:
\begin{equation}
     Q(\mathcal{N} \otimes \mathcal{N}_{SS}) = \sup_{S \in \mathcal{S}} I_c(\mathcal{N} \otimes S)
    \label{eq:5.4bis}
\end{equation}
where $\mathcal{N}_{SS}$ is the channel of unbounded dimension satisfying the optimization over the convex set $\mathcal{S}$ of symmetric side channels. In particular, it satisfies the following relation \cite{4}:
\begin{equation}
     Q(\mathcal{N} \otimes \mathcal{N}_{SS}) \geq \frac{1}{2}P(\mathcal{N})
    \label{eq:5.5}
\end{equation}
with $P(\mathcal{N})$ denoting the private capacity\footnote{In a nutshell, the private capacity defines the rate at which the channel can be used to send classical data that is secure against eavesdropper with access to the environment of the channel.} of channel $\mathcal{N}$.

Combined with the fact that the known Horodecki channels have a non-vanishing private capacity -- i.e., $P(\mathcal{N}) > 0$ -- this key result demonstrates that the capacity of Horodecki channels together with symmetric channels is non-vanishing. Namely, there exists a zero-quantum-capacity symmetric channel that, when used with a zero-quantum-capacity Horodecki channel, leads to a positive capacity. 
However, this result involves symmetric channels $\mathcal{N}_{SS}$ with infinite dimensional input, given the $\sup$ in \eqref{eq:5.4}. Hence, further bounds for symmetric side channels with finite dimensional inputs are needed.

Accordingly, it has been shown that when Alice and Bob use a 4-dimensional Horodecki channel $\mathcal{N}_H$ given in \eqref{eq:5.2} to communicate quantum messages with a symmetric channel given by 50\% two-qubit erasure channel $\mathcal{N}_{E_{50\%}}$ given in \eqref{eq:5.1}, the startling effect of superactivation occurs \cite{4}. When these two channels are combined, in fact, they satisfy \cite{4}:
\begin{align}
   & 0.1 < I_c(\rho,\mathcal{N}_H \otimes \mathcal{N}_{E_{50\%}}) \leq \nonumber \\
   & \quad I_c(\mathcal{N}_H\otimes \mathcal{N}_{E_{50\%}}) \leq Q(\mathcal{N}_H \otimes \mathcal{N}_{E_{50\%}})
    \label{eq:5.6}
\end{align}
where $I_c(\rho,\mathcal{N}_H \otimes \mathcal{N}_{E_{50\%}})$ is the coherent information of the channel $\mathcal{N}_H \otimes \mathcal{N}_{E_{50\%}}$ over a particular input state $\rho$ whose expression can be found in \cite{4}.

The two channels $\mathcal{N}_H \otimes \mathcal{N}_{E_{50\%}}$ are neither antidegradable nor PPT, having quantum capacity greater than zero. Therefore, we might interpret the gained capacity $Q(\mathcal{N}_H \otimes \mathcal{N}_{E_{50\%}}) > 0.1$ as the symmetry of the erasure channel being somehow \textit{broken} as an effect of the private information leaked through the Horodecki channel \cite{16}.

\subsection{Non-Convexity of Quantum Capacity}
\label{Sec:5.3}

Astoundingly, another form of superactivation for the previous channels has been revealed, in terms of the non-convexity property of the quantum capacity.  A channel, that is a flagged convex combination of the two zero capacity channels, can be constructed, and is given by
\begin{equation}
    \label{eq:5.7}
    \mathcal{M}_p=p\mathcal{N}_H\otimes\ket{0}\bra{0}+(1-p)\mathcal{N}_{E_{50\%}}\otimes\ket{1}\bra{1}
\end{equation}
It is a flagged\footnote{The flagged extension of quantum channels plays an essential role for finding tight bounds for quantum channel capacities that cannot be expressed as single-letter formulae.  Particular examples are the depolarizing channel and the generalized amplitude damping channel, whose capacity bounds are still an open problem for particular ranges of their noise parameters. Interested readers might be referred to the following recent results \cite{kianvash2020bounding, giovanetti,wang2020optimizing}.}, convex combination which can be switched between acting as $\mathcal{N}_H$ and $\mathcal{N}_{E_{50\%}}$ with the aid of an ancillary qubit degree of freedom. 

To better understand the capabilities of this channel for transmitting quantum information, one would calculate coherent capacity over multiple uses, as its one shot coherent capacity clearly vanishes because $\mathcal{N}_H\otimes\ket{0}\bra{0}$ and $\mathcal{N}_{E_{50\%}}\otimes\ket{1}\bra{1}$ are both zero-capacity channels. Subsequently, its two-shot coherent capacity is given by \cite{4}:
\begin{align}
    \label{eq:5.8}
    I_c({\rho},\mathcal{M}_p\otimes\mathcal{M}_p)&=p^2I_c({\rho},\mathcal{N}_H\otimes\mathcal{N}_H)\nonumber\\
    &+p(1-p)I_c({\rho},\mathcal{N}_H\otimes\mathcal{N}_{E_{50\%}})\nonumber\\&+p(1-p)I_c(\rho,\mathcal{N}_{E_{50\%}}\otimes\mathcal{N}_H)\nonumber\\
    &+(1-p)^2I_c({\rho},\mathcal{N}_{E_{50\%}}\otimes\mathcal{N}_{E_{50\%}})
\end{align}
Under symmetry restrictions of the input state $\rho$, the two-shot coherent capacity is not vanishing over a given region of the convexity parameter $p$ \cite{4,16}. 

This new channel, contrary to its constituent channels, has a non-vanishing capacity, exhibiting an extreme form of superactivation.
This confirms that the communication potential of a channel depends on the context in which it is used or on what other channels are available with it. This claim, will be further supported by the phenomenon of causal activation. 

\subsection{Classical Capacity}
\label{Sec:5.4}
As discussed in Section~\ref{Sec:5.1}, quantum channels can have zero capacity due of different reasons. This allows to categorize zero-capacity channels into different classes \cite{80,81,82}. Hence, if we use independently two quantum channels of different classes, the entanglement and coherence that might be available in the input state allow the channels to interfere with each other. Consequently, each one can leak some amount of information that the other channel does not allow. This interference between the two channels gives an equivalent channel that is of neither class, resulting in a noise reduction that beats the vanishing capacity of the individual quantum channels.

This cannot happen when quantum channels are used to transmit classical information, because only a channel, whose output is the same regardless of the input message, can have zero classical capacity \cite{0}. Hence, there exists only a single class of channels with zero classical capacity, and it is not possible to exploit channels of different classes to superactivate their classical capacities \cite{7,Gyongyosi_2012}.

\section{Causal Activation of Quantum Channel Capacities}
\label{Sec:6}

In ordinary quantum Shannon theory, although the information carriers obey the laws of quantum mechanics, the treatment of their propagation remains classical. Indeed, the informational carriers are transmitted through a well-defined trajectory which is assumed a-priori or can be chosen randomly, for example, by tossing a coin. Recent works proposed to generalise the framework of quantum Shannon theory \cite{3,13,9} such as, not only the information or the channels, but also the \textit{placement} of the channels -- i.e., the trajectories along with the carriers propagate -- can be treated as quantum and being subjected to the superposition principle.

In this section, we will review the possible advantages following the extension of quantum Shannon theory to include quantum trajectories, which is considered as the second quantization of classical Shannon theory \cite{3,13}.

\subsection{Quantum Switch}
\label{Sec:6.1}

\begin{figure}
    \centering
    \includegraphics[width=0.9\columnwidth]{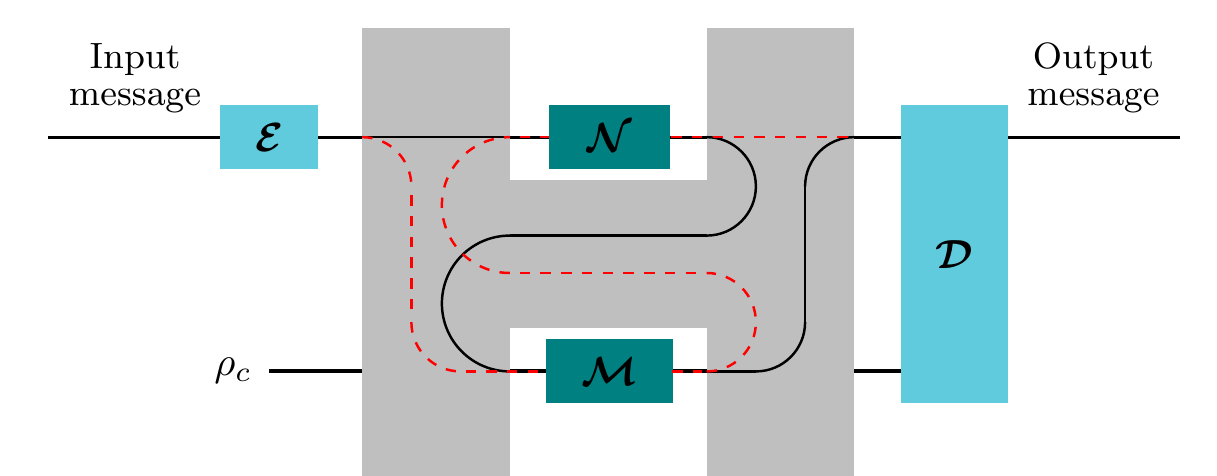}
    \caption{The quantum switch supermap, where two quantum channels $\mathcal{N}$ and $\mathcal{M}$ are placed in a genuinely quantum configuration given by a coherent superposition of causal orders between the two channels \cite{13}. Within the figure, $\rho_c$ denotes the \textit{control system}, part of the switch supermap, controlling the causal order between the two channels. Whenever the control qubit is initialized in a superposed state, the two channels are placed in a coherent superposition of the two different causal orders $\mathcal{M} \circ \mathcal{N}$ and $\mathcal{N} \circ \mathcal{M}$.}
    \label{Fig:12}
    \hrulefill
\end{figure}

A key example of quantum trajectories, which has been proven to be useful for communication, is given by the quantum switch\footnote{When the complete positivity of quantum combs or process tensors is restricted to non-signaling channels only, a wider class of supermaps emerges, which includes the quantum switch as a particular instance. The quantum switch supermap cannot be described by any form of a temporal process tensor or quantum comb, unless postselection on some degree of freedom of the environment is allowed \cite{83,Milz_2018}.} \cite{18,19,marcello,83,10,11,20,procopio-2020,cacciapuoti2019capacity}, illustrated in Figure~\ref{Fig:12}. Such a supermap, given two quantum channels $\mathcal{N}$ and $\mathcal{M}$, generates a new configuration in which the two channels are in a coherent superposition of two different causal orders, namely, $ \mathcal{M} \circ \mathcal{N}$ and $\mathcal{N} \circ \mathcal{M}$.

Formally, the quantum switch maps the two original channels $\mathcal{N}$ and $\mathcal{M}$ into a new quantum channel $\mathcal{S}_{\rho_c}(\mathcal{N},\mathcal{M})(\cdot)$, whose output is given by:
\begin{equation}
    \mathcal{S}_{\rho_c}(\mathcal{N},\mathcal{M})(\rho) = \sum_{ij} S_{ij} (\rho \otimes \rho_c) S^{\dagger}_{ij}
  \label{eq:6.1}  
\end{equation}
where $\rho$ is the input state, $\rho_c$ is the state controlling the causal order between the two channels in hand, and $\{ S_{ij} \}$ denotes the Kraus operators of the switch, given by:
\begin{equation}
    S_{ij} = N_i M_j \otimes \ket{0}\bra{0}_c + M_j N_i \otimes \ket{1}\bra{1}_c
    \label{eq:6.2}
\end{equation}
with $\{ N_i \}$ and $\{ M_j \}$ denoting the Kraus operators of the respective channels. We should note that the structure of the switch is independent of the Kraus representation of the individual channels.

This new resource has proven to provide advantages over the classical placement of quantum channels, violating the bottleneck inequality \eqref{eq:app.3.9} \cite{18,19,20,cacciapuoti2019capacity}. The rationale for this astonishing violation is that the coherent control within the quantum switch allows for the order -- in which the channels act on the information carrier -- to be entangled with a control degree of freedom. As a consequence, a constructive interference results from the coherent superposition of the causal order between the channels, allowing for a \textit{reduction} of the overall noise affecting the information carrier.

It is worth noting that the control system, whose state is fixed a-priory, is crucial in the switch. Indeed, it seems to be locking a considerable amount of information present in the coherent superposition of the orders. Clearly, with no access to the measurement outcome of the control qubit at the receiver -- hence, by ``tracing'' it out -- we cannot recover that amount of information. Furthermore, given that the control qubit embeds a fixed and a-priori determined quantum state, it can not be exploited by the sender to encode information\footnote{The control qubit might be regarded as the degree of freedom of the environment that is responsible for the order of the channels. This degree of freedom is accessible by the communication provider, who communicates measurement outcomes on it to the receiver Bob \cite{Globecom2021}.}, i.e., it does not constitute a side channel \cite{13}.

\subsection{Causal Activation of Holevo Information}
\label{Sec:6.2}

\begin{figure}
    \centering
    \includegraphics[width=0.9\columnwidth]{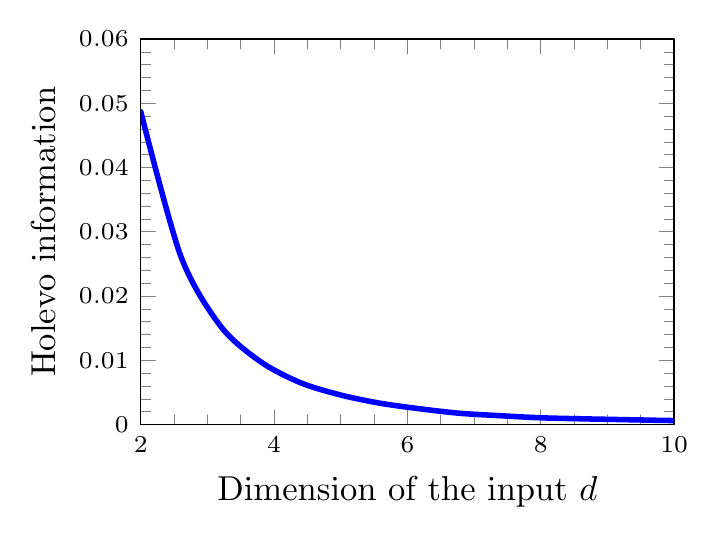}
    \caption{The Holevo information of the effective channel $\mathcal{S}_{\rho_c}(\mathcal{N}_{CD},\mathcal{N}_{CD})(\cdot)$ implemented through the quantum switch when $\rho_c$ is the density matrix for the state $\frac{1}{\sqrt{2}} \big( \ket{0}+\ket{1})$. The plot is an illustration of the results derived in \cite{18}.}
    \label{fig:13}
	\hrulefill
\end{figure}

The use of the quantum switch for the transfer of classical information over quantum channels has been shown to outperform the usual communication setups -- namely, sequential or parallel placement of channels in a causal order -- of quantum Shannon theory.

Specifically, when two completely depolarizing channels -- each with vanishing Holevo information, prohibiting them from transmitting classical messages whatever classical configuration they are used in -- are combined together in the quantum switch, they can deliver a non-vanishing amount of classical information \cite{18}. The completely depolarizing channel $\mathcal{N}_{CD}$ for a $d$-dimensional input $\rho$ is described by a mixture of $d^2$ mutually orthogonal unitaries\footnote{The completely depolarizing channel $\mathcal{N}_{CD}$ and its Kraus representation are discussed in Appendix~\ref{App:1}.} $\{U_i\}_{i=1}^{d^2}$ so that:
\begin{equation}
    \mathcal{N}_{CD}(\rho)=\frac{1}{d^2}\sum_{i=1}^{d^2}U_i\rho U_i^{\dagger}
    \label{eq:6.2bis}
\end{equation}
with the Kraus operators in \eqref{eq:6.2} describing the quantum switch supermap given by:
\begin{equation}
    S_{ij} = \frac{1}{d^2} \big( U_iU_j \otimes \ket{0}\bra{0}_c + U_jU_i \otimes \ket{1}\bra{1}_c \big)
    \label{eq:6.2ter}
\end{equation}

When the controller is initialized in the state $\rho_c=\sqrt{p}\ket{0}\bra{0}+\sqrt{1-p}\ket{1}\bra{1}+\sqrt{p(1-p)}(\ket{0}\bra{1}+\ket{1}\bra{0})$, the output \eqref{eq:6.1} of the quantum switch is given explicitly by:
\begin{align}
    \mathcal{S}_{\rho_c}(\mathcal{N}_{CD},\mathcal{N}_{CD})(\rho) &= \frac{I}{d} \otimes \big( p\ket{0}\bra{0}_c + (1-p)\ket{1}\bra{1}_c \big) \nonumber\\
    & \quad +\frac{\rho}{d^2}\otimes\sqrt{p(1-p)}(\ket{0}\bra{1}_c+\ket{1}\bra{0}_c)
    \label{eq:6.2quater}
\end{align}
where $I$ is the $d \times d$ identity matrix. By accounting for \eqref{eq:6.2quater} with $p = \frac{1}{2}$, the Holevo information achievable through the quantum switch $\mathcal{S}_{\rho_c}(\mathcal{N},\mathcal{M})(\cdot)$ is given by:
\begin{align}
    \chi \big( \mathcal{S}_{\ket{+}\bra{+}}(\mathcal{N}_{CD},\mathcal{N}_{CD}) \big) &= \log d+S(\tilde{\rho_c}) + \nonumber \\
    & \quad - H_{min}(\mathcal{S}_{\rho_c}(\mathcal{N}_{CD},\mathcal{N}_{CD}))
    \label{eq:6.2-5}
\end{align}
where $S(\tilde{\rho}_c)$ is the von Neumann entropy of the reduced state of the control system $\tilde{\rho_c} = \frac{1}{2}\ket{0}\bra{0}+\frac{1}{2}\ket{1}\bra{1}+\frac{1}{2d^2}(\ket{0}\bra{1}+\ket{1}\bra{0})$, and $H_{min}(\mathcal{S}_{\rho_c}(\mathcal{N}_{CD},\mathcal{N}_{CD}))$ is the minimum output entropy of the effective channel $\mathcal{S}_{\rho_c}(\mathcal{N}_{CD},\mathcal{N}_{CD})$, given by:
\begin{align}
    & H_{min}(\mathcal{S}_{\rho_c}(\mathcal{N}_{CD},\mathcal{N}_{CD})) = -\Big[ \frac{d+1}{2d^2} \log \frac{d+1}{2d^2} + \nonumber \\
    & \quad \quad \quad \quad \quad + \frac{2(d-1)}{2d} \log \frac{1}{2d} + \frac{d-1}{2d^2} \log \frac{d-1}{2d^2}\Big]
    \label{eq:6.2-6}
\end{align}
A plot of the Holevo information $\chi(\mathcal{S}_{\rho_c}(\mathcal{N}_{CD},\mathcal{N}_{CD}))$ in \eqref{eq:6.2-5}, characterizing the capability of the quantum switch to transfer classical information, is given in Figure~\ref{fig:13}. It is clear from the plot that the completely depolarizing channel, which has vanishing classical capacity over arbitrary many uses, gains a non vanishing Holevo information\footnote{Indeed, it has been shown that -- through $n>13$ channels placed in a superposition of cyclic causal orders, the quantum switch can activate the coherent information of the fully depolarizing channel as well \cite{10}, resulting in a non-vanishing quantum capacity.} whenever two instances of the channels are used within the quantum switch. It is worthwhile to mention that the Holevo information represents just a lower bound on the regularized classical capacity achievable with the quantum switch, which is non-vanishing as well.

This result, although moderate in terms of capacity improvement as shown in the figure, is of crucial importance from a communication engineering perspective, since it violates one of the fundamental bounds for classical trajectories, namely the bottleneck inequality given in \eqref{eq:app.3.9}.

\subsection{Causal Activation of Quantum Capacity}
\label{Sec:6.3}

As for classical capacity, there exists -- as well -- quantum channels with vanishing quantum capacity that, when combined within the quantum switch, gain a non-vanishing quantum capacity \cite{19}.

An illustrative example is the entanglement breaking channel $\mathcal{N}_{EB}$ characterized by the Kraus operators $\{X,Y\}$, and whouse ouput state is given by:
\begin{equation}
    \mathcal{N}_{EB}(\rho)=\frac{1}{2}(X\rho X+Y\rho Y)
    \label{eq:6.2-7}
\end{equation}
with $X$ and $Y$ denoting $2\times2$ Pauli matrices.

This channel has vanishing quantum capacity $Q(\mathcal{N}_{EB})=0$, regardless of the adopted classical (serial or parallel) configuration, since it is anti-degradable, i.e., the output on the receiver can be obtained by post-processing the output of the environment, resulting in a violation of no-cloning theorem as mentioned in Section~\ref{Sec:5.1}.

However, the quantum switch activates its capacity to its maximum\footnote{We further note that the qubit channels that might witness such perfect activation of the quantum capacity are the only ones unitary equivalent to the entanglement breaking channel given in \eqref{eq:6.2-7} \cite{19}.} whenever the control qubit places the channels in an equal superposition of orders, that is \cite{19}:
\begin{equation}
    Q \big( \mathcal{S}_{\ket{+}\bra{+}}(\mathcal{N}_{EB},\mathcal{N}_{EB}) \big) = 1
    \label{eq:6.2-8}
\end{equation}

This astonishing result can be easily understood by considering the output of the quantum switch, given by:
\begin{equation}
    \mathcal{S}_{\ket{+}\bra{+}}(\mathcal{N}_{EB},\mathcal{N}_{EB})(\rho) = \frac{1}{2}\rho\otimes \ket{+}\bra{+}_c + \frac{1}{2}Z\rho Z\otimes \ket{-}\bra{-}_c
    \label{eq:6.4}
\end{equation}
We can see that the outcome in \eqref{eq:6.4} is equivalent to a convex combination of two flagged channels $\mathcal{I}$ and $\mathcal{Z}$ and, hence, the coherent information of the equivalent channel is simply the convex sum of the coherent information of the two flagged channels:
\begin{equation}
    I_c \big( \mathcal{S}_{\ket{+}\bra{+}}(\mathcal{N}_{EB},\mathcal{N}_{EB}) \big)=\frac{1}{2}\times 1+\frac{1}{2}\times 1=1
\end{equation}

This result is astonishing, since it non only violates the bottleneck inequality given in \eqref{eq:app.3.9} as discussed in the previous subsection, but it activates the capacity to its maximum value, starting from zero-capacity channel.

Although our previous discussion explicitly shows the advantages of the quantum trajectories for communications, closed-form expressions of the ultimate capacities achieved through the quantum switch are yet to be solved for generic quantum channels. In this direction, many efforts are made to obtain tight upper and lower bounds on the quantum switch capacity. In particular, it has been shown \cite{procopio-2020} that the use of the three copies of the completely depolarizing channel outperforms the bound given in \eqref{eq:6.2-5}. This has been extended to show that the asymptotic use of many copies of the completely depolarizing channel in a superposition of cyclic orders achieves perfect transmission of classical information \cite{10,11}. Furthermore, upper and lower bounds of the quantum switch capacity have been obtained for different types of channels \cite{18,19,10,20,cacciapuoti2019capacity}.

\section{Conclusions and Future Perspective}
\label{Sec:7}

\begin{table*}
	\centering
	\begin{tabular}{| p{0.15\textwidth} | p{0.35\textwidth} | p{0.45\textwidth}|}
		\toprule
		    \textbf{} & \textbf{Superactivation} & \textbf{Causal activation}\\
		\midrule
		    \textbf{Entanglement} & Yes: within the encoding & Yes: between the causal order of the channels and the control system\\
        \midrule
            \textbf{Type of the channels} & Two different channels belonging to different zero-capacity classes & Two different or identical channels, as long as their Kraus operators do not pairwise commute/anti-commute with each others \\ 
        \midrule
            \textbf{Channel placement} & Classical & Quantum: superposition of relative orders \\
        \midrule
            \textbf{Channels with zero classical capacity} & Not activated & Activated\\
        \midrule
            \textbf{Channels with zero quantum capacity} & Activated & Activated \\
        \midrule
            \textbf{Noise Reduction} & Always & Not always\\
		\bottomrule
	\end{tabular}
	\caption{Superactivation vs causal activation. Although both superactivation and causal activation arise from the phenomenon of entanglement, and they both enable information transmission even through channels with zero capacity, they exhibit fundamental differences as summarized within the table.}
	\label{Tab:01}
	\hrulefill
\end{table*}

In classical communications, which are based on classical information flowing through classical channels, it is widely known that the channel capacity is additive. Namely, whenever a channel cannot transmit classical information over a single use, it can never gain potential to transmit information over multiple uses or when assisted by other zero-capacity classical channels.

Conversely, the weird unconventional phenomena of \textit{superadditivity},  \textit{superactivation} and \textit{causal activation} of quantum channel capacities violate known bounds and assumptions of classical Shannon theory, boosting -- sometimes with astonishing gains such as in Section~\ref{Sec:5} and Section~\ref{Sec:6.2} -- both the classical and the quantum capacities.

Hence, it is of paramount importance to i) discuss the rationale for these phenomena to appear in the quantum realm, and ii) highlight open problems and research directions, both from a communication engineering perspective.

\subsection{Discussion}
\label{Sec:7.1}

\subsubsection*{A.1) The role of quantum signatures}

As thoroughly discussed in the previous sections, the advantage that the phenomena of superadditivity, superactivation and causal activation provide for communications is based on the presence of entanglement, though in different disguises.

In superadditivity and superactivation, entanglement is exploited in the used codewords, enabling information carriers to be correlated while each traverses one channel. If the sender use separable codewords, as shown in Figure~\ref{Fig:10-a} with reference to the superactivation phenomenon, these phenomena do not occur. Conversely, for causal activation the entanglement is manifested in the correlation between: i) the order in which the channels acts on the information carrier, and ii) the degree of freedom of the control system, which necessarily does not carry any information.

Similarly to this key difference in the exploitation of the key-resource represented by entanglement there exists another distinction in terms of channel placement between superadditivity/superactivation and causal activation, as summarized with Table~\ref{Tab:01} for super vs causal activation. Specifically, the former two phenomena occur with a classical placement of channels -- either through i) multiple uses of the same channel, or ii) use of different zero-capacity channels from different classes -- whereas the latter occurs when a quantum trajectory is exploited -- with the only restrictions to have the channel Kraus operators not commuting or anti-commuting pairwise.

It is worthwhile to underline that -- regardless of the differences between the three phenomena -- quantum channels are a fundamental constraint for this marvels to occur. Meaning that these phenomena do not have any classical counterpart when classical channels are used for communication. 

\subsubsection*{A.2) Difference between causal activation and superadditivity/superactivation}

Furthermore, it is tempting to believe that quantum channels placed in quantum trajectories provide stronger advantages with respect to classical configurations such as those exploited by superadditivity and superactivation. However, this is not the case. Indeed, in the case of the causal activation, the information carrier undergoes a superposition of two sequences of channels with different causal orders, which might result in an overall noise addition instead of reduction. And the rationale for this is due to the fact that a destructive interference -- rather than a constructive one -- can take place. Differently, in superadditivity/superactivation, the information carriers are split between the different uses of the same channel or the different channels such as each carrier undergoes a single operation, which can only induce a noise reduction, and never a noise amplification. 

Finally, an interesting intersection between the two kinds of channels placement might be found by considering the family of flagged channels. In fact, a similarity between the phenomenon of superactivation in flagged convex combinations of zero-capacity channels, discussed in Section~\ref{Sec:5.3} and the phenomenon of causal activation in the quantum switch arises. This similarity becomes clear by noticing that the resulting channel from the quantum switch of two channels or more -- such as the one given in \eqref{eq:6.4} -- is nothing else than a quantum-flagged convex combination of two channels, which might have zero capacity in particular cases.

\subsection{Open Problems}
\label{Sec:7.2}
Besides the marvelous communication advantages that the discussed phenomena enable, there are relevant issues -- from the engineering perspective -- that we should point out and properly discuss.

\vspace{6mm}
\subsubsection*{B.1) Superactivation}

Primarily, superactivation is not yet fully understood, and many questions in this direction are yet to be answered. Basically, it is still important to understand whether there exist other families of zero-capacity quantum channels, besides the antidegradable and PPT families. Indeed, whether the superactivation holds for other Horodecki channels without positive private capacity, or whether there are other pairs of channels that witness such effect -- besides the 50\% erasure and the 4-dimensional Horodecki discussed in the text -- is still not answered.

Furthermore, besides the mentioned issues arising in discrete quantum channels, there is much more to discover and to investigate in the continuous domain \cite{Pirandola-2012,giovanetti-2018}. Recently, it has been showed that superactivation can be revealed in a broad range of thermal attenuator channels, even when the transmissivity is quite low, or the thermal noise is high \cite{adesso-2019}. This urges further investigations of whether superactivation might occur in physically relevant circumstances of quantum Gaussian channels \cite{Smith_2011,Lercher_2013}. This would be a triumph for future quantum communications based on quantum properties of light.

\vspace{6mm}
\subsubsection*{B.2) Superadditivity}

With reference to the superadditivity phenomenon, it has been proved for channels which might be relevant in realistic scenarios. Indeed, superadditivity has been shown for a given range of the depolarizing channel. Furthermore, a recent superadditivity phenomenon of the coherent information has been shown for the dephasure channel, which is a concatenation of an erasure and a dephasing channel. This erasure channel can be seen as a pure-loss bosonic channel on a dual-rail qubit system, which is a good model for optical fibers.

A strong superadditivity phenomenon has been revealed in quasi-zero-capacity channels. Specifically, quantum rocket channel -- namely, a channel with a $2\log d$ input qubits with private capacity less than 2 -- combined with the $d$-dimensional 50\% erasure channel -- which has zero private capacity -- can achieve high capacity in the order of $\frac{1}{2}\log d$ \cite{26}, hence, significantly larger than the capacity of the former channel. Consequently, intensive efforts are devoted to further investigations on the superadditivity of useful channels, both i) from a theoretical point of view, to serve as a laboratory for understanding quantum capacities, and ii) from a practical point of view, to harness the effect of superadditivity in near term quantum communication technologies.

However, and differently from quantum capacities, practical and concrete examples of superadditivity of the Holevo capacity are still missing, leaving an open door for future research to reveal the usefulness of superadditivity for the transmission of classical information over quantum channels. Moreover, a full understanding of the gap between capacities of quantum channels under different constraints -- namely, classical encoding-quantum decoding and quantum encoding-classical decoding -- is still missing. This urges further investigation of finite blocklength coding and decoding strategies \cite{Chung-16}, and the comparison between collective measurements and LOCC (local operations and classical communication) strategies on the discrimination of product states. The later has been thoroughly investigated recently in \cite{shor-2022}. We should highlight that we have omitted in this manuscript the discussion of superadditivity in trade-off capacities of quantum channels. This is the capacity given by a trade-off region considering a limited assistance of quantum communication by classical communication and entanglement. It has been shown that this kind of quantum capacity exhibits a superadditivity phenomenon. Interested reader is referred to \cite{zhuang-2017,zhuang-2018,zhuang-2021}.

Finally, a key issue is constituted by the fact that capacities of realistic channels, which models practical quantum communication scenarios on different platforms, are still not known. In particular, the capacities of the generalized amplitude damping channel is still not fully understood \cite{giovanetti-2018,Khatri_2020}. This channel can be seen as the qubit analogue of the bosonic thermal noise channel, and it models some of the sources of noise in superconducting circuit-based quantum computing. To this aim, many techniques for obtaining upper bounds of quantum channel capacities have been chased.  For upper bounds on the classical capacity of quantum channels, the reader can be referred to \cite{LedKauDat-18,Wang_2018,Filippov_2018,Filippov_2018-2,60}. In equal footing, for upper bounds of quantum capacities of quantum channels the reader is referred to \cite{Wolf_2007,smith-UB,Sutter_2017,Tomamichel_2017}.

\vspace{6mm}
\subsubsection*{B.3) Causal activation}

Not very different from the previous two phenomena, there is a lot to be understood in causal activation. This phenomenon has been shown to be advantageous for some practical channels, like the entanglement breaking channel in the case of quantum information transmission discussed in Section~\ref{Sec:6.3} and the completely depolarizing channel when it comes to the classical information transmission discussed in Section~\ref{Sec:6.2}. Nevertheless, causal activation for continuous variable channels is still missing, which would be of paramount importance for photonic-based future quantum communications.

Another issue that might face the engineering of causal activation is represented by the coherent control of realistic channels. Basically, to be able to perform coherent control, all we need to know is the properties of the quantum channels themselves, which are -- not easily -- obtained by quantum process tomography \cite{Chuang_1997,Bisio_2009}. Even more challenging, it has been shown that there are processes revealed to break one of the key properties of quantum channels, which is complete positivity \cite{sudarchan,Milz_2021,white2021nonmarkovian}. These processes cannot be described by Kraus operators and, hence, the quantum switch paradigm fails in this regard. 

A possible link between superactivation and causal activation of quantum channels might be tackled through the environment-assisted communication paradigm \cite{Werner-2004,hayden-2004,winter-2005}. On one hand, it has been shown that the quantum switch can be viewed as a one-way LOCC environment-assisted strategy \cite{Globecom2021}, where the environment is controlled by a helper. In this context, the control qubit of the switch arises as a residual degree of freedom of the environment. This particular strategy (the quantum switch) perfectly corrects the noisy channels when it is optimal, otherwise, the quantum switch fails to perfectly mitigate the noise. It is worth mentioning that optimality is with regard to the one-way LOCC strategy maximizing the environment-assisted capacity of the corrected channel. On the other hand, it has been shown that, when the helper is allowed to use entangled states of the environment, two useless channels with zero capacity under environment assistance might activate their joint capacity \cite{Mancini-2016,ManciniWinter-2016}. This opens a future direction for the investigation of the link between correlated control degrees of freedom among multiple quantum switches, and the possible superactivation therein. This will help better characterizing and understanding the capacity of the quantum channels used in the quantum switch.

Besides the advantages that the quantum switch can bring to point-to-point communications -- namely, mitigating the noise of quantum channels by placing them in coherent superposition of relative orders -- it would be quite valuable to find practical applications for quantum networks. A first contribution toward this issue has been proposed in \cite{Lab-2021}, where the indefinite causal order framework has been used to generate multipartite entanglement. Importantly, it has been shown that the application of the quantum switch can be advantageous for the achievement of distributed multipartite entanglement generation between remote nodes of a quantum network. Consequently, the quantum switch may play the missing part in achieving reliable photonic multi-qubit gates or, at least, a quantum interface between different qubit technologies, mapping entangled states engineered in a particular platform -- i.e., superconducting entanglement -- to photonic flying qubits used for long distance point-to-point communication \cite{rubino-2022}.

These different advantages of the quantum switch suggest a new way of looking at quantum networks. Namely, new quantum internet protocol stacks are yet to be proposed \cite{Jessica-2022}, taking into account the coherent control in general, and the superposition of causal order paradigm in particular, laying the ground for a complete understanding of the full potential of future communication networks.

\appendices


\section{Quantum information basics: crash course}
\label{App:0.1}

\subsubsection{Quantum bit and superposition principle}
\label{AppA:1}
Information, either classical or quantum, can be encoded in the state of the simplest quantum mechanical system, namely, the quantum bit (qubit). Mathematically\footnote{Here we adopt the bra-ket notation, which is usually adopted to denote the vector representing the state of a qubit. Indeed, a ket $\ket{\cdot}$ represents a column vector, while a bra $\bra{\cdot} = \ket{\cdot}^\dagger$ represents its hermitian conjugate. A scalar product of two vectors $\ket{\psi}$ and $\ket{\phi}$ is then denoted as $\langle \psi | \phi \rangle$, whereas a direct product of a ket and a bra is given by $| \phi \rangle \langle \psi |$.}, the state of a qubit is defined as a vector $\ket{\psi}$ in a two-dimensional Hilbert space $\mathcal{H}$. Therein it is possible to choose a basis, as instance the computational basis $\{\ket{0}, \ket{1}\}$ which draws an analogy with the states $0$ and $1$ of a classical bit. Then, according to the superposition principle, an arbitrary state of a qubit can be expressed as a linear combination of the chosen basis states: \begin{equation}\label{QubitState}
|\psi\rangle = \alpha |0\rangle + \beta |1\rangle
\end{equation}
where $\alpha, \beta \in \mathbb{C}$, and $|\alpha|^2 + |\beta|^2 = 1$. The state $|\psi\rangle$ in \eqref{QubitState} is said to be in a superposition of the states $\ket{0}$ and $\ket{1}$.

\subsubsection{Unitary transformations}
\label{AppA:2}
If a quantum system (such as a qubit) is closed, it can evolve in time only under deterministic and reversible unitary transformations $U$, i.e., transformations satisfying:
\begin{equation}
    U^\dagger U = I_\mathcal{H}
\end{equation}
where $I_{\mathcal{H}}$ is an identity in the Hilbert space $\mathcal{H}$. This means that, given the state of the system at some initial time point $t_1$, its state at a certain time $t_2$ is fully determined by the corresponding unitary operator:
\begin{equation}
    |\psi(t_2)\rangle = U(t_2, t_1)|\psi(t_1)\rangle
\end{equation}
which depends only on times $t_1$ and $t_2$. Unitary transformations play a crucial role in quantum information and quantum communications since they can seen as gates acting on a qubit. In this picture, a quantum gate has input and output ports for a qubit, and the time evolution is hidden in the relationship between them, 
\begin{equation}
    |\psi\rangle_{out} = U|\psi\rangle_{in}
\end{equation}
Typical examples of quantum gates widely used in quantum information are the Pauli gates
\begin{equation}
    X = \begin{pmatrix} 0 & 1 \\ 1 & 0 \end{pmatrix} \;\; Z = \begin{pmatrix} 1 & 0 \\ 0 & -1 \end{pmatrix} \;\; Y = \begin{pmatrix} 0 & -i \\ i & 0 \end{pmatrix} 
\end{equation}
which flip the bit ($\ket{0} \rightarrow \ket{1}$), the phase ($\alpha\ket{0} + \beta \ket{1} \rightarrow \alpha\ket{0} - \beta \ket{1}$), or both, respectively.

An important consequence of the constraint on the transformations of a closed quantum system to be unitary is the celebrated no-cloning theorem (see Section~\ref{Sec:2.1}), which states the impossibility of creating an independent copy of an unknown quantum state. Indeed, there exists no unitary operator $U$ acting on two quantum systems able to transform state $\ket{\psi_1}$ of one system into state $\ket{\psi_2}$ of another one, regardless of $\ket{\psi_2}$.

\subsubsection{Projective measurements}
\label{AppA:3}
If the state of the system is unknown, certain information on it can be acquired by measuring some (observable) property of it. Mathematically\footnote{We first describe the projective measurement, and then at the end of this appendix we generalize it by considering the positive operator-valued measure (POVM). For an exhaustive treatise about the subject, the reader is referred to \cite{nielsen_chuang_2010}.}, any observable is described by an operator $A$ that is self-adjoint (i.e., $A^\dagger = A$) and can be expanded as:
\begin{equation}
    A = \sum_i a_i M_i
\end{equation}
where $\{a_i\}$ are its eigenvalues describing the possible outcomes of the measurement, and $M_i$ are the orthogonal projectors onto the eigenvectors associated with the corresponding eigenvalues:
\begin{eqnarray}\label{MeasurementOrthogonality}
    M_i M_j &=& \delta_{ij} M_i \\
    \sum_i M_i &=& I_{\mathcal{H}} \label{MeasurementCompleteness}
\end{eqnarray}
By measuring the observable $A$, a certain outcome $a_i$ is obtained. However,  accordingly to the quantum measurement postulate, after this measurement the system is left in the eigenstate associated with the projector $M_i$. With more details, when a measurement is performed on a system in the state $\ket{\psi}$, the outcome $a_i$ is obtained with probability calculated according to the Born's rule: 
\begin{equation}
    \mathbb{P}(a_i) = \langle \psi | M_i^\dagger M_i | \psi \rangle = \langle \psi | M_i | \psi \rangle
\end{equation}
After the measurement, the system collapses into the state
\begin{equation}
\label{eq:appe_a}
    \frac{M_i \ket{\psi}}{\sqrt{\mathbb{P}(a_i)}}.
\end{equation}
We note that any following measurement of the same observable reveals again the same outcome $a_i$ and state in \eqref{eq:appe_a}.

\begin{exmp}
Let us consider a simple example to better present the above concept related to the quantum measurement. Specifically, let us suppose to be interested in measuring the qubit state \eqref{QubitState} in the computational basis $\{\ket{0}, \ket{1}\}$. In this case, $M_0 = \ket{0}\bra{0}$ and, $M_1 = \ket{1}\bra{1}$. By measuring the considered state and according to the Born's rule, we obtain the outcome ``$0$'' with probability given by:
\begin{equation}
    \mathbb{P}(0) = (\alpha^* \bra{0} + \beta^* \bra{1}) \ket{0}\bra{0} (\alpha \ket{0} + \beta \ket{1}) = |\alpha|^2
\end{equation}
and the outcome ``$1$'' with the probability:
\begin{equation}
    \mathbb{P}(1) = (\alpha^* \bra{0} + \beta^* \bra{1}) \ket{1}\bra{1} (\alpha \ket{0} + \beta \ket{1}) = |\beta|^2
\end{equation}
and the system, after the measurement, is left in the state $|0\rangle$ or $|1\rangle$, respectively. We could measure the qubit in any other basis, for example, $\{ |\pm\rangle = \frac{|0\rangle \pm ||1\rangle}{\sqrt{2}} \}$. The corresponding observable can be constructed as
\begin{equation}
    A = a_+ M_+ + a_- M_- = \ket{+}\bra{+} - \ket{-}\bra{-}
\end{equation}
where $a_+ = 1$, $a_- = -1$, and $M_+ = \ket{+}\bra{+}$, $M_- = \ket{-}\bra{-}$. In this case, the measurement reveals both outcomes ``$+$'' or ``$-$'' with the same probability
\begin{eqnarray}
    \nonumber \mathbb{P}(\pm) &=& (\alpha^* \bra{0} + \beta^* \bra{1}) \ket{\pm}\bra{\pm} (\alpha \ket{0} + \beta \ket{1}) \\
    &=& \frac{|\alpha|^2 + |\beta|^2}{2}  = \frac{1}{2}
\end{eqnarray}
\end{exmp}

\subsubsection{Mixed states and density matrix}\label{AppA:4}
In the situations when the knowledge on the actual state is lacking (for example, if the system undergoes the action of noise), it cannot be described by a well-defined vector in Hilbert space. This means that the system is in a certain state with some probability, i.e., it has to be described by a statistical mixture of vectors in Hilbert space. Such a statistical mixture is called \textit{mixed state} (in contrast to a well-defined vector which represents a \textit{pure state}) and it can be defined formally by adapting the formalism of \textit{density matrix}. Indeed, if the system, with dimension $d$, is in one of the states $\{\ket{\psi_i}\}_i^d$ with corresponding probability $p_i$, the density matrix that describes its overall state is defined as
\begin{equation}\label{def:DensityOperator}
    \rho = \sum_i^d p_i |\psi_i\rangle\langle\psi_i|
\end{equation}
For a pure state $\ket{\phi}$, the density matrix reduces to $\rho=\ket{\phi}\bra{\phi}$. Generally speaking, any operator $\rho$ can be a density operator and describe a state of the system, as long as it fulfills the following conditions,
\begin{enumerate}
    \item $\rho^\dagger = \rho$, i.e., $p_i \in \mathbb{R}$ for all $i$,
    \item $\rho \geq 0$, i.e., $p_i \geq 0$ for all $i$,
    \item $\operatorname{Tr}(\rho) = 1$, i.e., $\sum_i^d p_i = 1$.
\end{enumerate}
These conditions ensure that the eigenvalues of $\rho$ can be interpreted as probabilities, namely, they are real, positive, and sum up to the unity. It is necessary to stress out the crucial difference between these ``classical'' probabilities $p_i$ and the ``quantum'' ones $\mathbb{P}(i)$. The probabilities $\mathbb{P}(i)$ appear when one performs a measurement on the (well-defined) system's state due to the Born's rule, whereas the probabilities $p_i$ describe our a priori knowledge of the actual system's state independently on measurement. Indeed, when a measurement of an observable $A$ is performed on a qubit being in the state $\rho$, an outcome $a_i$ is revealed with the probability
\begin{equation}
    \mathbb{P}(a_i) = \operatorname{Tr}(M_i \rho)
\end{equation}
leaving the system in the state
\begin{equation}
    \frac{M_i \rho M_i}{\mathbb{P}(a_i)}
\end{equation}
 
\begin{exmp}
Being the state in (\ref{QubitState}) a pure state, its density matrix can be evaluated as
\begin{eqnarray}
    \nonumber \rho_\psi &=& |\psi\rangle \langle\psi| \\
    \nonumber  &=& |\alpha|^2 |0\rangle\langle 0| + \alpha\beta^* |0\rangle\langle 1|  + \alpha^*\beta |1\rangle\langle 0| + |\beta|^2 |1\rangle\langle 1| \\
    &=& \begin{pmatrix} |\alpha|^2 & \alpha\beta^* \\ \alpha^*\beta & |\beta|^2 \end{pmatrix}
\end{eqnarray}
On the other hand, the classical mixture of the states $0$ and $1$ with the probabilities $|\alpha|^2$ and $|\beta|^2$ is described by the mixed state
\begin{eqnarray}
    \nonumber \rho &=& |\alpha|^2 |0\rangle\langle 0| + |\beta|^2 |1\rangle\langle 1| \\
    &=& \begin{pmatrix} |\alpha|^2 & 0 \\ 0 & |\beta|^2 \end{pmatrix}
\end{eqnarray}
When measured in computational basis, in both cases, the qubit can be found in the state $0$ and $1$ with probabilities $|\alpha|^2$ and $|\beta|^2$, respectively. However, if the measurement is performed in a basis which includes $|\psi\rangle$, i.e., it answer the question ``Is the qubit in the state $|\psi\rangle$ or not?'', then, in the first case, the answer is always ``Yes'', and the measurement does not change the state of the system. In the second case, the outcome ``Yes'' is obtained indeed with the probability
\begin{equation}
    \mathbb{P}(\text{\rm Yes}) = |\alpha|^4 + |\beta|^4
\end{equation}
As discussed, the crucial difference is that, in the first case, the qubit stays in a well-defined state $|\psi\rangle$ which is revealed, as we have seen, when a suitable measurement is performed. In the second case, however, the qubit a priori stays in one of the states $|0\rangle$ or $|1\rangle$ with the corresponding probabilities.
\end{exmp}

\subsubsection{POVM}\label{AppA:5}
Before ending the appendix it is important to highlight that the projective measurements introduced above and described by a set of orthogonal projectors $\{M_i\}$, which satisfy conditions (\ref{MeasurementOrthogonality}) and (\ref{MeasurementCompleteness}), represent a special case of the general quantum measurement postulate. However, there are important problems in quantum computation and quantum information, such as the optimal way to distinguish a set of quantum states, which require a more general tool, as the positive operator-valued measure (POVM) formalism \cite{nielsen_chuang_2010}, where the measurement operators $M_i$ are not necessarily orthogonal.

\begin{exmp}
An important example of using POVM in quantum communications is given by the problem of distinguishing between non-orthogonal states. Given a set of $N$ linearly independent states $\{\ket{\psi_i}\}$, no projective measurement can tell with a certainty that a qubit has been in one of them before measurement if they are not orthogonal. However, a wisely chosen POVM allows to perfectly distinguish between them by paying the price that sometimes no information about the state can be revealed at all. Indeed, it can be achieved by considering a set of states $\{\ket{\tilde{\psi}_i}\}$ such that a state $\ket{\tilde{\psi}_i}$ is orthogonal to all the states under interest but $\ket{\psi_i}$~\cite{Barnett-09}, i.e.,
\begin{equation}
    \langle \tilde{\psi}_{i} | \psi_j \rangle = \delta_{ij}
\end{equation}
where $\delta_{ij}$ is unity for $i=j$ and zero otherwise. Then the POVM consisting of $N$ projectors 
\begin{equation}
M_i = \frac{1}{|\langle \tilde{\psi}_{i} | \psi_i \rangle|^2} |\tilde{\psi}_{i}\rangle \langle \tilde{\psi}_{i}|    
\end{equation}
and the operator $M_{N+1} = I - \sum_i M_i$ allows to distinguish perfectly between $\{\ket{\psi_i}\}$. Indeed, finding an outcome $i \in \{1,...,N\}$ suggests that the system has been in the state $|\psi_i\rangle$ before measurement. However, finding the outcome $N+1$ associated with the operator $M_{N+1}$ does not give any information about the state of the system at all. For example, let us assume that we have a qubit and want to distinguish between two states, $|\psi_1\rangle = |0\rangle$ and $|\psi_2\rangle = |+\rangle$. In this case, a POVM consisting of operators
\begin{eqnarray}
    M_1 &=& 2|-\rangle\langle -| \\
    M_2 &=& 2|1\rangle\langle 1| \\
    M_3 &=& I - 2|-\rangle\langle -| - 2|1\rangle\langle 1|
\end{eqnarray}
does the job.
\end{exmp}

\subsubsection{Composite systems and entanglement}
\label{AppA:6}
A generic pure uncorrelated state of a composite system of $n$ qubits $\{\ket{\psi_i}_{i=1}^{n}\}$ is described by a joint quantum state 
\begin{equation}
    \ket{\psi}=\otimes_{i=1}^n\ket{\psi_i}=\ket{\psi_1}\otimes\ket{\psi_2}\otimes\cdots\otimes\ket{\psi_n} \label{eq:3.3.5.1}
\end{equation}
belonging to a $2^n$-dimensional complex Hilbert space. To simply illustrate this, we consider a two qubit system $A$ and $B$. The two systems are described individually in the basis $\{\ket{0},\ket{1}\}$. Accordingly, their joint state would be described by the tensor product basis given as $\{\ket{00},\ket{01},\ket{10},\ket{11}\}$. Any state $\ket{\psi}^{AB}$ of the joint system would be given explicitly by 
\begin{equation}
    \ket{\psi}^{AB}=\alpha_0\ket{00}+\alpha_1\ket{01}+\alpha_2\ket{10}+\alpha_3\ket{11} \label{eq:3.3.5.2}
\end{equation}
with $\alpha_i\in \mathbb{C}$ and $\sum_i|\alpha_i|^2=1$. Any joint state of this composite system that cannot be written in a product form as in \eqref{eq:3.3.5.1} should present some form of correlations between systems $A$ and $B$. This form of correlations is called \textit{entanglement}, and the corresponding state is deemed \textit{entangled state}. A famous example of an entangled state in two qubit systems is the set of Bell pairs given by:
\begin{align}
    \ket{\Phi^\pm}&=\frac{1}{\sqrt{2}}(\ket{00}\pm\ket{11})\\
    \ket{\Psi^\pm}&=\frac{1}{\sqrt{2}}(\ket{01}\pm\ket{10})
\end{align}
More generally, any bipartite quantum system, no matter its state is pure or mixed, is said to be entangled if it cannot be written as a convex combination (hence, probabilistic mixture) of product states in the form:
\begin{equation}
    \rho^{AB}=\sum_ip_i\rho^A_i\otimes\rho^B_i
\end{equation}
A joint state that can be written in this form is called \textit{separable}. It is worth noting that separable states can have classical correlations between the systems $A$ and $B$.

\section{Quantum channels}
\label{App:1}

A quantum communication channel $\mathcal{N}$ is described mathematically by a completely positive trace-preserving (CPTP) map $\mathcal{C}: \rho_A \to \rho_B$ from states $\rho \eqdef \rho^A \in \mathcal{L}(\mathcal{H}^{A})$ belonging to the set of density operators $\mathcal{L}(\mathcal{H}^{A})$ over the input Hilbert space $\mathcal{H}^{A}$ to states $\rho^B\in\mathcal{L}(\mathcal{H}^{B})$ on an output Hilbert space $\mathcal{H}^{B}$. The condition of CPTP assures that the output of the map $\mathcal{C}$ is a valid density operator. In fact, it assures that
\begin{itemize}
    \item $\mathcal{C}$ outputs a positive operator (positivity),
    \item for any $n$, $\mathcal{I}_n \otimes \mathcal{C}$ -- with $\mathcal{I}_n$ denoting an identity map on $n$-dimensional operators -- outputs a positive operator (complete positivity),
    \item $\operatorname{Tr}(\mathcal{C}[\rho_A]) = \operatorname{Tr}(\rho_A)$ (trace preservation).
\end{itemize}
In the following we provide two simple examples to better understand the above concepts.

\begin{exmp}
The completely depolarizing channel is a widely-used quantum channel model, and it is described by the following input-output relationship:
\begin{equation}
    \mathcal{N}_{CD}(\rho_A) = \frac{\operatorname{Tr}(\rho_A)}{d} I_d
\end{equation}
where $\rho_A$ is the input state belonging to a $d$-dimensional Hilbert space $\mathcal{H}^A$, and $I_d$ is the $d$-dimensional identity matrix. The output state $\mathcal{N}_{CD}(\rho_A)$ has a unique $d$-degenerate positive eigenvalue $\frac{1}{d}$ and $\operatorname{Tr}(\mathcal{N}_{CD}(\rho_A)) = \operatorname{Tr}(\rho_A)$. As a consequence, $\mathcal{N}_{CD}$ is a positive and trace-preserving map. Moreover, it is a completely positive map. In fact, by adding an ancilla system $E$, we can consider the action of the map $\mathcal{I}_n \otimes \mathcal{N}_{CD}$ on the entire state $\rho_{EA} \in \mathcal{L}(\mathcal{H}^E \otimes \mathcal{H}^A)$ with $\rho_A = \operatorname{Tr}_E(\rho_{EA})$. Accordingly, we obtain that:
\begin{equation}
    (\mathcal{I}_n \otimes \mathcal{N}_{CD})(\rho_{EA}) = \operatorname{Tr}_A(\rho_{EA}) \otimes \frac{I_d}{d}
\end{equation}
has positive eigenvalues since $\operatorname{Tr}_A(\rho_{EA})$ is a state. Hence, completely depolarizing map $\mathcal{N}_{CD}$ is a quantum channel.
\end{exmp}

\begin{exmp}
Let us consider a map $\mathcal{T}$ that transposes a state $\rho_A = \sum_{ij} p_{ij} |i\rangle \langle j|$ of the system $A$, where we fix $\{|i\rangle\}$ as a computational basis in $\mathcal{H}^A$. Its output, given by:
\begin{equation}
    \mathcal{T}(\rho_A) = \sum_{ij} p_{ij} |j\rangle \langle i|
\end{equation}
with $T$ denoting the matrix transpose, exhibits obviously the same eigenvalues and trace as $\rho_A$. Hence, $\mathcal{T}$ is positive and trace-preserving. However, let us add another system $B$ in order to check whether $\mathcal{T}$ is completely positive. The state of the entire system reads:
\begin{equation}
    \rho_{BA} = \sum_{ijkl} p_{klij} \ket{k} \bra{l} \otimes \ket{i} \bra{j}
\end{equation}
and, if $\mathcal{T}$ acts on $A$, it becomes:
\begin{eqnarray}
    \nonumber (\mathcal{I}_n \otimes \mathcal{T})(\rho_{BA}) &=& \sum_{ijkl} p_{klij} \ket{k} \bra{l} \otimes \ket{j} \bra{i}
\end{eqnarray}
Now let us assume $A$ and $B$ to be maximally entangled qubits (hence, $n=2$),
\begin{equation}
    \rho_{BA} \equiv \Phi^{BA} = \frac{1}{2}\begin{pmatrix} 1 & 0 & 0 & 1 \\ 0 & 0 & 0 & 0 \\ 0 & 0 & 0 & 0 \\ 1 & 0 & 0 & 1 \end{pmatrix}
\end{equation}
and consider again the action of the map $\mathcal{T}$ on $A$. The entire output is given by
\begin{equation}
    (\mathcal{I}_2 \otimes \mathcal{T})(\rho_{BA}) = \frac{1}{2}\begin{pmatrix} 1 & 0 & 0 & 0 \\ 0 & 0 & 1 & 0 \\ 0 & 1 & 0 & 0 \\ 0 & 0 & 0 & 1 \end{pmatrix}
\end{equation}
whose eigenvalues are $\pm \frac{1}{2}$. Since one of its eigenvalues is negative, $(\mathcal{I}_2 \otimes \mathcal{T})(\rho_{BA})$ is not positive and, therefore, not a state. This means that the transpose map $\mathcal{T}$ is not completely positive and, hence, it cannot represent any quantum channel. Nevertheless, the partial transpose map $(\mathcal{I}_n \otimes \mathcal{T})$ plays an important role in quantum communications lying in the core of the PPT- or Peres-Horodecki criterion for determining entanglement.
\end{exmp}

There are several ways of representing quantum channels formally, some of which will be useful in our discussion. 

\subsection{Kraus Representation}
\label{AppB:A}
A quantum channel $\mathcal{N}$ is described by an operator sum decomposition in Kraus operators as follows \cite{42,43,Wilde}:
\begin{equation}
    \label{eq:app.1.1}
    \mathcal{N}(\rho)=\sum_{i=1}^kA_i\rho A_i^{\dagger}
\end{equation}
with $\rho\in \mathcal{L}(\mathcal{H}^A)$ and with $\{A_i\}_{i=1}^k$ being linear operators from $\mathcal{L}(\mathcal{H}^{A})$ to $\mathcal{L}(\mathcal{H}^{B})$ satisfying the normalization condition:
\begin{equation}
    \sum_{i=1}^k A_iA_i^{\dagger}=I_{\mathcal{H}^{A}}
\end{equation}

\begin{exmp}\label{ExampleKrausCompDepChan}
The completely depolarizing channel $\mathcal{N}_{CD}$ has the following Kraus representation,
\begin{equation}
\mathcal{N}_{CD}(\rho)=\frac{1}{d^2}\sum_{i=1}^{d^2}\hat{U}_i \rho \hat{U}_i^\dagger
\end{equation}
with $\{\hat{U}_i\}$ being a set of unitary operators that are mutually orthogonal, i.e., $\operatorname{Tr}(\hat{U}_i^\dagger \hat{U}_j) = d\delta_{ij}$. For a qubit ($d=2$), a set of Pauli operators with identity can be chosen, $\{ \hat{U}_i \} = \{I_2, X, Y, Z\}$, leading to the Kraus representation
\begin{equation}\label{QubitDepChan}
\mathcal{N}_{CD}(\rho)=\frac{1}{4}\Bigl[ \rho + X\rho X + Y\rho Y + Z\rho Z \Bigr]
\end{equation}
In this representation, the channel can be interpreted as a noisy channel that causes a bit error ($X$), a phase error ($Z$), both errors ($Y$), or no error ($I_2$) with the same probability $p=\frac{1}{4}$.
\end{exmp}

\begin{exmp}
The qubit completely depolarizing channel (\ref{QubitDepChan}) can be naturally generalized to the Pauli channel that causes the mentioned above errors with the corresponding probabilities $p_X, p_Y, p_Z$. This is the quantum channel usually adopted in quantum communications to model a noisy qubit channel, and it has the following Kraus representation,
\begin{equation}
\mathcal{P}(\rho) = (1 - \Sigma_{XYZ})\rho + p_X X\rho X + p_Y Y\rho Y + p_Z Z\rho Z
\end{equation}
where $\Sigma_{XYZ} = p_X + p_Y + p_Z$. Obviously, for $p_X = p_Y = p_Z = \frac{1}{4}$ the Pauli channel reduces to the completely depolarizing channel (\ref{QubitDepChan}). On the other hand, the choice $p_X = p_Y = p_Z = \frac{q}{3}$ leads to the depolarizing channel
\begin{equation}
    \mathcal{N}_D(\rho) = (1 - q)\rho + \frac{q}{3} \Bigl( X\rho X + Y\rho Y + Z\rho Z \Bigr)
\end{equation}
\end{exmp}

\subsection{Isometric extension (Stinespring dilation)}
\label{AppB:B}

\begin{figure*}[t]
   \centering
    \includegraphics[width=0.9\textwidth]{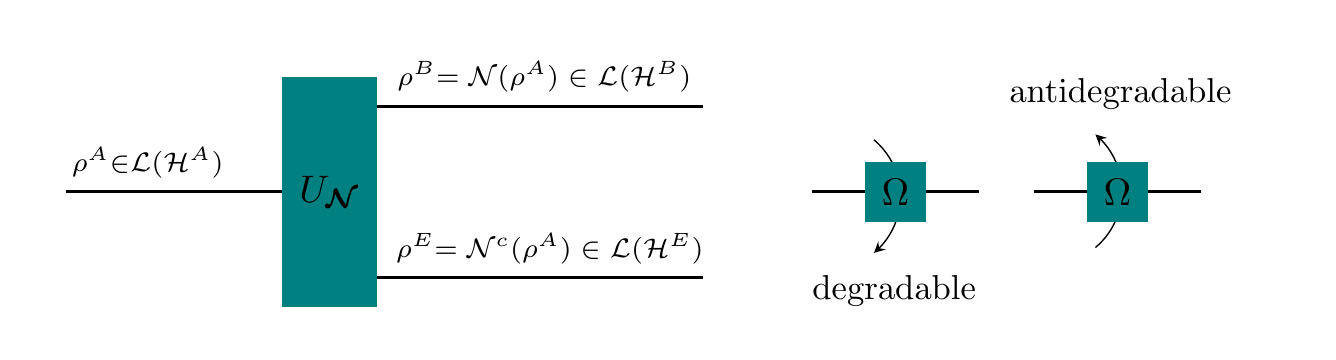}
    \caption{A scheme depicting channel $\mathcal{N}: \mathcal{L}(\mathcal{H}^A) \rightarrow \mathcal{L}(\mathcal{H}^B)$ as the reduced dynamics of an isometry describing the joint evolution of the system \textit{source}-\textit{receiver} and the environment $E$. Clearly, $\rho^B = \operatorname{Tr}_{E}[U_{\mathcal{N}} \rho^A U_{\mathcal{N}}^{\dagger}]$ and $\rho^E = \operatorname{Tr}_{B}[U_{\mathcal{N}} \rho^A U_{\mathcal{N}}^{\dagger}]$. The figure depicts also the relations holding for degradable/antidegradable channels.} 
    \label{Fig:A.1}
    \hrulefill
\end{figure*}

A quantum channel $\mathcal{N}$ can be described -- as shown in Figure~\ref{Fig:A.1} -- by a reduced dynamics $\mathrm{Tr}_{E}(\cdot)$ on the isometry (i.e., a map that preserves the inner product) $U_{\mathcal{N}}$ simulating the joint evolution of the system $A$ and environment $E$ together as \cite{Wilde,44}:
\begin{equation}
    \label{eq:app.1.2}
    \mathcal{N}(\rho)=\mathrm{Tr}_{E}(U_{\mathcal{N}}\rho U_{\mathcal{N}}^{\dagger})
\end{equation}
where $U_{\mathcal{N}}$ is a linear operator that maps $\mathcal{H}^A$ onto $\mathcal{H}^{B} \otimes \mathcal{H}^E$ such that $U_{\mathcal{N}}^{\dagger}U_{\mathcal{N}} = I_{\mathcal{H}^{A}}$. The two descriptions \eqref{eq:app.1.1} and \eqref{eq:app.1.2} are equivalent in the sense that if we know one Kraus decomposition $\{A_i\}_{i=1}^k$ of the channel, given an orthogonal basis of $\mathcal{H}^{E}$ as $\{\ket{i}\}^{E}$, the isometric extension $U_{\mathcal{N}}$ is given by:
\begin{equation}
    \label{eq:app.1.3}
    U_{\mathcal{N}}=\sum_{i=1}^k A_i\otimes \ket{i}^{E}
\end{equation}

\begin{exmp}
For the Pauli channel $\mathcal{P}$ introduced in the previous example, the set of Kraus operators is $\{A_i\} = \{ \sqrt{1 - \Sigma_{XYZ}} I_2, \sqrt{p_X} X, \sqrt{p_Y} Y, \sqrt{p_Z} Z \}$. Therefore, its isometric extension reads
\begin{equation}
U_\mathcal{P} = \begin{pmatrix} \sqrt{1 - \Sigma_{XYZ}} & 0 \\ 0 & \sqrt{p_X} \\ 0 & -i\sqrt{p_Y} \\ \sqrt{p_Z} & 0 \\ 0 & \sqrt{1 - \Sigma_{XYZ}} \\ \sqrt{p_X} & 0 \\ i\sqrt{p_Y} & 0 \\ 0 & -\sqrt{p_Z} \end{pmatrix}.
\end{equation}
In particular, for the completely depolarizing channel $\mathcal{N}_{CD}$, the isometric extension reduces to
\begin{equation}
U_{\mathcal{N}_{CD}} = \frac{1}{2}\begin{pmatrix} 1 & 0 \\ 0 & 1 \\ 0 & -i \\ 1 & 0 \\ 0 & 1 \\ 1 & 0 \\ i & 0 \\ 0 & -1 \end{pmatrix}.
\end{equation}
\end{exmp}

It is important to underline that the Kraus decomposition of a quantum channel is not unique, thus the construction of the isometric extension of the channel is not unique as well. Another important concept associated to the Stinespring dilation is the complementary channel $\mathcal{N}^c$ of a quantum channel $\mathcal{N}$. The complementary channel describes the channel transmitting information to the environment rather than transmitting information to the output Hilbert space $\mathcal{H}^{B}$, and it is given by:
\begin{equation}
    \label{eq:app.1.4}
    \mathcal{N}^c(\rho)=\mathrm{Tr}_{B}(U_{\mathcal{N}}\rho U_{\mathcal{N}}^{\dagger})
\end{equation}

\subsection{Choi state of a quantum channel}
\label{AppB:C}

\begin{figure}[t]
   \centering
    \includegraphics[width=0.9\columnwidth]{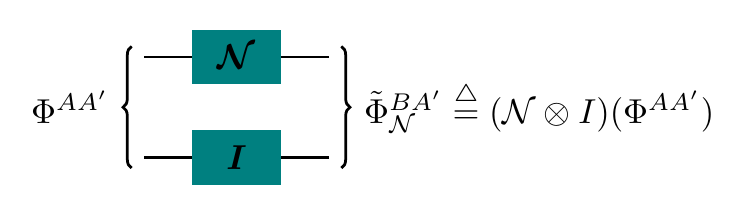}
    \caption{A scheme depicting the Choi state $\tilde{\Phi}^{BA'}_{\mathcal{N}}$ of the arbitrary channel $\mathcal{N}$, obtained by sending i) one part of the maximally entangled state $\Phi^{AA'}$ through channel $\mathcal{N}$, and ii) the other part through the identity channel $\mathcal{I}$.}
    \label{Fig:A.2}
    \hrulefill
\end{figure}

A fundamental relation between quantum channels and states is the Choi-Jamio\l{}kowski isomorphism. This isomorphism enables a one-to-one map between an arbitrary quantum channel $\mathcal{N}$ and a density operator -- referred to as $\tilde{\Phi}^{BA'}_{\mathcal{N}}$ in the following -- in $\mathcal{L}(\mathcal{H}^B\otimes \mathcal{H}^{A'})$ on the Hilbert space $\mathcal{H}^B\otimes \mathcal{H}^{A'}$ of the joint system $BA'$, with $A'$ denoting the auxiliary system showed in Figure~\ref{Fig:A.2}.

The connection results from a direct application of the map $\mathcal{N}$ on one part of a maximally entangled state\footnote{Where $\mathcal{H}^{A'}$ is isomorphic to the input Hilbert space $\mathcal{H}^A$ with dimension $d$ that, generally speaking, might be different from the dimension of $\mathcal{H}^B$.} of $AA'$ such as $\ket{\Phi}^{AA'}=\frac{1}{\sqrt{d}}\sum_i\ket{ii}^{AA'}$ with density matrix $\Phi^{AA'}$, in order to create what is known as the \textit{Choi-Jamio\l{}kowski state} (CJ) of the channel $\mathcal{N}$:
\begin{equation}
    \label{eq:app.1.5}
    \tilde{\Phi}^{BA'}_{\mathcal{N}} \eqdef \tilde{\Phi}^{BA'}_{\mathcal{N} \otimes \mathcal{I}_d} = (\mathcal{N}\otimes \mathcal{I}_d) (\Phi^{AA'})
\end{equation}

\begin{exmp}
The Choi state of the completely depolarizing channel $\mathcal{N}_{CD}$ reads
\begin{equation}
    \tilde{\Phi}^{BA'}_{\mathcal{N}_{CD}} = \frac{I_{d^2}}{d^2}
\end{equation}
\end{exmp}

\begin{exmp}
The Choi state of the qubit Pauli channel $\mathcal{P}$ reads
\begin{equation}
    \tilde{\Phi}^{BA'}_{\mathcal{P}} = \frac{1}{2} \begin{pmatrix} 1 - \Sigma_{XY} & 0 & 0 & 1 - \tilde{\Sigma}_{XYZ} \\ 0 & \Sigma_{XY} & \Delta_{XY} & 0 \\ 0 & \Delta_{XY} & \Sigma_{XY} & 0 \\ 1 - \tilde{\Sigma}_{XYZ} & 0 & 0 & 1 - \Sigma_{XY} \end{pmatrix}
\end{equation}
where $\Sigma_{XY} = p_X + p_Y$, $\Delta_{XY} = p_X - p_Y$, and $\tilde{\Sigma}_{XYZ} = p_X + p_Y + 2p_Z$.
\end{exmp}

\section{Degradability/anti-degradability of quantum channels}
\label{App:2}
The definition of the complementary channel given in \eqref{eq:app.1.4} allows us to introduce the notion of degradability of a quantum channel \cite{Wilde,46,45}. 

A channel $\mathcal{N}$ is said to be degradable if the final state obtained by the environment can be obtained by postprocessing the state at the receiver by applying a third channel (CPTP) map, as shown in Figure~\ref{Fig:A.1}. Formally, the channel $\mathcal{N}$ is degradable if there exists a CPTP map $\Omega:\mathcal{L}(\mathcal{H}^B)\rightarrow \mathcal{L}(\mathcal{H}^E)$ satisfying the relation:
\begin{equation}
    \label{eq:app.2.1}
    \mathcal{N}^c=\Omega\circ\mathcal{N}
\end{equation}
Similarly, a channel is said to be anti-degradable \cite{47} if there exists a CPTP map $\Omega:\mathcal{L}(\mathcal{H}^E)\rightarrow \mathcal{L}(\mathcal{H}^B)$  satisfying:
\begin{equation}
    \label{eq:app.2.2}
    \mathcal{N}=\Omega\circ\mathcal{N}^c
\end{equation}
Many channels are neither degradable nor anti-degradable. However, it was shown that qubit channels with one qubit environment are always either degradable or anti-degradable or both (symmetric) \cite{84}. A particular example of anti-degradable channels is the set of entanglement breaking channels \cite{85} mentioned in Section~\ref{Sec:5.1}. These are the channels whose Choi state given in \eqref{eq:app.1.5} is separable \cite{85}. It is known that the set of anti-degradable channels is convex, that is, any convex combination on anti-degradable channels is an anti-degradable channel, but surprisingly, the set of degradable channels is not convex \cite{81}.

\section{Entropic quantities}
\label{App:3}
Entropic quantities play an essential role in the study of quantum communications, as they characterizes the performance of quantum channels. The von Neumann (quantum) entropy $S(\rho)$ of a quantum state $\rho$ is given by \cite{Wilde,47}:
\begin{eqnarray}
    \nonumber S(\rho)&=&-\mathrm{Tr}(\rho\log_2(\rho))\ \\
    \label{eq:app.3.1} &=& -\sum_i \lambda_i \log_2(\lambda_i)
\end{eqnarray}
where $\{\lambda_i\}$ is the set of eigenvalues of $\rho$, i.e., ``classical'' probabilities $p_i$ in its expansion (\ref{def:DensityOperator}). This generalizes the classical Shannon entropy of a random variable $X$ defined as \cite{41}:
\begin{equation}
    \label{eq:app.3.2}
    H(X) = - \sum_xp(x)\log_2 p(x)
\end{equation}

\begin{exmp}
To grasp better the introduced entropy concept, let us consider the pure state $\rho_\psi = \begin{pmatrix} |\alpha|^2 & \alpha\beta^* \\ \alpha^*\beta & |\beta|^2 \end{pmatrix}$. Its von Neumann entropy is given by:
\begin{eqnarray}
    \nonumber S(\rho_\psi) &=& 0 \cdot \log_2(0) + 1 \cdot \log_2(1) \\
    &=& 0,
\end{eqnarray}
being pure states characterized by only one eigenvalue different by zero and equal to one. Differently, for the probabilistic mixture $\rho_\psi = \begin{pmatrix} |\alpha|^2 & 0 \\ 0 & |\beta|^2 \end{pmatrix}$ the von Neumann entropy coincides with the classical Shannon entropy,
\begin{equation}
    S(\rho) = H(|\alpha|^2, |\beta|^2),
\end{equation}
which is maximal for $|\alpha| = |\beta| = \frac{1}{\sqrt{2}}$.
\end{exmp}

Let $\mathcal{N : \mathcal{H}^A \rightarrow \mathcal{H}^B}$ be a quantum channel and let $A'$ be an auxiliary system evolving through $\mathcal{I}$ as shown in Figure~\ref{Fig:A.2}, with the additional property of being a purifying system for $\rho^A$. Specifically, the auxiliary system $A'$ is chosen so that the joint state $\rho^{AA'}$, satisfying
\begin{equation}
    \label{eq:app.3.3}
    \rho^{A} = \mathrm{Tr}_{A'}(\rho^{AA'}).
\end{equation}
is a pure state, regardless of $\rho^A$ being a pure or a mixed state:

Also, let us denote the entropy of the input state $\rho^A$ as:
\begin{equation}
    \label{eq:app.3.4}
    S(A) \eqdef S(\rho^A)
\end{equation}
with a slight abuse of notation, given the dependence of $S(A)$ on the input state $\rho^A$, but being consistent with the literature \cite{Wilde,0,Caruso_2014}. Similarly, the entropy of the output state  $\rho^B \eqdef \mathcal{N}(\rho^A)$ of the channel as:
\begin{equation}
    \label{eq:app.3.5}
    S(B) \eqdef S(\mathcal{N}(\rho))
\end{equation}
Accordingly, the entropy of the output of the complementary channel $\mathcal{N}^c$ can be written as: \begin{align}
    \label{eq:app.3.6}
     S(E) & \eqdef S\big(\mathcal{N}^c(\rho)\big) \nonumber \\
     & = S\big( (\mathcal{N}\otimes \mathcal{I}) (\rho^{AA'})\big) \nonumber \\
     & \eqdef S\big(BA'\big)
\end{align}
This quantity is known as the \textit{entropy of exchange} \cite{49}, which refers to the amount of information leaking to the environment instead of being reliably transferred to the receiver. The relation between the different states in \eqref{eq:app.3.6} is better understood through the isometric representation of the channel $\mathcal{N}$ as:
\begin{align}
    \mathcal{N}^c(\rho^A)&=\mathrm{Tr}_{BA'}\Bigl((\mathcal{U}_{\mathcal{N}}\otimes\mathcal{I})(\rho^{AA'})\Bigr)\nonumber\\
    \mathcal{N}\otimes\mathcal{I}(\rho^{AA'})&=\mathrm{Tr}_{E}\Bigl((\mathcal{U}_{\mathcal{N}}\otimes\mathcal{I})(\rho^{AA'})\Bigr)
\end{align}
with $(\mathcal{U}_{\mathcal{N}}\otimes\mathcal{I})(\rho^{AA'}) = (U_{\mathcal{N}}\otimes\mathcal{I})\rho^{AA'} (U_{\mathcal{N}}^{\dagger}\otimes\mathcal{I})$ and $U_{\mathcal{N}}$ is the isometric extension given by \eqref{eq:app.1.3}.

Moreover, the equality between the first and the last line in \eqref{eq:app.3.6} results from the fact that the state of the global system, given by the environment $E$, the receiver $B$ and the purifying system $A'$, is a pure state. This purity of the joint system $EBA'$ can be easily observed from the fact that the joint evolution on the input system $\rho^{AA'}$ -- which is pure by definition -- is in fact an isometry given by $U_{\mathcal{N}}\otimes\mathcal{I}$. Indeed, this isometry preserves purity by definition. As a consequence, the entropies of any complementary bi-partitions on the joint output pure system should be equal. Hence, the equality $S(E) = S(BA')$ holds.

The previous entropies -- input, output and exchanged -- are the essential building blocks for many information measures in quantum communications\footnote{For a complete understanding on the relation between them, we refer the reader to \cite{Wilde,51,52}.}. In the following, we focus on the measures used within the paper.

\vspace{3pt}

A key measure needed for our discussions in Section~\ref{Sec:4.1} is the \textit{Holevo information} \cite{57}. This is a functional $\chi(\cdot,\cdot)$ of an input ensemble of states $\{p_x,\rho_x\}$ that the sender Alice inputs to the channel $\mathcal{N}$ for transmitting classical information through a quantum channel. Formally, the Holevo information of channel $\mathcal{N}$ with respect to the arbitrary input $\rho=\sum_xp_x\rho_x$ is given by:
\begin{equation}
    \label{eq:app.3.10}
    \chi(\{p_x,\rho_x\},\mathcal{N})=S(\mathcal{N}(\rho))-\sum_xp_xS(\mathcal{N}(\rho_x))
\end{equation}
where $\rho$ is the quantum ensemble encoding the classical message given by the alphabet $\mathcal{X}$ over which the random variable $X$ takes values.

It has been shown that the Holevo information provides an upper bound on the mutual information $I(X:Y)$, given by:
\begin{equation}
    \label{eq:app.3.7b}
    I(X:Y)=H(X)+H(Y)-H(X,Y)
\end{equation}
where $X$ is the random variable describing the message $x$ to be transferred by Alice, and $Y$ is the random variable referring to the output, after a POVM is applied by Bob to estimate the value $x$. This is known as the Holevo bound \cite{58}, and is given by:
\begin{equation}
    \label{eq:app.3.11}
    I(X:Y) \leq \chi(\{p_x,\rho_x\},\mathcal{N})
\end{equation}
It is worth mentioning, that the Holevo information is useful for many tasks in quantum estimation and quantum discrimination, for which it has been derived.

Another key measure is the \textit{quantum mutual information} of channel $\mathcal{N}$ with respect to the arbitrary state $\rho \eqdef \rho^A$ as: 
\begin{equation}
    \label{eq:app.3.7}
    I(\rho,\mathcal{N}) = S(A) + S(B) - S(E)
\end{equation}
which is the quantum version of Shannon's mutual information given in \eqref{eq:app.3.7b}.

Similarly, a measure needed for our discussions in Section~\ref{Sec:4.2} is the \textit{coherent information} of channel $\mathcal{N}$ with respect to the arbitrary state $\rho$, given by \cite{51,52,Wilde}:
\begin{align}
    \label{eq:app.3.8}
    I_c(\rho,\mathcal{N}) &= S(\mathcal{N}(\rho)) - S(\mathcal{N}^c(\rho))  \nonumber\\
    &= S(B)-S(E) \nonumber\\
    & = -S(A'|B)
\end{align}
with $S(A'|B) \eqdef S(BA') - S(B)$ denoting the \textit{conditional von Neumann entropy} and the last identity following from \eqref{eq:app.3.6}.

It can be easily seen from the second line of \eqref{eq:app.3.8} that the coherent information is the difference between the amount of information arriving to the receiver given by the output entropy, and the amount of information leaked to the environment given by the entropy of exchange. Furthermore, from the third line of the same equation, we see that the coherent information is the negative of the conditional quantum entropy. This latter quantity can be negative, in contrast to its classical counterpart, namely, the conditional entropy $H(X|Y)$. An interpretation of the negativity of this quantity has been given in the context of quantum state merging \cite{86}, where it has been shown that the negativity of the quantum conditional entropy relates to the fact that the sender and the receiver gain a potential for future quantum communications. For extensive details on the properties of the quantum mutual information and the coherent information the reader is referred to \cite{0,Wilde,56,53,54,55}.

We further note that both the Holevo information and the coherent information satisfy a data processing inequality. Specifically, whenever two arbitrary channels $\mathcal{N}$ and $\mathcal{M}$ are placed sequentially, they satisfy the following bottleneck inequalities:
\begin{align}
    \label{eq:app.3.9}
    f(\rho, \mathcal{M}\circ\mathcal{N}) &\leq \min\big\{ f(\rho, \mathcal{M}), f(\rho, \mathcal{N})\big\}
\end{align}
with $\circ$ denoting the dequential concatenation operator and $f(\cdot,\cdot)$ denoting either $\chi(\cdot,\cdot)$ or $I_c(\cdot,\cdot)$.

\section{Quantum codes and rates}
\label{App:4}

An important notion both practically and theoretically is the notion of a code. Generally, if Alice and Bob want to communicate a message, they choose appropriate encoding and decoding strategies, allowing them to reach their ultimate rate of communication, by counteracting the effect of noise of the communication line. Formally, this consists of an encoding map $\mathcal{E}$:
\begin{equation}
    \label{eq:app.4.1}
    \mathcal{E}: \mathcal{M}\rightarrow\mathcal{L}(\mathcal{H}^{\otimes n})
\end{equation}
from the alphabet of classical messages $\mathcal{M}$ with $k = \log_2 |\mathcal{M}|$ to a large state space of $n$ quantum carriers of information, and a decoding map $\mathcal{D}$:
\begin{equation}
    \label{eq:app.4.2}
    \mathcal{D}:\mathcal{L}(\mathcal{H}^{\otimes n})\rightarrow \mathcal{M}
\end{equation}
from the joint state of the $n$-carriers to the alphabet $\mathcal{M}$. This is summarized in Figure~\ref{Fig:05}. In the case of communicating quantum messages, the alphabet $\mathcal{M}$ above is replaced by the set of quantum states $\mathcal{L}(\mathcal{H})$ over a Hilbert space $\mathcal{H}$ of dimension $d$, and $k=\log_2 d$.

Each element of the image set $\mathcal{L}(\mathcal{H}^{\otimes n})$  is called a codeword and the rate of the code is given by the non-negative number $R \eqdef \frac{k}{n}$. Clearly, a rate is achievable if there exists code -- i.e., an encoder $\mathcal{E}$ and a decoder $\mathcal{D}$ -- so that the probability of decoding the message erroneously vanishes as $n$ goes to infinity.

\bibliographystyle{IEEEtran}
\bibliography{ref}

\begin{IEEEbiography}
[{\includegraphics[width=1in,height=1.25in,clip,keepaspectratio]{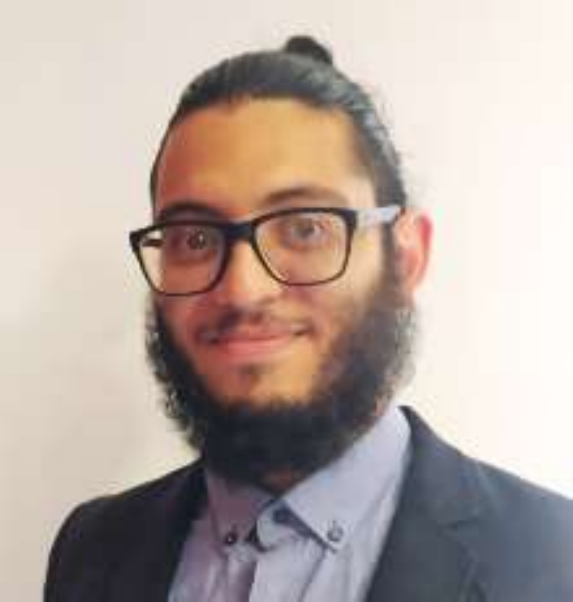}}]{Seid Koudia} Received the B.Sc degree in fundamental physics in 2015 and the M.Sc degree in theoretical physics with distinction in 2017 from the University of Sciences and Technology Houari Boumedien (USTHB). Currently, he is pursuing a PhD degree in Quantum technologies with the Future Communications Laboratory (FLY), Department of Electrical Engineering and Information Technology (DIETI). His research interests include quantum information theory, quantum communications, quantum networks and quantum coding theory.
\end{IEEEbiography}

\begin{IEEEbiography}
[{\includegraphics[width=1in,height=1.25in,clip,keepaspectratio]{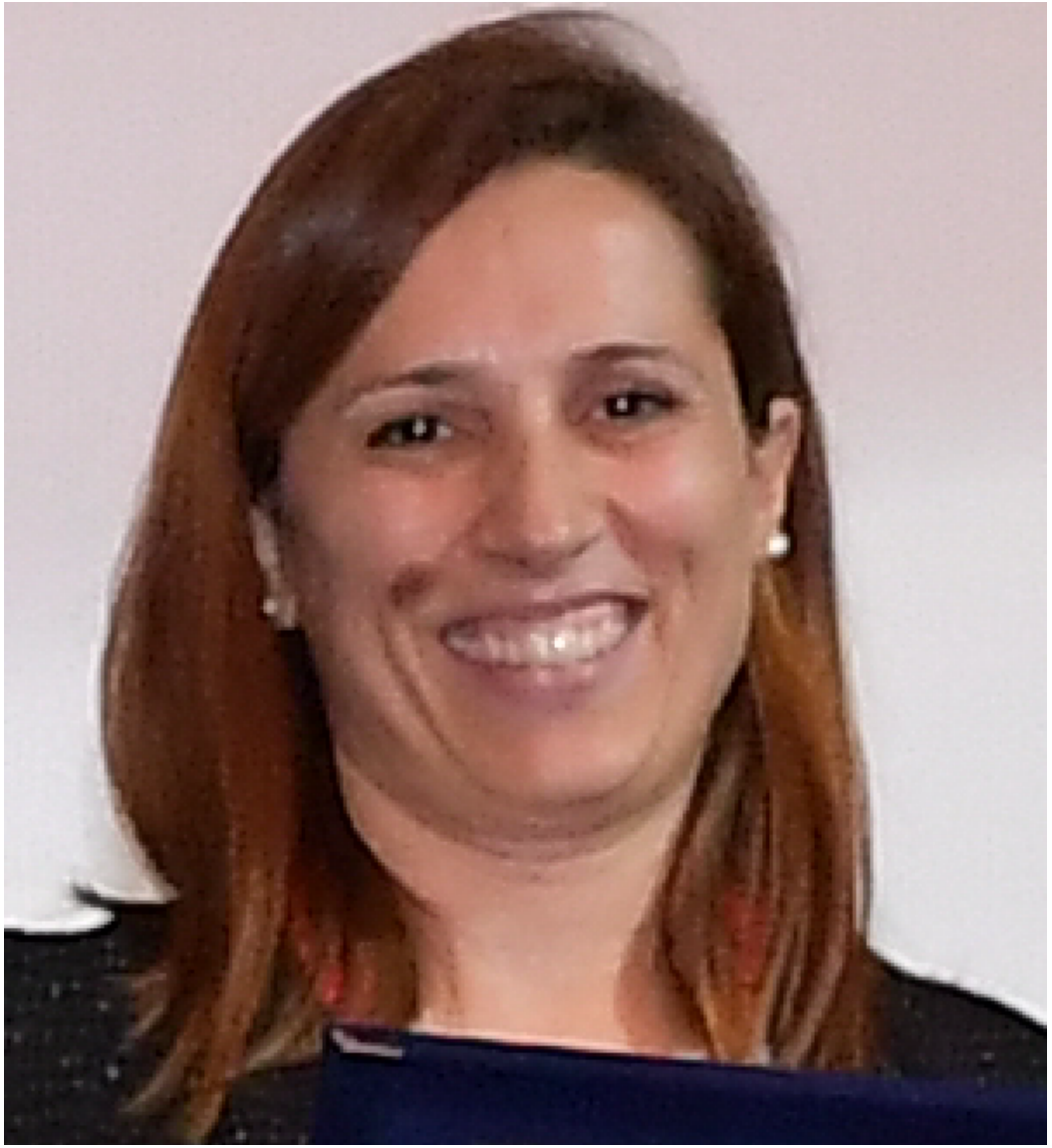}}]
{Angela Sara Cacciapuoti} (M'10, SM'16) is a professor at the University of Naples Federico II (Italy). Since July 2018 she held the national habilitation as “Full Professor” in Telecommunications Engineering. Her work has appeared in first tier IEEE journals and she has received different awards and recognition, including the \textit{``2022 IEEE ComSoc Best Tutorial Paper Award''} and \textit{``2021 N2Women: Stars in Networking and Communications''}. For the Quantum Internet topics, she is a \textit{IEEE ComSoc Distinguished Lecturer}, class of 2022-2023. Currently, Angela Sara serves as \textit{Area Editor} for IEEE Communications Letters, and as \textit{Editor/Associate Editor} for the journals: IEEE Trans. on Communications, IEEE Trans. on Wireless Communications, IEEE Trans. on Quantum Engineering, IEEE Network. She was the recipient of the \textit{2017 Exemplary Editor Award} of the IEEE Communications Letters. From 2020 to 2021, Angela Sara was the Vice-Chair of the IEEE ComSoc Women in Communications Engineering (WICE). Previously, she has been appointed as Publicity Chair of WICE. From 2016 to 2019 she has been an appointed member of the IEEE ComSoc Young Professionals Standing Committee. From 2017 to 2020, she has been the Treasurer of the IEEE Women in Engineering (WIE) Affinity Group of the IEEE Italy Section. Her current research interests are mainly in Quantum Communications, Quantum Networks and Quantum Information Processing.
\end{IEEEbiography}

\begin{IEEEbiography}
[{\includegraphics[width=1in,height=1.25in,clip,keepaspectratio]{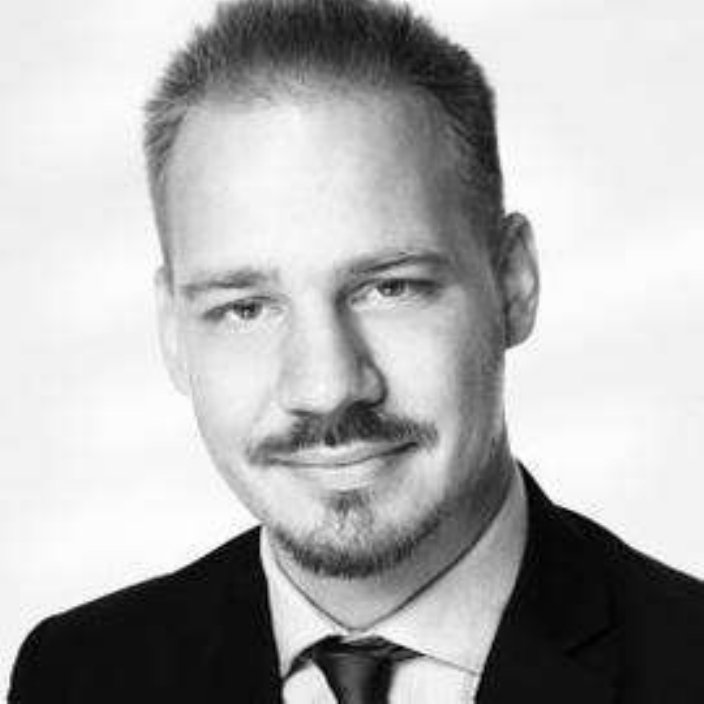}}]{Kyrylo Simonov} received the M.Sc. degree in physics in 2014 from the Taras Shevchenko National University of Kyiv (Ukraine) with a thesis on physics of DNA and the Ph.D. degree in physics in 2018 from the University of Vienna (Austria) with a thesis on quantum foundations. Since 2018 he worked at the Faculty of Mathematics of the University of Vienna (Austria) on mathematical foundations of quantum mechanics and applications of nonstandard analysis. His research interests include quantum information theory, quantum communications, quantum foundations, quantum thermodynamics, and mathematical foundations of quantum theory.
\end{IEEEbiography}

\begin{IEEEbiography}
[{\includegraphics[width=1in,height=1.25in,clip,keepaspectratio]{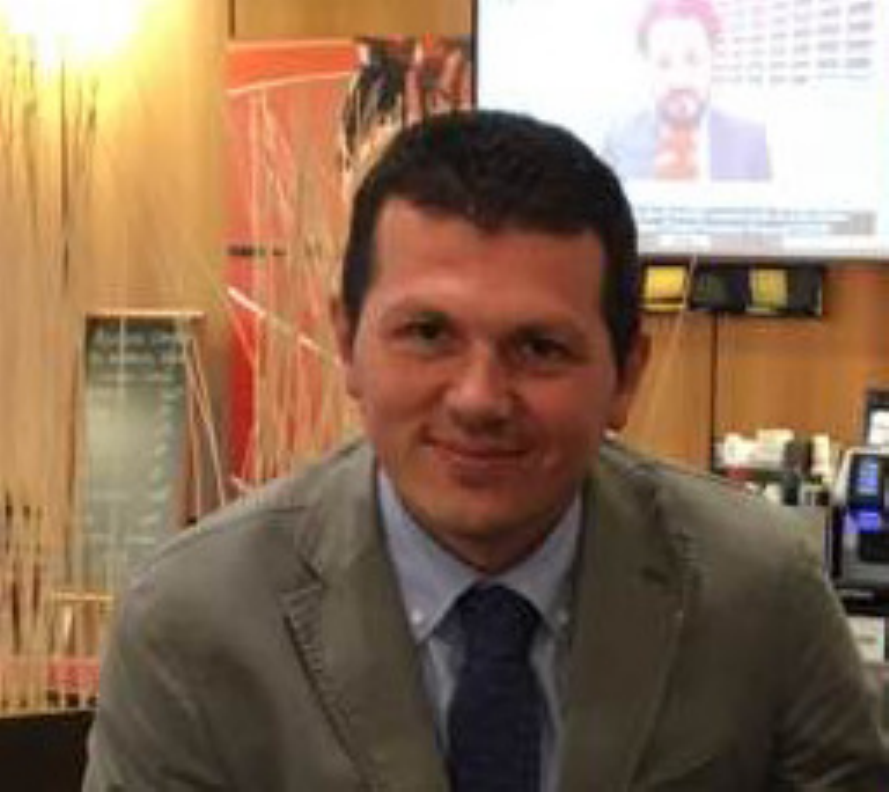}}]{Marcello Caleffi} (M'12, SM'16) received the M.S. degree with the highest score (summa cum laude) in computer science engineering from the University of Lecce, Lecce, Italy, in 2005, and the Ph.D. degree in electronic and telecommunications engineering from the University of Naples Federico II, Naples, Italy, in 2009.  Currently, he is Associate professor at the DIETI Department, University of Naples Federico II. From 2010 to 2011, he was with the Broadband Wireless Networking Laboratory at Georgia Institute of Technology, Atlanta, as visiting researcher. In 2011, he was also with the NaNoNetworking Center in Catalunya (N3Cat) at the Universitat Politecnica de Catalunya (UPC), Barcelona, as visiting researcher. Since July 2018, he held the Italian national habilitation as \textit{Full Professor} in Telecommunications Engineering. His work appeared in several premier IEEE Transactions and Journals, and he received multiple awards, including \textit{best strategy} award, \textit{most downloaded article} awards and \textit{most cited article} awards. Currently, he serves as \textit{associate technical editor} for IEEE Communications Magazine and as \textit{associate editor} for IEEE Trans. on Quantum Engineering and IEEE Communications Letters. He served as Chair, TPC Chair, Session Chair, and TPC Member for several premier IEEE conferences. In 2016, he was elevated to IEEE Senior Member and in 2017 he has been appointed as Distinguished Lecturer from the \textit{IEEE Computer Society}. In December 2017, he has been elected Treasurer of the Joint \textit{IEEE VT/ComSoc Chapter Italy Section}. In December 2018, he has been appointed member of the IEEE \textit{New Initiatives Committee}.
\end{IEEEbiography}

\balance

\end{document}